\documentclass[12pt,notitlepage,fleqn]{article}%
\usepackage{graphicx}
\usepackage{amsmath}
\usepackage{indentfirst}
\usepackage{amsfonts}
\usepackage{amssymb}%
\newtheorem{problem}{Hypothesis}

\begin{document}

\begin{center}
{\Huge The Body Center Cubic Quark Lattice Model}

\bigskip

\bigskip

{\normalsize Jiao Lin Xu}

{\small The Center for Simulational Physics, The Department of Physics and
Astronomy}

{\small University of Georgia, Athens, GA 30602, USA}

E- mail: {\small \ Jxu@Hal.Physast.uga.edu}

\bigskip

\textbf{Abstract}
\end{center}

{\small \ The Standard Model while successful in many ways is incomplete
\cite{S-Model}; many questions remain. The origin of quark masses and
hadronization of quarks are awaiting an answer. From the Dirac sea concept, we
infer that two kinds of elementary quarks (u(0) and d(0)) constitute a body
center cubic (BCC) quark lattice with a lattice constant \textit{a }%
}${\small \leq}${\small 10}$^{-18}${\small m} {\small in the vacuum. Using
energy band theory and the BCC quark lattice, we can deduce the rest masses
and the intrinsic quantum numbers (I, S, C, b and Q) of quarks. With the quark
spectrum, we deduce a baryon spectrum. The theoretical spectrum is in
agreement well with the experimental results.} {\small Not only will this
paper provide a physical basis for the Quark Model, but also it will open a
door to study the more fundamental nature at distance scales }$\leq
${\small 10}$^{-18}${\small m. This paper predicts some new quarks u}$_{c}%
${\small (6490) and d}$_{b}${\small (9950) , and new baryons }$\Lambda_{c}%
^{+}${\small (6500), }$\Lambda_{\text{b}}^{0}${\small (9960). }



\section{Introduction}

The Standard Model of particle physics has been enormously successful in
explaining and predicting a wide range of phenomena. For example, the
discovery of weak vector bosons W$^{\pm},$ Z$^{0}$ at the predicted masses and
so forth. In spite of the successes, the Standard Model is incomplete
\cite{S-Model}, and many questions remain. One is how many quarks are there?
QCD does not throw any light on how many quark flavors there should be. The
origin of quark masses is unknown. The rigorous basis of confinement and
hadronization of quarks are other questions which await answers
\cite{Fayyazuddin}, \cite{Confinement}. The minimal version of the Standard
Model has 19 arbitrary parameters. This high degree of arbitrariness suggests
that a more fundamental theory underlies the Standard Model. As M. K.
Gaillard, P. D. Grannis and F. J \ Sciulli have pointed out \ \cite{S-Model}:

\begin{quotation}
\textquotedblleft We do not expect the Standard Model to be valid at
arbitrarily short distances. However, its remarkable success strongly suggests
that the Standard Model will remain an excellent approximation to nature at
distance scales as small as 10$^{-18}$ m\textquotedblright\ .
\end{quotation}

Thus, from the Dirac sea concept \cite{D-Sea}, we have proposed a accompanying
excitation concept. This concept can give a rigorous basis for the confinement
of quarks \cite{Confine}. Then, from the Dirac sea concept, we have infered
(see Appendix A) that two kinds elementary quarks (u(0) and d(0)) constitute a
body center cubic quark lattice \cite{BCC} with a lattice constant\ a $\leq$
10$^{-18}$m \cite{S-Model} in the vacuum (the BCC Quark Lattice Model). Using
only two kinds of elementary quarks (u(0) and d(0)), we have deduced the rest
masses and the intrinsic quantum numbers (including I, S, C , b and Q) of the
ground quarks (u,d, s, c and b) \cite{Q-Mass}. With the sum law, we found all
important baryons that agree well with the experimental results
\cite{Baryon02}.\ 

In terms of the BCC Quark Lattice Model, using only two kinds elementary
quarks (u(0) and d(0)), we will deduce the rest masses and the intrinsic
quantum numbers (I, S, C, b and Q) of the low energy quark spectrum first.
Then, using the sum laws, we deduce the baryon spectrum that agrees well with
the experimental results in this paper. Not only will this paper answer
\textquotedblleft the origin of quark masses\textquotedblright\ and
\textquotedblleft hadronization of quarks\textquotedblright\ (baryon part, we
can deduce the meson spectrum also-see our next paper), two important problems
not answered by the Standard Model, but also it will provide a door to study
the nature at distance scales $\leq$ 10$^{-18}$ m and to look for a more
fundamental theory.

\section{Fundamental Hypotheses}

\begin{problem}
: There are only two kinds of elementary quarks, u(0) and d(0), with S=C=b=0
in the vacuum state. There are super strong attractive interactions among the
quarks (colors). These forces make and hold an infinite body center cubic
(BCC) quark lattice with a periodic constant \textit{a} $\leq$ 10$^{-18}m$ in
the vacuum. (see Appendix A)
\end{problem}

\begin{problem}
: Due to the effect of the vacuum quark lattice, fluctuations of energy
$\varepsilon${\LARGE \ }and intrinsic quantum numbers (such as the Strange
number $S$) of an excited quark exist. The fluctuation of the Strange number
is always $\Delta$S = $\pm$1 \cite{RealS}.
\end{problem}

\ \ \ \ \ \ \ \ \ \ \ \ \ \ \ \ \ \ \ \ \ \ \ \ \ \ \ \ \ \ \ \ \ \ \ \ \ \ \ \ \ \ \ \ \ \ \ \ \ \ \ \ \ \ \ \ \ \ \ \ \ \ \ \ \ \ \ \ \ \ \ \ \ \ \ \ \ \ \ \ \ \ \ \ \ \ \ \ \ \ \ \ \ \ \ \ \ \ \ \ \ \ \ \ \ \ \ \ \ \ \ \ \ \ \ \ \ \ \ \ \ \ \ \ \ \ \ \ \ \ \ \ \ \ \ \ \ \ \ \ \ \ \ \ \ \ \ \ \ \ \ \ \ \ \ \ \ \ \ \ \ \ \ \ \ \ \ \ \ \ \ \ \ \ \ \ \ \ \ \ \ \ \ \ \ \ \ \ \ \ \ \ \ \ \ \ \ \ \ \ \ \ \ \ \ \ \ \ \ \ \ \ \ \ \ \ \ \ \ \ \ \ \ \ \ \ \ \ \ \ \ \ \ \ \ \ \ \ \ \ \ \ \ \ \ \ \ \ \ \ \ \ \ \ \ \ \ \ \ \ \ \ \ \ \ \ \ \ \ \ \ \ \ \ \ \ \ \ \ \ \ \ \ \ \ \ \ \ \ \ \ \ \ \ \ \ \ \ \ \ \ \ \ \ \ \ \ \ \ \ \ \ \ \ \ \ \ \ \ \ \ \ \ \ \ \ \ \ \ \ \ \ \ \ \ \ \ \ \ \ \ \ \ \ \ \ \ \ \ \ \ \ \ \ \ \ \ \ \ \ \ \ \ \ \ \ \ \ \ \ \ \ \ \ \ \ \ \ \ \ \ \ \ \ \ \ \ \ \ \ \ \ \ \ \ \ \ \ \ \ \ \ \ \ \ \ \ \ \ \ \ \ \ \ \ \ \ \ \ \ \ \ \ \ \ \ \ \ \ \ \ \ \ \ \ \ \ \ \ \ \ \ \ \ \ \ \ \ \ \ \ \ \ \ \ \ \ \ \ \ \ \ \ \ \ \ \ \ \ \ \ \ \ \ \ \ \ \ \ \ \ \ \ \ \ \ \ \ \ \ \ \ \ \ \ \ \ \ \ \ \ \ \ \ \ \ \ \ \ \ \ \ \ \ \ \ \ \ \ \ \ \ \ \ \ \ \ \ \ \ \ \ \ \ \ \ \ \ \ \ \ \ \ \ \ \ \ \ \ \ \ \ \ \ \ \ \ \ \ \ \ \ \ \ \ \ \ \ \ \ \ \ \ \ \ \ \ \ \ \ \ \ \ \ \ \ \ \ \ \ \ \ \ \ \ \ \ \ \ \ \ \ \ \ \ \ \ \ \ \ \ \ \ \ \ \ \ \ \ \ \ \ \ \ \ \ \ \ \ \ \ \ \ \ \ \ \ \ \ \ \ \ \ \ \ \ \ \ \ \ \ \ \ \ \ \ \ \ \ \ \ \ \ \ \ \ \ \ \ \ \ \ \ \ \ \ \ \ \ \ \ \ \ \ \ \ \ \ \ \ \ \ \ \ \ \ \ \ \ \ \ \ \ \ \ \ \ \ \ \ \ \ \ \ \ \ \ \ \ \ \ \ \ \ \ \ \ \ \ \ \ \ \ \ \ \ \ \ \ \ \ \ \ \ \ \ \ \ \ \ \ \ \ \ \ \ \ \ \ \ \ \ \ \ \ \ \ \ \ \ \ \ \ \ \ \ \ \ \ \ \ \ \ \ \ \ \ \ \ \ \ \ \ \ \ \ \ \ \ \ \ \ \ \ \ \ \ \ \ \ \ \ \ \ \ \ \ \ \ \ \ \ \ \ \ \ \ \ \ \ \ \ \ \ \ \ \ \ \ \ \ \ \ \ \ \ \ \ \ \ \ \ \ \ \ \ \ \ \ \ \ \ \ \ \ \ \ \ \ \ \ \ \ \ \ \ \ \ \ \ \ \ \ \ \ \ \ \ \ \ \ \ \ \ \ \ \ \ \ \ \ \ \ \ \ \ \ \ \ \ \ \ \ \ \ \ \ \ \ \ \ \ \ \ \ \ \ \ \ \ \ \ \ \ \ \ \ \ \ \ \ \ \ \ \ \ \ \ \ \ \ \ \ \ \ \ \ \ \ \ \ \ \ \ \ \ \ \ \ \ \ \ \ \ \ \ \ \ \ \ \ \ \ \ \ \ \ \ \ \ \ \ \ \ \ \ \ \ \ \ \ \ \ \ \ \ \ \ \ \ \ \ \ \ \ \ \ \ \ \ \ \ \ \ \ \ \ \ \ \ \ \ \ \ \ \ \ \ \ \ \ \ \ \ \ \ \ \ \ \ \ \ \ \ \ \ \ \ \ \ \ \ \ \ \ \ \ \ \ \ \ \ \ \ \ \ \ \ \ \ \ \ \ \ \ \ \ \ \ \ \ \ \ \ \ \ \ \ \ \ \ \ \ \ \ \ \ \ \ \ \ \ \ \ \ \ \ \ \ \ \ \ \ \ \ \ \ \ \ \ \ \ \ \ \ \ \ \ \ \ \ \ \ \ \ \ \ \ \ \ \ \ \ \ \ \ \ \ \ \ \ \ \ \ \ \ \ \ \ \ \ \ \ \ \ \ \ \ \ \ \ \ \ \ \ \ \ \ \ \ \ \ \ \ \ \ \ \ \ \ \ \ \ \ \ \ \ \ \ \ \ \ \ \ \ \ \ \ \ \ \ \ \ \ \ \ \ \ \ \ \ \ \ \ \ \ \ \ \ \ \ \ \ \ \ \ \ \ \ \ \ \ \ \ \ \ \ \ \ \ \ \ \ \ \ \ \ \ \ \ \ \ \ \ \ \ \ \ \ \ \ \ \ \ \ \ \ \ \ \ \ \ \ \ \ \ \ \ \ \ \ \ \ \ \ \ \ \ \ \ \ \ \ \ \ \ \ \ \ \ \ \ \ \ \ \ \ \ \ \ \ \ \ \ \ \ \ \ \ \ \ \ \ \ \ \ \ \ \ \ \ \ \ \ \ \ \ \ \ \ \ \ \ \ \ \ \ \ \ \ \ \ \ \ \ \ \ \ \ \ \ \ \ \ \ \ \ \ \ \ \ \ \ \ \ \ \ \ \ \ \ \ \ \ \ \ \ \ \ \ \ \ \ \ \ \ \ \ \ \ \ \ \ \ \ \ \ \ \ \ \ \ \ \ \ \ \ \ \ \ \ \ \ \ \ \ \ \ \ \ \ \ \ \ \ \ \ \ \ \ \ \ \ \ \ \ \ \ \ \ \ \ \ \ \ \ \ \ \ \ \ \ \ \ \ \ \ \ \ \ \ \ \ \ \ \ \ \ \ \ \ \ \ \ \ \ \ \ \ \ \ \ \ \ \ \ \ \ \ \ \ \ \ \ \ \ \ \ \ \ \ \ \ \ \ \ \ \ \ \ \ \ \ \ \ \ \ \ \ \ \ \ \ \ \ \ \ \ \ \ \ \ \ \ \ \ \ \ \ \ \ \ \ \ \ \ \ \ \ \ \ \ \ \ \ \ \ \ \ \ \ \ \ \ \ \ \ \ \ \ \ \ \ \ \ \ \ \ \ \ \ \ \ \ \ \ \ \ \ \ \ \ \ \ \ \ \ \ \ \ \ \ \ \ \ \ \ \ \ \ \ \ \ \ \ \ \ \ \ \ \ \ \ \ \ \ \ \ \ \ \ \ \ \ \ \ \ \ \ \ \ \ \ \ \ \ \ \ \ \ \ \ \ \ \ \ \ \ \ \ \ \ \ \ \ \ \ \ \ \ \ \ \ \ \ \ \ \ \ \ \ \ \ \ \ \ \ \ \ \ \ \ \ \ \ \ \ \ \ \ \ \ \ \ \ \ \ \ \ \ \ \ \ \ \ \ \ \ \ \ \ \ \ \ \ \ \ \ \ \ \ \ \ \ \ \ \ \ \ \ \ \ \ \ \ \ \ \ \ \ \ \ \ \ \ \ \ \ \ \ \ \ \ \ \ \ \ \ \ \ \ \ \ \ \ \ \ \ \ \ \ \ \ \ \ \ \ \ \ \ \ \ \ \ \ \ \ \ \ \ \ \ \ \ \ \ \ \ \ \ \ \ \ \ \ \ \ \ \ \ \ \ \ \ \ \ \ \ \ \ \ \ \ \ \ \ \ \ \ \ \ \ \ \ \ \ \ \ \ \ \ \ \ \ \ \ \ \ \ \ \ \ \ \ \ \ \ \ \ \ \ \ \ \ \ \ \ \ \ \ \ \ \ \ \ \ \ \ \ \ \ \ \ \ \ \ \ \ \ \ \ \ \ \ \ \ \ \ \ \ \ \ \ \ \ \ \ \ \ \ \ \ \ \ \ \ \ \ \ \ \ \ \ \ \ \ \ \ \ \ \ \ \ \ \ \ \ \ \ \ \ \ \ \ \ \ \ \ \ \ \ \ \ \ \ \ \ \ \ \ \ \ \ \ \ \ \ \ \ \ \ \ \ \ \ \ \ \ \ \ \ \ \ \ \ \ \ \ \ \ \ \ \ \ \ \ \ \ \ \ \ \ \ \ \ \ \ \ \ \ \ \ \ \ \ \ \ \ \ \ \ \ \ \ \ \ \ \ \ \ \ \ \ \ \ \ \ \ \ \ \ \ \ \ \ \ \ \ \ \ \ \ \ \ \ \ \ \ \ \ \ \ \ \ \ \ \ \ \ \ \ \ \ \ \ \ \ \ \ \ \ \ \ \ \ \ \ \ \ \ \ \ \ \ \ \ \ \ \ \ \ \ \ \ \ \ \ \ \ \ \ \ \ \ \ \ \ \ \ \ \ \ \ \ \ \ \ \ \ \ \ \ \ \ \ \ \ \ \ \ \ \ \ \ \ \ \ \ \ \ \ \ \ \ \ \ \ \ \ \ \ \ \ \ \ \ \ \ \ \ \ \ \ \ \ \ \ \ \ \ \ \ \ \ \ \ \ \ \ \ \ \ \ \ \ \ \ \ \ \ \ \ \ \ \ \ \ \ \ \ \ \ \ \ \ \ \ \ \ \ \ \ \ \ \ \ \ \ \ \ \ \ \ \ \ \ \ \ \ \ \ \ \ \ \ \ \ \ \ \ \ \ \ \ \ \ \ \ \ \ \ \ \ \ \ \ \ \ \ \ \ \ \ \ \ \ \ \ \ \ \ \ \ \ \ \ \ \ \ \ \ \ \ \ \ \ \ \ \ \ \ \ \ \ \ \ \ \ \ \ \ \ \ \ \ \ \ \ \ \ \ \ \ \ \ \ \ \ \ \ \ \ \ \ \ \ \ \ \ \ \ \ \ \ \ \ \ \ \ \ \ \ \ \ \ \ \ \ \ \ \ \ \ \ \ \ \ \ \ \ \ \ \ \ \ \ \ \ \ \ \ \ \ \ \ \ \ \ \ \ \ \ \ \ \ \ \ \ \ \ \ \ \ \ \ \ \ \ \ \ \ \ \ \ \ \ \ \ \ \ \ \ \ \ \ \ \ \ \ \ \ \ \ \ \ \ \ \ \ \ \ \ \ \ \ \ \ \ \ \ \ \ \ \ \ \ \ \ \ \ \ \ \ \ \ \ \ \ \ \ \ \ \ \ \ \ \ \ \ \ \ \ \ \ \ \ \ \ \ \ \ \ \ \ \ \ \ \ \ \ \ \ \ \ \ \ \ \ \ \ \ \ \ \ \ \ \ \ \ \ \ \ \ \ \ \ \ \ \ \ \ \ \ \ \ \ \ \ \ \ \ \ \ \ \ \ \ \ \ \ \ \ \ \ \ \ \ \ \ \ \ \ \ \ \qquad

\section{The Spectrum of the Quarks$\ $}

\subsection{The Motion Equation\ \ \ \ \ \ \ }

According to the Fundamental \textbf{Hypothesis I}, there is a body center
cubic quark lattice in the vacuum. When an excited quark (q) is moving inside
the vacuum quark lattice. It obeys the special quark Dirac equation
\cite{Q-Mass}:
\begin{equation}
(\text{i}\hbar\frac{\partial}{\partial t}\text{+i}\hbar\text{c}\overrightarrow
{\alpha}\cdot\overrightarrow{\nabla}\text{-}\beta\text{ m}_{q}\text{c}%
^{2}\text{-V}_{0}\text{)}\psi\text{(}\overrightarrow{r}\text{,t) = 0,}
\label{S-Q-Dir}%
\end{equation}
where m$_{q}$ is the bare mass \cite{BareMass} of the elementary quark u(0)
(or d(0)), V$_{0}$ is a constant in any time, any location and any reference
frame. Since $($i$\hbar\frac{\partial}{\partial t}$+ i$\hbar$c$\overrightarrow
{\alpha}\cdot\overrightarrow{\nabla}$-$\beta$m$_{q}$c$^{2}$) $\psi
$($\overrightarrow{r}$,t) = 0 is a free particle Dirac equation and V$_{0}$ is
only a constant, the above equation (\ref{S-Q-Dir}) is Lorentz-invarient.

The purpose of this paper is to deduce the rest masses and the intrinsic
quantum numbers of the quarks. These results are independet from the reference
frame.$\ $Thus we can use the low energy approximation to deduce them.

\subsection{The Low Energy Approximation}

We use a classic approximation of the special quark Dirac equation--the
Schr\"{o}dinger equation \cite{Schrod} (our results will show that this is a
very good approximation) to deduce the rest masses and the intrinsic quantum
numbers of the quarks:
\begin{equation}
\frac{\hslash^{2}}{2m_{q}}\nabla^{2}\Psi\text{+(}{\huge \varepsilon}%
\text{-V}_{0}\text{)}\Psi=0, \label{Schrod}%
\end{equation}
where V$_{0}$ is an average field approximation of the strong interaction
periodic field\ of the vacuum quark lattice with body center cubic periodic
symmetries and $m_{q}$ is the bare mass \cite{BareMass} of the elementary
quark u(0) (or d(0)) in the pure vacuum. This Schr\"{o}dinger equation is a
strong interactions motion equantion of a elementary quark u(0) (or d(0)) with
body center cubic (BCC) periodic symmetries.

\subsection{The Free Particle Approximation}
The solution of the above equation (\ref{Schrod}) is a free particle
plane wave function of the u(0)-quark or the d(0)-quark:%

\begin{equation}
\varepsilon_{\overrightarrow{k}}\text{ =}\frac{\text{(}h/2\pi\text{)}^{2}%
}{2m_{q}}\text{(k}_{x}^{2}\text{+k}_{y}^{2}\text{+k}_{z}^{2}\text{)+V}_{0},
\label{Free-E}%
\end{equation}%
\begin{equation}
\Psi_{k}=e^{(i\overrightarrow{\text{k}}\cdot\overrightarrow{r})}%
,\ \overrightarrow{\text{k}}=\text{(k}_{x}\text{,k}_{y}\text{,k}_{z}\text{).}
\label{F-Wave}%
\end{equation}
They represent running waves and carry momentum $\overrightarrow{\text{P}}$ =
$\hslash\overrightarrow{\text{k}}$ of the excited quarks (u or d) with S = C =
b =0 since u(0) and d(0) have S = C = b =0. When $\overrightarrow{\text{k}}$=
(k$_{x}$,k$_{y}$,k$_{z}$) = 0, from (\ref{Free-E}), we get the rest masses of
excited state u (or d) of the elementary quark u(0) (or d(0))
\[
\text{m}_{u}\text{ = m}_{d}\text{ = }\frac{\text{V}_{0}}{\text{c}^{2}}.
\]
Using the accompanying excited concept \cite{Confine} and the masses of p and
n, we get%

\begin{align}
\text{M}_{\text{p}}  &  =\text{m}_{u}\text{+m}_{u\text{'}}\text{+m}%
_{d\text{'}}\nonumber\\
&  \thickapprox\text{M}_{\text{n}}\text{=m}_{d}\text{+m}_{u\text{'}}%
\text{+m}_{d\text{'}}\nonumber\\
&  \text{= }\text{940 Mev/c.} \label{MpMn}%
\end{align}
From \cite{Confine} and \cite{Q-Mass02}
\begin{equation}
\text{m}_{u\text{'}}\text{ = 3 Mev and m}_{d\text{'}}\text{ = 7Mev,}
\label{Mu'd'}%
\end{equation}
we have
\begin{equation}
\text{m}_{u}\text{C}^{2}\text{ =m}_{d}\text{C}^{2}\text{ = V}_{0}\text{ = 930
Mev.} \label{Vo}%
\end{equation}
Thus the u-quark has%
\begin{equation}
\text{S = C = b =0, I = I}_{z}\text{ = s = }\frac{1}{2}\text{, Q=}\frac{2}%
{3}\text{, m = 930Mev,}\ \ \label{u}%
\end{equation}
the d-quark has%
\begin{equation}
\text{S = C = b = 0, I = -I}_{z}\text{= s = }\frac{1}{2}\text{, Q = }\frac
{-1}{3}\text{, m = 930Mev.} \label{d}%
\end{equation}

The flavored particles are the tracer particles of \ the BCC quark lattice \cite{Q-Mass}.
In order to deduce the rest masses of the flavored quarks, we have to consider
the BCC symmetreries.

\subsection{The Energy Bands}

According to the energy band theory \cite{E-Band}, considing the BCC
symmetries, the wave functions of the Schr\"{o}dinger equation have to satisfy
the body center cubic periodic symmetries. The solution will divide into
energy bands. If we take
\[
\vec{k}\text{ = (2}\pi\text{/a)(}\xi\text{, }\eta\text{, }\zeta\text{),}%
\]
the energy of the Schr\"{o}dinger equation is \cite{Q-Mass}
\begin{equation}
\varepsilon\text{(}\vec{k}\text{,}\vec{n}\text{) =V}_{0}\text{+}%
\alpha\text{E(}\vec{k}\text{,}\vec{n}\text{),} \label{mass}%
\end{equation}%
\begin{equation}
\alpha\text{ = h}^{2}\text{/2m}_{q}\text{a}^{2}\text{,} \label{Alphar}%
\end{equation}%
\begin{equation}
\text{E(}\vec{k}\text{,}\vec{n}\text{) = (n}_{1}\text{-}\xi\text{)}%
^{2}\text{+(n}_{2}\text{-}\eta\text{)}^{2}\text{+(n}_{3}\text{-}\zeta
\text{)}^{2}\text{.} \label{Energy}%
\end{equation}
and the plane wave function is%

\begin{equation}
\Psi_{\vec{k}}\text{(}\vec{r}\text{) = }\exp\text{\{(- i2}\pi\text{/\textit{a}%
)[(n}_{1}\text{-}\xi\text{)x+(n}_{2}\text{-}\eta\text{)y+(n}_{3}\text{-}%
\zeta\text{)z]\},} \label{P-Wave}%
\end{equation}
where \textit{a} is the periodic constant of the quark lattice, \textit{a
}$\leq$ 10$^{-18}$m; and n$_{1}$, n$_{2}$ and n$_{3}$ are integers n$_{1}$ =
l$_{2}$+l$_{3}$, n$_{2}$ = l$_{3}$+l$_{1}$, n$_{3}$ = l$_{1}$+l$_{2},$
\begin{equation}%
\begin{tabular}
[c]{l}%
l$_{1}$ =1/2( -n$_{1}$+n$_{2}$+n$_{3}$),\\
l$_{2}$ =1/2( +n$_{1}$-n$_{2}$+n$_{3}$),\\
l$_{3}$ =1/2( +n$_{1}$+n$_{2}$-n$_{3}$),
\end{tabular}
\ \ \ \ \ \label{L-N}%
\end{equation}
satisfying the condition that only those values of $\overrightarrow{n}$ =
(n$_{1}$, n$_{2}$, n$_{3}$) are allowed, which make $\overrightarrow{l}$ =
(l$_{1}$, l$_{2}$, l$_{3}$\ ) an intergral vector \cite{L-Value}. Condition
(\ref{L-N}) implies that the vector $\vec{n}=(n_{1},n_{2},n_{3})$ can only
take certain values. For example, $\vec{n}$ cannot take $(0,0,1)$ or $\left(
1,1,-1\right)  $, but can take $(0,0,2)$ and $(1,-1,2)$. This is a resut of
BCC symmetries.

\subsection{How to find the energy bands}

Now we will demonstrate how to find the energy bands.

The first Brillouin zone \cite{Brillouin} of the body center cubic lattice is
shown in Fig.1. (depicted from \cite{E-Band} (Fig.1)) and \cite{Brillouin}
(Fig. 8. 10).

\begin{figure}[h]
\vspace{4.8in}
\includegraphics{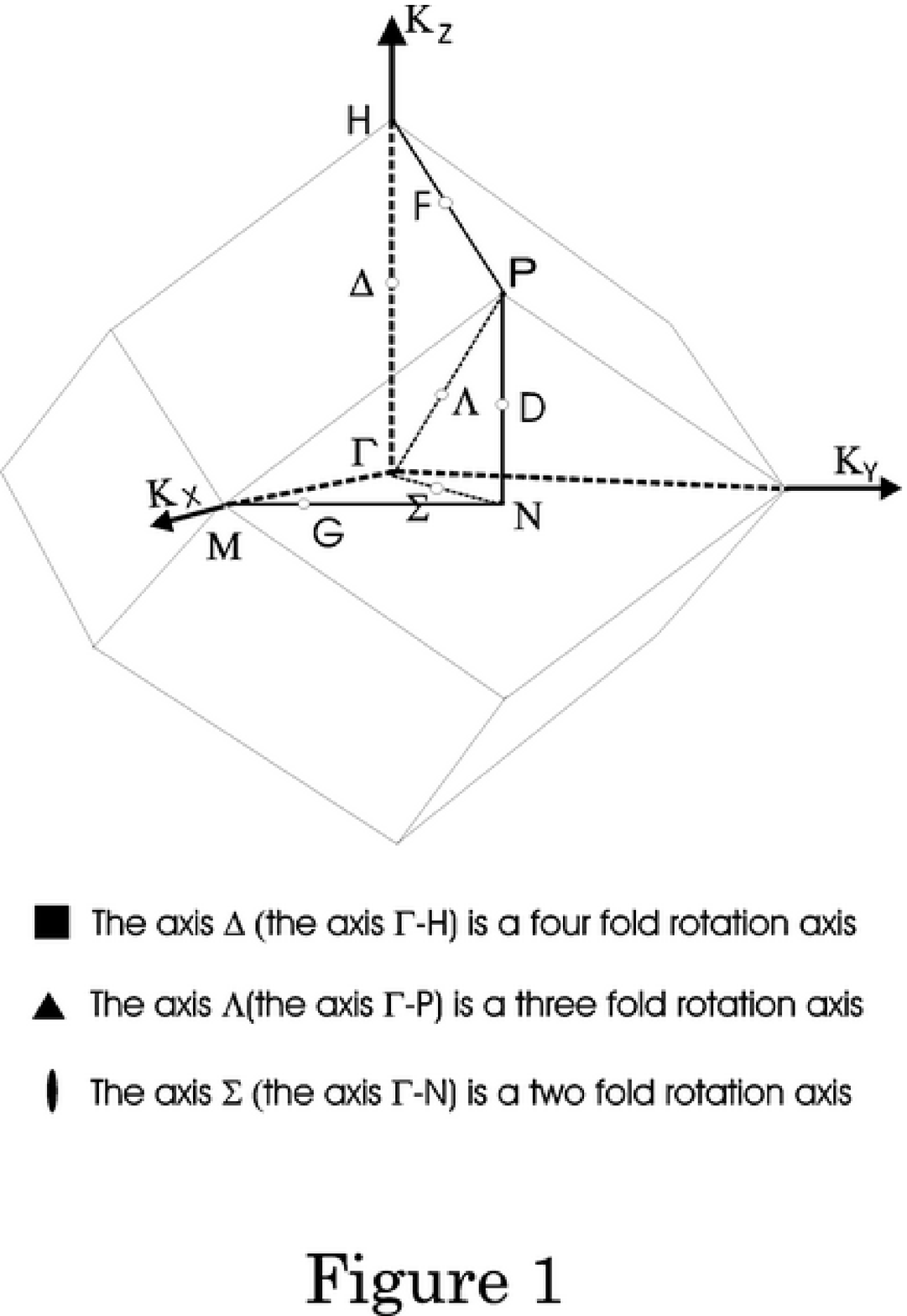}
\label{Fig1}
\caption{\small The first Brillouin zone of the body center cubic lattice.
The symmetry points and axes are indicated. The$\Delta$ -axis is a
four-fold rotation axis, S = 0, the quark family
q$_{\Delta}$ will be born on the axis.
The axes $\Lambda$ and F are three-fold rotation axes, 
S =\ -1, the quark family q$_{\Sigma}$ 
will be born on the axes. The axes $\Sigma$ and G are two-fold
rotation axes, S = -2, the quark family q$_{\Xi}$ will be born on the
axes. The D-axis is parallel to the axis $\Delta$ , S = 0. The
axis is a two-fold rotation axis, the quark family q$_{N}$ will be born on the axis. }
\end{figure}

In Fig.1, the ($\xi$, $\eta$, $\varsigma$) coordinates of the symmetry points
are:
\begin{gather}
\Gamma\text{ = (0, 0, 0), H = (0, 0, 1), P = (1/2, 1/2, 1/2),}\nonumber\\
\text{N = (1/2, 1/2, 0), M = (1, 0, 0).} \label{Points}%
\end{gather}
and the ($\xi$, $\eta$, $\zeta$) coordinates of the symmetry axes are:
\begin{align}
\Delta\text{ }  &  \text{= (0, 0, }\zeta\text{),\ 0
$<$%
}\zeta\text{
$<$%
1; \ \ \ \ \ \ \ \ \ \ }\Lambda\text{ = (}\xi\text{, }\xi\text{, }\xi\text{),
0
$<$%
}\xi\text{
$<$%
1/2;}\nonumber\\
\Sigma\text{{}}  &  \text{= (}\xi\text{, }\xi\text{, 0), 0
$<$%
}\xi\text{
$<$%
1/2; \ \ \ \ \ \ \ D = (1/2, 1/2, }\xi\text{), 0
$<$%
}\xi\text{
$<$%
1/2;}\nonumber\\
\text{G}  &  \text{= (}\xi\text{, 1-}\xi\text{, 0), 1/2
$<$%
}\xi\text{
$<$%
1; \ \ \ F = (}\xi\text{, }\xi\text{, 1-}\xi\text{), 0
$<$%
}\xi\text{
$<$%
1/2.} \label{Axes}%
\end{align}

For any valid value of the vector $\vec{n}$, substituting the ($\xi$, $\eta$,
$\zeta$) coordinates of the symmetry points (\ref{Points}) or the symmetry
axes\ (\ref{Axes}) into Eq. (\ref{Energy}) and Eq. (\ref{P-Wave}), we can get
the E($\vec{k}$, $\vec{n}$) values and the wave functions at the symmetry
points and on the symmetry axes. In order to show how to calculate the energy
bands, we give the calculation of some low energy bands in the symmetry axis
$\Delta$ as an example (the results are illustrated in Fig. 2(a)).

First, from (\ref{Energy}) and (\ref{P-Wave}) we find the formulae for the
E($\vec{k}$, $\vec{n}$) values and the wave functions at the end points
$\Gamma$ and H of the symmetry axis $\Delta$, as well as on the symmetry axis
$\Delta$ it self:
\begin{equation}
\text{E}_{\Gamma}\text{ = n}_{1}^{2}\text{ + n}_{2}^{2}\text{ + n}_{3}%
^{2}\text{,} \label{Gama}%
\end{equation}%
\begin{equation}
\Psi_{\Gamma}=\exp\text{\{-i(2}\pi\text{/a}_{x}\text{)[n}_{1}\text{x+n}%
_{2}\text{y+n}_{3}\text{z]\}.} \label{Ga-W}%
\end{equation}%
\begin{equation}
\text{E}_{\text{H}}\text{ = n}_{1}^{2}\text{ + n}_{2}^{2}\text{ + (n}%
_{3}\text{-1)}^{2}, \label{H-E}%
\end{equation}%
\begin{equation}
\Psi_{\text{H}}\text{ = }\exp\text{\{-i(2}\pi\text{/a}_{x}\text{)[n}%
_{1}\text{x+n}_{2}\text{y+(n}_{3}\text{-1)z]\}.} \label{H-W}%
\end{equation}%
\begin{equation}
\text{E}_{\Delta}\text{= n}_{1}^{2}\text{ + n}_{2}^{2}\text{ + (n}_{3}%
\text{-}\zeta\text{)}^{2}\text{,} \label{Dalta-E}%
\end{equation}%
\begin{equation}
\Psi_{\Delta}\text{= }\exp\text{\{-i(2}\pi\text{/a}_{x}\text{)[n}%
_{1}\text{x+n}_{2}\text{y+(n}_{3}\text{-}\zeta\text{)z]\}.} \label{Dalta-W}%
\end{equation}

Then, using (\ref{Gama})--(\ref{Dalta-W}) and beginning from the lowest
possible energy, we can obtain the corresponding integer vectors $\vec{n}$ =
(n$_{1}$, n$_{2}$, n$_{3}$) (satisfying (\ref{L-N})) and the wave functions:

1. The lowest E($\vec{k}$, $\vec{n}$) is at ($\xi$, $\eta$, $\zeta$) = 0 (the
point $\Gamma$) and with only one value of $\vec{n}$ = (0,0,0) \ \ (see
(\ref{Gama}) and (\ref{Ga-W})):
\begin{equation}
\vec{n}\text{ = (0, 0, 0), \ \ \ \ \ E}_{\Gamma}\text{ = 0, \ \ \ \ \ }%
\Psi_{\Gamma}\text{ = 1}.
\end{equation}

2. Starting from E$_{\Gamma}$=0, along the axis $\Delta$, there is one energy
band (the lowest energy band) E$_{\Delta}$ = $\zeta^{2}$, with n$_{1}$ =
n$_{2}$ = n$_{3}$ = 0 (see(\ref{Dalta-E}) and (\ref{Dalta-W})) ended at the
point E$_{H}$=1:
\begin{gather}
\vec{n}\text{ = (0, 0, 0), \ \ \ \ }\nonumber\\
\text{\ \ E}_{\Gamma}\text{ = 0}\rightarrow\text{ \ E}_{\Delta}\text{ = }%
\zeta^{2}\rightarrow\text{ \ E}_{H}\text{=1,}\nonumber\\
\Psi_{\Delta}\text{ =}\exp\text{[i(2}\pi\text{/a}_{x}\text{)(}\zeta\text{z)].
\ \ \ \ \ \ \ \ \ }%
\end{gather}

3. At the end point H of the energy band E$_{\Gamma}$ =0 $\rightarrow
\ $\ E$_{\Delta}$= $\zeta^{2}\rightarrow$ \ \ E$_{H}=1$, the energy $E_{H}=1$.
Also at point $H$, $E_{H}=1$ when $n=(\pm1,0,1)$, $(0,\pm1,1)$, and $(0,0,2)$
(see (\ref{H-E}) and (\ref{H-W})):
\begin{equation}
\text{E}_{\text{H}}\text{ = 1, \ \ }\Psi_{\text{H}}\text{ = [e}^{[\text{i(2}%
\pi\text{/a}_{x}\text{)(}\pm\text{x})]}\text{, e}^{[\text{i(2}\pi\text{/a}%
_{x}\text{)(}\pm\text{y)]}}\text{, e}^{[\text{i(2}\pi\text{/a}_{x}\text{)(}%
\pm\text{z})]}\text{].}%
\end{equation}

4. Starting from $E_{H}=1$, along the axis $\Delta$, there are three energy
bands ended at the points $E_{\Gamma}=0$, $E_{\Gamma}=2$ and $E_{\Gamma}=4 $,
respectively:
\begin{gather}
\vec{n}=(\text{0,0,0})\text{, \ \ \ }\nonumber\\
E_{H}=1\rightarrow E_{\Delta}=\zeta^{2}\rightarrow E_{\Gamma}=0,\text{
}\nonumber\\
\Psi_{\Delta}=\exp[i(2\pi/a_{x})(\zeta z)]\text{. \ \ \ \ \ \ \ \ \ }%
\end{gather}%
\begin{gather}
\vec{n}=(\text{0,0,2})\text{, \ \ }\nonumber\\
E_{H}=1\rightarrow E_{\Delta}=\text{(2-}\zeta\text{)}^{2}\rightarrow
E_{\Gamma}=4\text{,}\nonumber\\
\text{ }\Psi_{\Delta}=\exp{[i(2\pi/a_{x})(2-\zeta)z)]}\text{.
\ \ \ \ \ \ \ \ }%
\end{gather}%
\begin{gather}
\vec{n}=(\pm\text{1,0,1})(\text{0,}\pm\text{1,1})\text{, \ \ }\nonumber\\
E_{H}=1\rightarrow E_{\Delta}=\text{1+(1-}\zeta\text{)}^{2}\rightarrow
E_{\Gamma}=2\text{,}\nonumber\\
\text{ }\Psi_{\Delta}=e^{{\{-i(2\pi/a_{x})[\pm x+(1-\zeta)z]\}}},\text{
}e^{{\{-i(2\pi/a_{x})[\pm y+(1-\zeta)z]\}}}\text{. \ \ \ \ \ }%
\end{gather}

5. The energy bands with four sets of values $\vec{n}\ (\vec{n}=(\pm$1,0,1$),$
$($0,$\pm$1,1$))$ ended at $E_{\Gamma}=2$. From (\ref{Gama}), $E_{\Gamma}=2$
also when $\vec{n}$ takes other eight sets of values: $\vec{n}=(1,\pm1,0)$,
$(-1,\pm1,0)$, $(\pm1,0,-1)$ and $(0,\pm1,-1)$. Putting the $12$ sets of
$\vec{n}$ values into Eq. (\ref{Ga-W}), we can obtain $12$ plane wave
functions:
\begin{equation}
\text{E}_{\Gamma}\text{ = 2, }\Psi_{\Gamma}\text{ = [e}^{\text{i(2}%
\pi\text{/a}_{x}\text{)(}\pm\text{x}\pm\text{y})}\text{, e}^{\text{i(2}%
\pi\text{/a}_{x}\text{)(}\pm\text{y}\pm\text{z)}}\text{, e}^{\text{i(2}%
\pi\text{/a}_{x}\text{)(}\pm\text{z}\pm\text{x})}\text{].}%
\end{equation}

6. Starting from $E_{\Gamma}=2$, along the axis $\Delta$, there are three
energy bands ended at the points $E_{H}=1$, $E_{H}=3$, and $E_{H}=5$,
respectively:
\begin{equation}
\vec{n}\text{ = (}\pm\text{ 1,0,1)(0,}\pm\text{ 1,1),\ E}_{\Gamma}\text{ =
2}\rightarrow\text{ \ E}_{\Delta}\text{ = 1+ (1-}\zeta\text{)}^{2}%
\rightarrow\text{ \ E}_{\text{H}}\text{ = 1,}%
\end{equation}%
\begin{equation}
\vec{n}\text{ = (1,}\pm\text{ 1, 0)(-1,}\pm\text{ 1,0),\ E}_{\Gamma}\text{ =
2}\rightarrow\text{ \ E}_{\Delta}\text{= 2+}\zeta^{2}\rightarrow\text{
\ E}_{\text{H}}\text{ = 3,}%
\end{equation}%
\begin{equation}
\vec{n}\text{= (}\pm\text{ 1, 0, -1)(0,}\pm\text{ 1, -1),\ E}_{\Gamma}\text{ =
2}\rightarrow\text{ \ E}_{\Delta}\text{= 1+(}\zeta\text{+1)}^{2}%
\rightarrow\text{ \ E}_{H}\text{=5.}%
\end{equation}
Continuing the process, we can find the low energy bands and the corresponding
wave functions. The wave functions are not needed for this paper, so we only
show the energy bands in Fig. 2-4. There are six small figures in Fig. 2-4.
Each of them shows the energy bands in one of the six axes in Fig. 1. Each
small figure is a schematic one where the straight lines that show the energy
bands shold be parabolic curves. The numbers above the lines are the values of
$\vec{n}$ = ($n_{1}$, $n_{2}$, $n_{3}$). The numbers (deg) under the lines are
the fold numbers of the energy bands with the same energy (the free particle
approximation). The numbers beside both ends of an energy band (the
intersection of the energy band line and the vertical lines) represent the
highest and lowest E($\vec{k}$,$\vec{n}$) values (see Eq. (\ref{Energy})) of
the band.

\newpage
\begin{figure}[h]
\vspace{4.8in}
\includegraphics{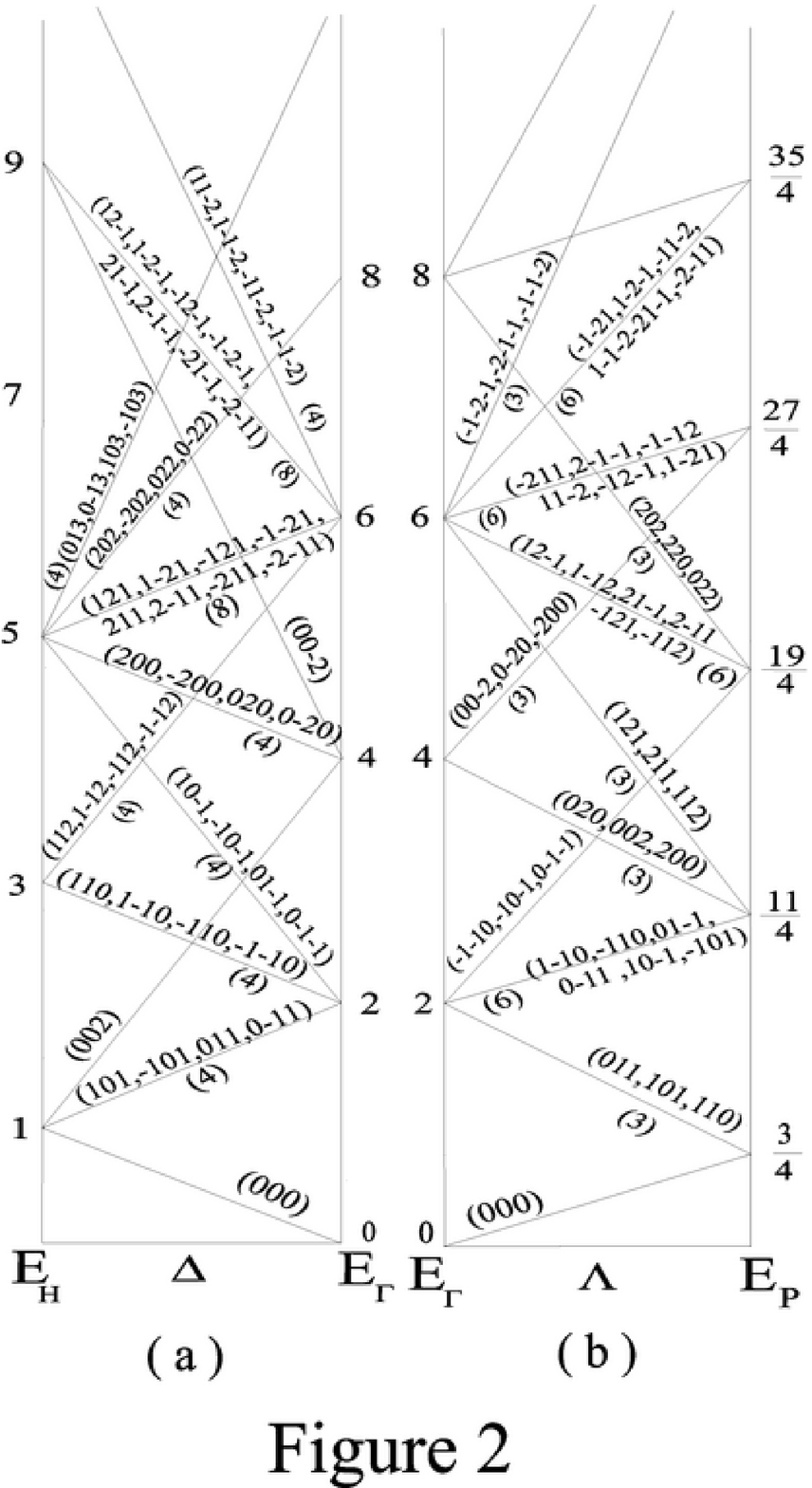}
\label{Fig2}
\caption {\small (a) The energy bands on the $\Delta$-axis. E$_{\Gamma}$ is the
value of E($\vec{k}$, $\vec{n}$) (see Eq. (\ref{Energy})) at the end point
$\Gamma,$ while E$_{H}$ is the value of E($\vec{k}$, $\vec{n}$) at other end
point H. \ \ (b) The energy bands on the $\Lambda$-axis. E$_{\Gamma}$ is the
value of E($\vec{k}$, $\vec{n}$) (see Eq. (\ref{Energy})) at the end point
$\Gamma,$ while E$_{P}$ is the value of E($\vec{k}$, $\vec{n}$) at other end
point P.}
\end{figure}

\ \ \ \ \ \ \ \ \ \ \ \ \ \ \ \ \ \ \ \ \ \ \ \ \ \ \ \ \ \ \ \ \ \
\newpage
\begin{figure}[h]
\vspace{4.8in}
\includegraphics{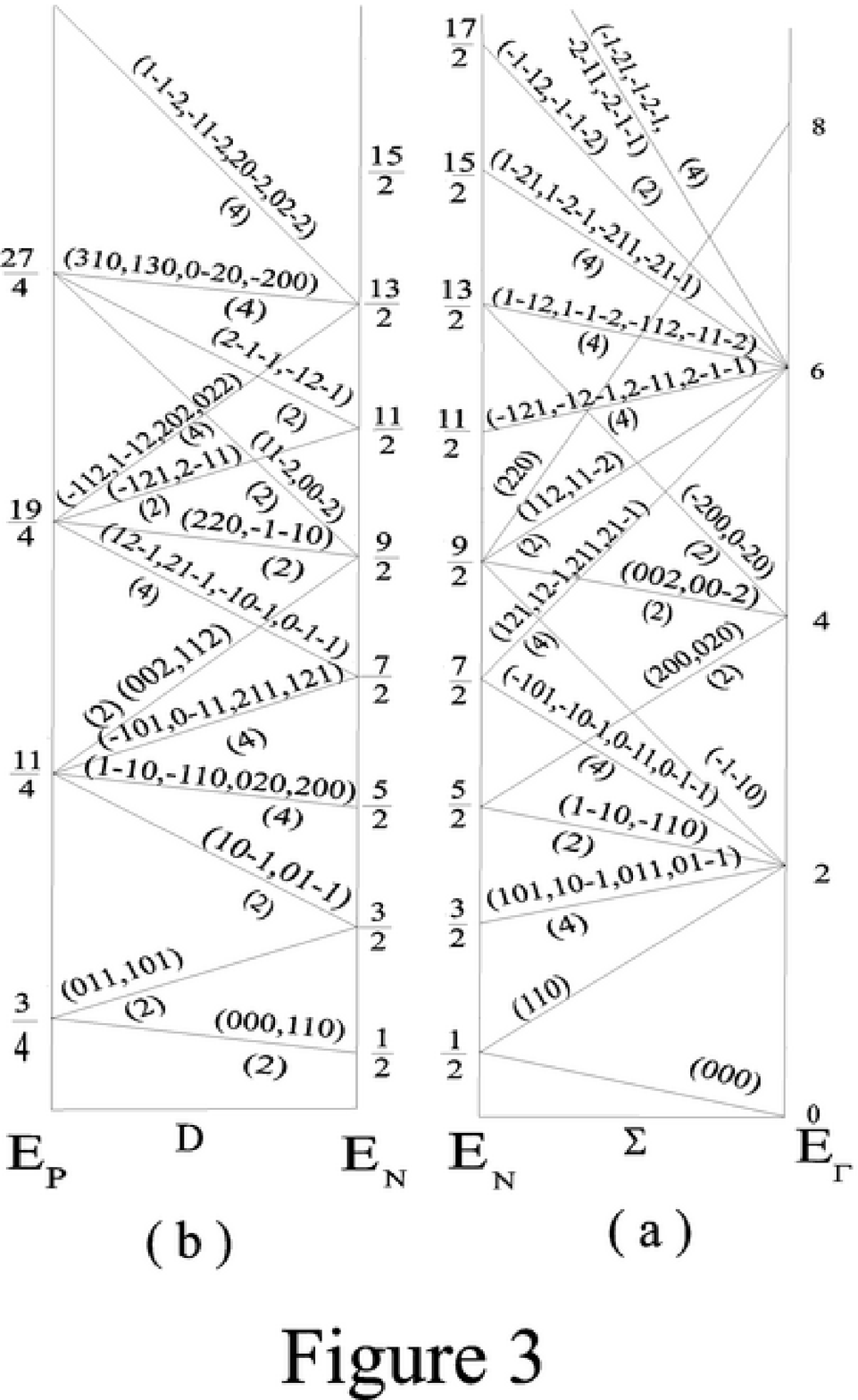}
\label{Fig3}
\caption{\ (a) The energy bands on the\ $\Sigma$-axis . E$_{\Gamma}$ is the
value of E($\vec{k}$, $\vec{n}$) (see Eq. (\ref{Energy})) at the end point
$\Gamma,$ while E$_{N}$ is the value of E($\vec{k}$, $\vec{n}$) at other end
point N. \ (b) The energy bands on the axis $D$. E$_{P}$ is the value of
E($\vec{k}$, $\vec{n}$) (see Eq. (\ref{Energy})) at the end point P$,$ while
E$_{N}$ is the value of E($\vec{k}$, $\vec{n}$) at other end point N.}
\end{figure}

\ \ \ \ \ \ \ \ \ \ \ \ \ \ \ \ \ \ \ \ \ \ \ \ \ \ \ \ \ \ \ \ \ \ \
\newpage
\begin{figure}[h]
\vspace{4.8in}
\includegraphics{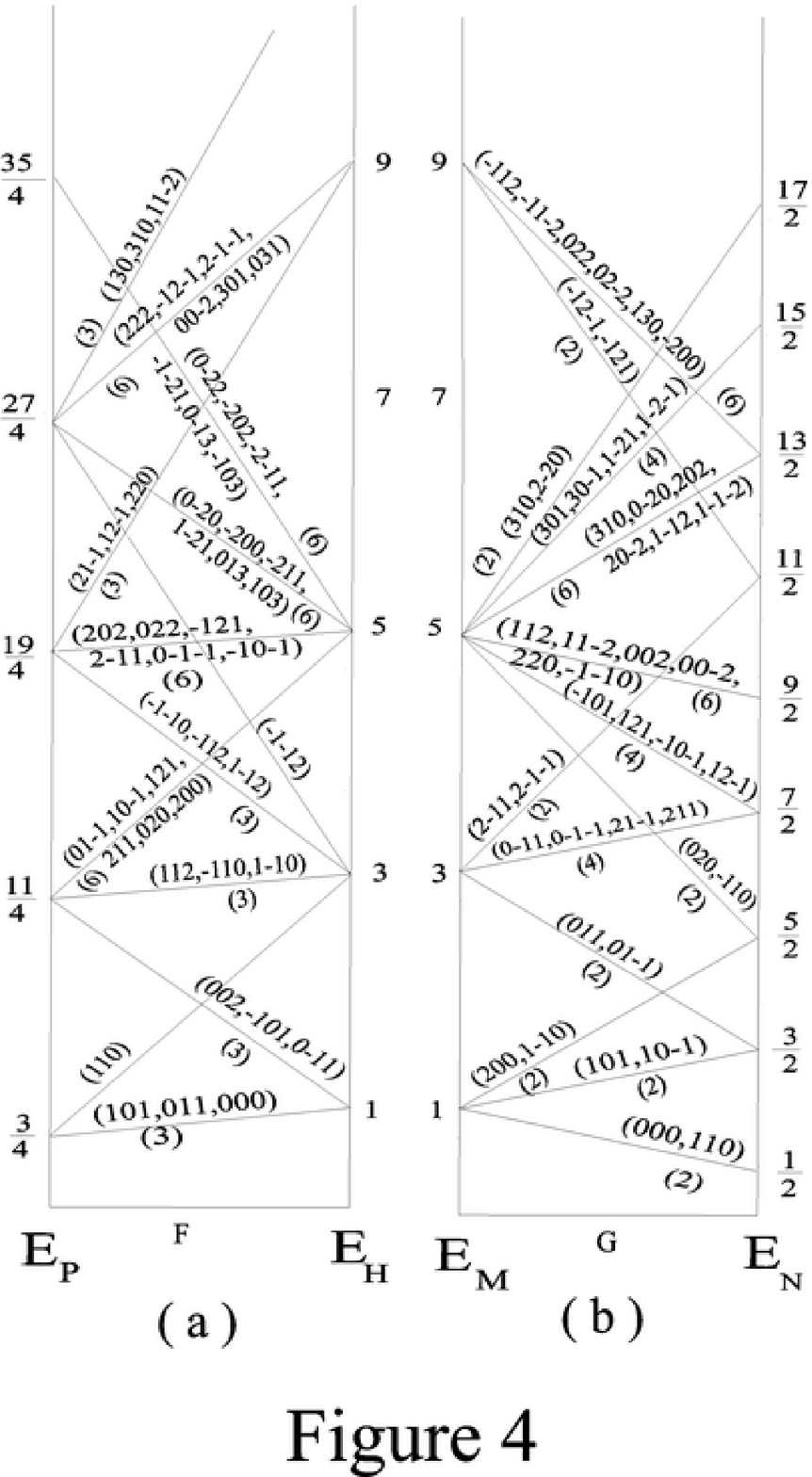}
\label{Fig4}
\caption{\ (a) The energy bands on the $F$-axis. E$_{P}$ is the value of
E($\vec{k}$, $\vec{n}$) (see Eq. (\ref{Energy})) at the end point $P,$ while
E$_{H}$ is the value of E($\vec{k}$, $\vec{n}$) at other end point H.\ \ (b)
The energy bands on the axis $G$.\ E$_{M}$ is the value of E($\vec{k}$,
$\vec{n}$) (see Eq. (\ref{Energy})) at the end point M$,$ while E$_{N}$ is the
value of E($\vec{k}$, $\vec{n}$) at other end point N.}
\end{figure}
\ \ \ \ \ \ \ \ \ \ \ \ \ \ \ \ \ \ \ \ \ \ \ \ \ \ \ \ \ \ \ \ \ \ \
\newpage
\begin{figure}[h]
\vspace{4.8in}
\includegraphics{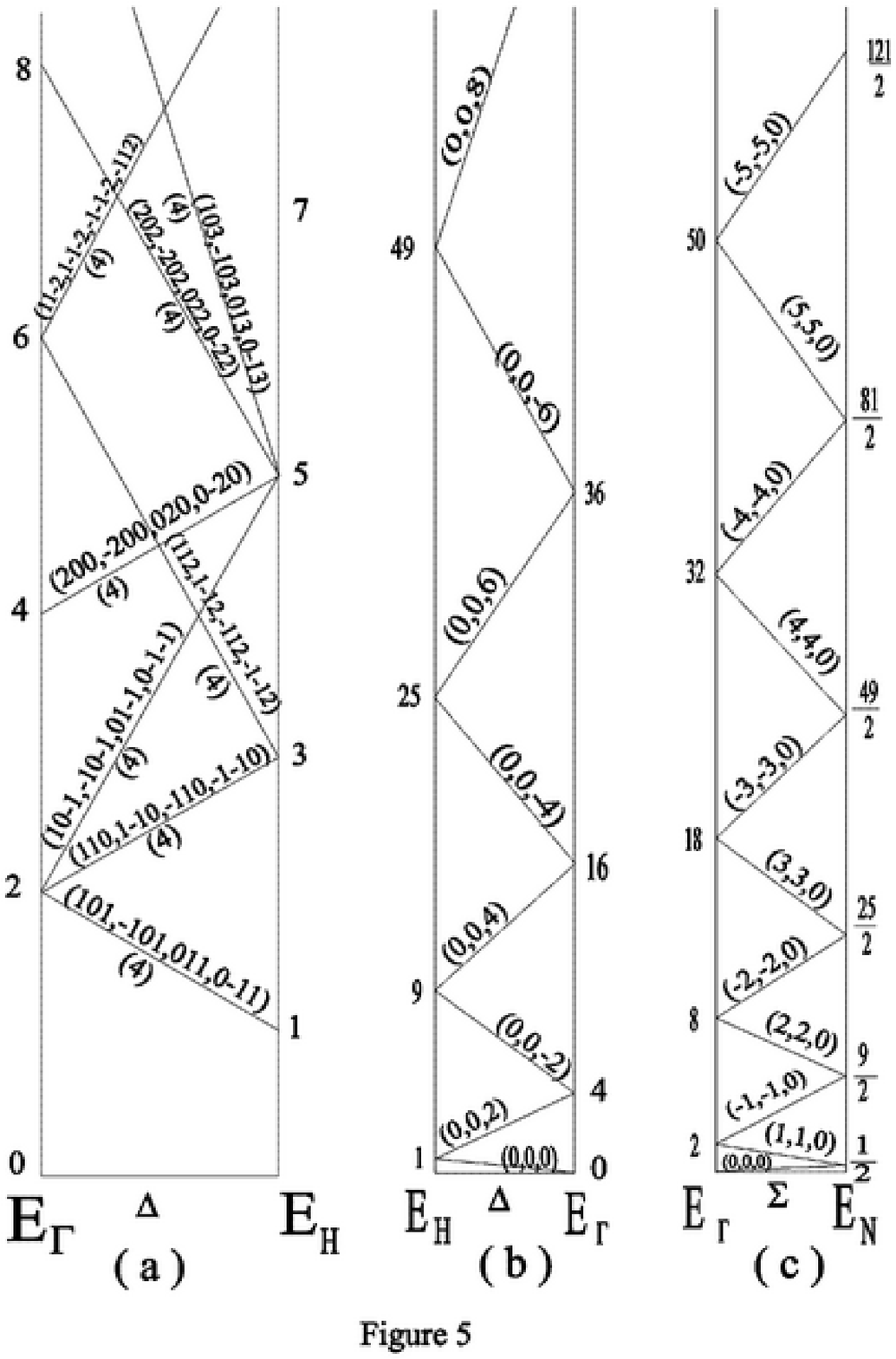}
\label{Fig5}
\caption{\ (a) The four-fold degenerate energy bands (selected from Fig.
2(a)) on the $\Delta$-axis. \ \ (b) The single energy bands (selected from
Fig. 2(a)) on the $\Delta$-axis. The numbers above the lines are the values of
$\vec{n}$ (n$_{1}$, n$_{2}$, n$_{3}$). \ \ (c) The single energy band
(selected from Fig. 3(a)) on the $\Sigma$-axis .}
\end{figure}

\bigskip

\bigskip

Putting the values of the E($\vec{k}$,$\vec{n}$) into Eq. (\ref{mass}), we get
the energy values (in Mev) in Table 7-12 of the Appendix B.

\subsection{The Recognization of the Quarks \ \ \ \ \ \ \ \ \ \ \textbf{\qquad
}}

We have already found the energy bands which are shown in Fig. 2---Fig. 5.
These energy bands represent the excited states of the elementary quarks u(0)
or d(0). They are the ground quarks and excited quarks. For the first
Brillouin zone, $\overrightarrow{n}$ = (0, 0, 0), it is a part of the free
quark solution (\ref{F-Wave}) and it has the lowest energy (mass). Thus it
represents the lowest mass u-quark (\ref{u}) and the d-quark (\ref{d}):
\begin{equation}
\overrightarrow{n}=(0,0,0),\text{ u(930) and d(930).} \label{n=000}%
\end{equation}
Now we deduce the quantum numbers and the rest masses of the energy bands.
Using the numbers and the rest masses, we can recognize the quarks.

\subsubsection{The Quantum Numbers and Energies of the Energy Bands
\textbf{\qquad}}

Now we find the formulae that we can use to deduce the quantum numbers and the
rest masses of the energy bands, as shown in the following:

1. An excited (from the vacuum) quark, q, has a baryon number
\begin{equation}
B=\frac{1}{3} \label{1/3}%
\end{equation}

2. The isospin $I$\ is determined by the energy band degeneracy deg
\cite{E-Band}, where
\begin{equation}
\text{deg = 2I + 1.} \label{IsoSpin}%
\end{equation}
If the deg
$>$%
the rotary fold $R$\ of the symmetry axis
\begin{equation}
\text{deg \
$>$
R,} \label{deg > R}%
\end{equation}
the deg will be divided $\gamma$-subdegeneracies first
\begin{equation}
\gamma\text{ = deg/R;} \label{Subdeg}%
\end{equation}
then, using (\ref{IsoSpin}), we can find the isospin values for each subdegeneracy.

The isospin I from (\ref{IsoSpin}), not only is the isospin of the q-quark
that the energy bands represent, but also is the isospin of the baryon
(qq'$_{1}$q'$_{2}$) that the energy bands represent. The components of the
isospin I:%
\begin{equation}
\text{I, I-1, I-2, ..., -I} \label{I-Component of B}%
\end{equation}
are the components of the baryon's isospin. The components of the quark are
deduced by (\ref{Iz})

3. Strange number $S$\ is determined by the rotary fold $R$\ of the symmetry
axis \cite{E-Band} with
\begin{equation}
S=R-4, \label{S-Number}%
\end{equation}
where the number $4$\ is the highest possible rotary fold number of the BCC
lattice.%
\begin{equation}
\text{For the }\Delta\text{-axis, R }\text{= 4, S = 0}, \label{S = 0}%
\end{equation}%
\begin{equation}
\text{for the }\Lambda\text{-axis, R }\text{= 3, S = -1;} \label{S = -1}%
\end{equation}%
\begin{equation}
\text{for the }\Sigma\text{-axis, R = 2, S = -2.} \label{S = - 2}%
\end{equation}
The three axes (the axes D, F and G) on the surface of the first Brillouin
zone have that the D-axis parallell the $\Delta$-axis, the the F-axis
parallell the $\Lambda$-axis and the the G-axis parallell the $\Sigma$-axis.
Thus%
\begin{equation}
\text{S}_{D}\text{ }\text{=}\text{ S}_{\Delta}\text{= 0,} \label{S(D)}%
\end{equation}%
\begin{equation}
\text{S}_{F}\text{=}\text{S}_{\Lambda}\text{= -1,} \label{S(F)}%
\end{equation}%
\begin{equation}
\text{S}_{G}\text{=}\text{S}_{\Sigma}\text{= -2.} \label{S(G)}%
\end{equation}
\qquad\qquad\qquad

4. Electric charge $Q$ of the exited quark, q, is determened completely by the
elementary quark u(0) (or d(0)) that is excited to produce the exited quark q.
Since the u(0)-quark has I$_{z}$ = 1/2
$>$
0 and the d(0)-quark has I$_{z}$ = -1/2
$<$
0, for the excited quark q with I$_{z}$
$>$
0, it is an excited state of the elementary quark u(0),\textbf{\ }
\begin{equation}
\text{Q}_{q}\text{ = Q}_{u}\text{ = +2/3;} \label{2/3}%
\end{equation}
for the excited quark q with I$_{z}$
$<$
0, it is an excited state of the elementary quark d(0),
\begin{equation}
\text{Q}_{q}\text{ = Q}_{d}\text{ = -1/3}. \label{-1/3}%
\end{equation}
For the excited quark q with I$_{z}$ = 0, considering the generalized
Gell-Mann-Nishijima relation \cite{GMN}, if \ q with S$_{G}$ (S+C+b)\
$>$
0, it is an excited state of the u(0)-quark
\begin{equation}
\text{Q}_{q}\text{ = Q}_{u}\text{ = +2/3}; \label{Q-C}%
\end{equation}
if \ q with S$_{G}$(S+C+b)\
$<$
0, it is an excited state of the d-quark
\begin{equation}
\text{Q}_{q}\text{ = Q}_{d}\text{ = -1/3}. \label{Q-D}%
\end{equation}
There is not any quark with I$_{z}$ = 0 and S$_{G}$ (S+C+b)\ = 0.

5. After we find the the electric charge Q and the strange number S (C, b) of
the excited quark q, using the generalized Gell-Mann-Nishijima relation
\cite{GMN}, we can find I$_{Z}$ of the excited quark q%
\begin{equation}
\text{I}_{z}\text{ = Q - }\frac{1}{2}\text{(B+S+C+b).} \label{Iz}%
\end{equation}
It is natural that the I$_{Z}$ of an energy band excited state of the
elementary quark u(0) (or d(0)) might be different from the I$_{Z}$ of the
elementary quark u(0) (or d(0)) in vacuum state, but their electric charges
will be the same.

6. If a degeneracy (or subdegeneracy) of a group of energy bands is smaller
than the rotary fold R
\begin{equation}
\ \deg\text{%
$<$%
\ R\ \ and\ \ R-}\deg\ \neq\text{ }2, \label{Condition}%
\end{equation}
\ then formula (\ref{S-Number}) will substituted by
\begin{equation}
\text{\={S} = R - 4.} \label{S-bar}%
\end{equation}
\ The real value of $S$\ is
\begin{equation}
\text{S = \={S} + }\Delta\text{S = S}_{Axis}\pm\text{1,} \label{S-fluc}%
\end{equation}
from Hypothesis II, $\Delta S=\pm1$. We have a\textbf{\ }formula to find
$\Delta$S,
\begin{equation}
\Delta\text{S = [1-2}\delta\text{(S)]Sign(}\vec{n}\text{),\ \ } \label{DaltaS}%
\end{equation}
where
\begin{equation}
\text{Sign(}\vec{n}\text{) = }\frac{\text{n}_{1}\text{+n}_{2}\text{+n}_{3}%
}{\left\vert \text{n}_{1}\right\vert \text{+}\left\vert \text{n}%
_{2}\right\vert \text{+}\left\vert \text{n}_{3}\right\vert }\text{.}
\label{Sign(n)}%
\end{equation}
For the $\Delta$-axis and the D-axis$,$ $\delta$(S) = 1, from (\ref{DaltaS}),
we get
\begin{equation}
\Delta\text{S = - Sign(}\vec{n}\text{).} \label{D-S-D}%
\end{equation}
For the $\Lambda$-axis, the $\Sigma$-axis, the F-axis and the G axis, $\delta
$(S) = 0, from (\ref{DaltaS}), we have
\begin{equation}
\Delta\text{S = Sign(}\vec{n}\text{).} \label{D-S-S}%
\end{equation}

7. The fluctuation of the strange number will be accompanied by an energy
change (\textbf{Hypothesis II}). We assume that the change of the energy
(perturbation energy) is proportional to (-$\Delta$S)\ and a number,
J,\ representing the energy order, as a phenomenological formula:
\begin{equation}
\Delta\varepsilon\text{ = (S+1)100(J+S)(-}\Delta\text{S),} \label{D-E}%
\end{equation}
for a single energy band, J will take 1, 2, ... from the lowest energy band to
higher ones for each of the two end points of the symmetry axes respectively. \ \ \ \ \ \ \ \ \ \ \ \ \ \ \ \ \ \ \ \ \ 

8. Charmed number $C$\ and Bottom number $b$: The \textquotedblleft Strange
number\textquotedblright, S, in (\ref{S-fluc}) is not completely the same as
the strange number in (\ref{S-Number}). In order to compare it with the
experimental results, we would like to give it a new name under certain
circumstances. Based on Hypothesis II, the new names will be the Charmed
number and the Bottom number: if S = +1, which originates from the fluctuation
$\Delta S=+1$, we call it the Charmed number C
\begin{equation}
C\text{ = +1;} \label{Charmed}%
\end{equation}
if S = -1$,$ which originates from the fluctuation $\Delta S=+1$ on a single
energy band, and there is an energy fluctuation,$\ $we call it the Bottom
number$\ b\ \ \ \ \ \ \ \ \ \ \ \ \ $%
\begin{equation}
\text{ }b=-1\text{.} \label{Battom}%
\end{equation}
Similarly, we can obtain charmed strange quarks q$_{\Xi_{C}}$\ and
q$_{\Omega_{C}}$ (Appendix C).

9. We assume that the excited quark's rest mass is the minimum energy of the
energy band that represents the excited quark: \ \ \ \ \ \ \ \ \ \ \ \ \ \ \ \ \ \ \ \ \ \ \ \ \ \ \ \ \ \ \ \ \ \ %

\begin{equation}
m(q)=\text{Minimum[V}_{0}\text{+ }\alpha\text{E( }\overrightarrow{k}%
\text{,}\overrightarrow{n}\text{)] + }\Delta\varepsilon\text{,} \label{U-Mass}%
\end{equation}
where V$_{0}$ = 930 Mev (\ref{Vo}); $\Delta\varepsilon$ is from (\ref{D-E});
fitting the energy band excited states to the experimental results (the baryon
spectrum and the meson spectrum), we find the $\alpha$ in (\ref{U-Mass})
\begin{equation}
\alpha\text{ = }\frac{\text{h}^{2}}{\text{2m}_{q}\text{a}^{2}}\text{ = 360
Mev.} \label{360}%
\end{equation}
This formula (\ref{U-Mass}) is the united mass formula that can give the
masses of all quarks.\ \ 

Using the above formulae, we can find the quark spectrum. We will start from
the $\Delta$-axis.

\subsubsection{The Quarks on the $\Delta$-Axis (the $\Gamma$-H
axis)\textbf{\qquad}}

Since the $\Delta$-axis is a four-fold rotatory axis (see Fig. 1), R = 4, from
(\ref{S-Number}), we have S = 0. Because the axis has $R=4$, from
(\ref{deg > R}) and (\ref{Subdeg}), the energy bands of degeneracy $8$ will be
divided into two four-fold degenerate bands. \ \ \ \ \ \ \ \ \ \ \ \ \ \ \ \ \ \ \ \ \ \ 

1. The four fold degenerate bands on the $\Delta$-axis ($\Gamma$-H)

For four-fold degenerate bands, using (\ref{IsoSpin}), we get $I=3/2$, and
I$_{Z\text{, B}}=3/2$, $1/2$, $-1/2$, $-3/2$ (\ref{I-Component of B}). Thus,
from (\ref{2/3}) and (\ref{-1/3}), each four-fold degenerate band represents a
four-fold quark family q$_{\Delta}$(q$_{\Delta}^{\frac{3}{2}}$, q$_{\Delta
}^{\frac{1}{2}}$, q$_{\Delta}^{\frac{-1}{2}}$, q$_{\Delta}^{\frac{-3}{2}}$)
with
\begin{equation}
\text{B = }\frac{\text{1}}{3}\text{, S = 0, I = }\frac{\text{3}}{2}\text{,
I}_{z,\text{ B}}\text{ =}\frac{\text{3}}{2}\text{, }\frac{\text{1}}{2}\text{,
-}\frac{\text{1}}{2}\text{, -}\frac{\text{3}}{2}\text{.} \label{3/2-Quark}%
\end{equation}
Using Fig.2(a) and Fig.5(a), we can get E$_{\Gamma}$, E$_{\text{H}}$, and
$\vec{n}$ values. Then, putting the values of E$_{\Gamma}$ and E$_{\text{H}}$
into the energy formula (\ref{Energy}) and (\ref{mass}), we can find
m$_{q_{\Delta}}$. Thus, we have [q$_{\Delta}$ = (q$_{\Delta}^{\frac{3}{2}}$,
q$_{\Delta}^{\frac{1}{2}}$, q$_{\Delta}^{\frac{-1}{2}}$, q$_{\Delta}%
^{\frac{-3}{2}}$)]:
\begin{equation}%
\begin{array}
[c]{lllll}%
\text{E} & \text{(n}_{1}\text{n}_{2}\text{n}_{3}\text{, ... )} &
{\small \varepsilon} & I & q_{\text{Name}}(m)\\
\text{E}_{H}\text{=1} & \text{(101,-101,011,0-11)} & \text{1290} & \text{3/2}
& \text{q}_{\Delta}\text{(1290)}\\
\text{E}_{\Gamma}\text{=2 } & \text{(110,1-10,-110,-1-10)} & \text{1650} &
\text{3/2} & \text{q}_{\Delta}\text{(1650}{\small )}\\
\text{E}_{\Gamma}\text{=2} & \text{(10-1,-10-1,01-1,0-1-1)} & \text{1650} &
\text{3/2} & \text{q}_{\Delta}\text{(1650)}\\
\text{E}_{H}\text{=3} & \text{(112,1-12,-112,-1-12)} & \text{2010} &
\text{3/2} & \text{q}_{\Delta}\text{(2010)}\\
\text{E}_{\Gamma}\text{=4} & \text{(200,-200,020,0-20)} & \text{2370} &
\text{3/2} & \text{q}_{\Delta}\text{(2370)}\\
\text{E}_{H}\text{=5} & \text{(121,1-21,-121,--1-21,} & \text{2730} &
\text{3/2} & \text{q}_{\Delta}\text{(2730)}\\
& \text{{\small \ }211,2-11,-211,-2-11)} & \text{2730} & \text{3/2} &
\text{q}_{\Delta}\text{(2730)}\\
\text{E}_{H}\text{=5} & \text{(202,-202,022,0-22)} & \text{2730} & \text{3/2}
& \text{q}_{\Delta}\text{(2730)}\\
\text{E}_{H}\text{=5} & \text{(013,0-13,103,-103)} & \text{2730} & \text{3/2}
& \text{q}_{\Delta}\text{(2730)}\\
\text{E}_{\Gamma}\text{=6} & \text{(12}\overline{\text{1}}\text{,1}%
\overline{\text{2}}\overline{\text{1}}\text{,}\overline{\text{1}}%
\text{21,}\overline{\text{1}}\overline{\text{2}}\overline{\text{1}}{\small ,}
& \text{3090} & \text{3/2} & \text{q}_{\Delta}\text{(3090)}\\
& \text{\ 21}\overline{\text{1}}\text{,2}\overline{\text{1}}\overline
{\text{1}}\text{,}\overline{\text{2}}\text{1}\overline{\text{1}}%
\text{,}\overline{\text{2}}\overline{\text{1}}\overline{\text{1}}\text{)} &
\text{3090} & \text{3/2} & \text{q}_{\Delta}\text{(3090)}\\
\text{E}_{\Gamma}\text{=6} & \text{(11}\overline{\text{2}}\text{,1}%
\overline{\text{1}}\overline{\text{2}}\text{,}\overline{\text{1}}%
\text{1}\overline{\text{2}}\text{,}\overline{\text{1}}\overline{\text{1}%
}\overline{\text{2}}\text{)} & \text{3090} & \text{3/2} & \text{q}_{\Delta
}\text{(3090).}%
\end{array}
\label{D-Quark}%
\end{equation}
\qquad\qquad

2. The single bands on the axis $\Delta$($\Gamma$-H) \cite{Q-Mass}. For the
single bands on the $\Delta$-axis, R =1, S$_{\Delta}$ = 0 from (\ref{S-Number}%
); d = 1, I =0 from (\ref{IsoSpin}). Since d = 1
$<$
R = 4 and R-d = 3 $\neq$ \ 2, according to (\ref{Condition}), we will use
(\ref{S-fluc}) instead of (\ref{S-Number}). Using (\ref{Charmed}) and
(\ref{D-E}), we have%

\begin{equation}%
\begin{array}
[c]{lllllllll}%
\text{ \ \ \ E} & \text{n}_{1,}\text{n}_{2,}\text{n}_{3} & \Delta\text{S} &
\text{ \ \ J} & \text{ }\Delta\epsilon & \text{I} & S & C & q_{\text{name}%
}\text{(m)}\\
\text{E}_{\Gamma}\text{=0} & \text{{\small 0, \ 0, \ 0}} & \text{{\small \ 0}}
& \text{J}_{\Gamma}\text{=0} & \text{{\small \ \ \ 0}} & \text{0} &
\text{{\small \ \ 0}} & \text{{\small 0}} & \text{q}_{N}\text{(930)}\\
\text{E}_{H}\text{=1} & \text{{\small 0, \ 0, \ 2}} & \text{{\small -1}} &
\text{J}_{\text{H}}\text{=1} & \text{{\small +100}} & 0 & \text{{\small \ -1}}
& \text{{\small 0}} & \text{d}_{S}\text{(1390)}\\
\text{E}_{\Gamma}\text{=4} & \text{{\small 0, 0, -2}} & \text{{\small +1}} &
\text{J}_{\Gamma}\text{=1} & \text{{\small -100}} & \text{0} &
\text{{\small \ \ 0}} & \text{{\small 1}} & \text{u}_{C}\text{(2270)}\\
\text{E}_{H}\text{=9} & \text{{\small 0, \ 0, \ 4}} & \text{{\small -1}} &
\text{J}_{\text{H}}\text{=2} & \text{{\small +200}} & \text{0} &
\text{{\small \ -1}} & \text{{\small 0}} & \text{d}_{S}\text{(4370)}\\
\text{E}_{\Gamma}\text{=16} & \text{{\small 0, \ 0, -4}} & \text{{\small +1}}
& \text{J}_{\Gamma}\text{=2} & \text{{\small -200}} & \text{0} &
\text{{\small \ \ 0}} & \text{{\small 1}} & \text{u}_{C}\text{(6490)}\\
\text{E}_{H}\text{=25} & \text{{\small 0, \ 0, \ 6}} & \text{{\small -1}} &
\text{J}_{\text{H}}\text{=3} & \text{{\small +300}} & \text{0} &
\text{{\small \ -1}} & \text{{\small 0}} & \text{d}_{S}\text{(10230)}\\
\text{E}_{\Gamma}\text{=36} & \text{{\small 0, \ 0, -6}} & \text{{\small +1}}
& \text{J}_{\Gamma}\text{=3} & \text{{\small -300}} & \text{0} &
\text{{\small \ \ 0}} & \text{{\small 1}} & \text{u}_{C}\text{(13590)}\\
\text{E}_{H}\text{=49} & \text{{\small 0, \ 0, \ 8}} & \text{{\small -1}} &
\text{J}_{\text{H}}\text{=4} & \text{{\small +400}} & \text{0} &
\text{{\small \ -1}} & \text{{\small 0}} & \text{d}_{S}\text{(18970)}\\
\text{E}_{\Gamma}\text{=64} & \text{{\small 0, \ 0, -8}} & \text{{\small +1}}
& \text{J}_{\Gamma}\text{=4} & \text{{\small -400}} & \text{0} &
\text{{\small \ \ 0}} & \text{{\small 1}} & \text{u}_{C}\text{(23570)}\\
\text{{\small \ldots\ .}} &  &  &  &  &  &  &  &
\end{array}
\label{D-ONE}%
\end{equation}

\subsubsection{The Quarks on the Axis $\Lambda$($\Gamma$-P)}

Since the $\Lambda$-axis is a three-fold rotatory axis (see Fig. 1), R = 3,
from (\ref{S-Number}), we have S = -1. From Fig. 2(b), we see that there is a
single energy band with $\vec{n}$ = (0, 0, 0), and all other bands are
three-fold degenerate energy bands (d = 3) and six-fold degenerate bands (d =
6). From (\ref{deg > R}) and (\ref{Subdeg}), the six-fold degenerate energy
bands will be divided into two three-fold energy bands.

For the three-fold degenerate energy bands, using (\ref{IsoSpin}),
(\ref{I-Component of B}), (\ref{2/3}), (\ref{-1/3}), and (\ref{Q-D}), we have
I = 1, and I$_{z,B}$ = 1, 0, -1. Thus, we get a three-member quark family
q$_{\Sigma}$(q$_{\Sigma}^{1}$, q$_{\Sigma}^{0}$, q$_{\Sigma}^{-1}$) with B =
1/3, S = -1,\ I = 1. Similar to (\ref{D-Quark}), using Fig. 2(b), we get
\begin{equation}%
\begin{array}
[c]{ccccc}%
\text{E} & \text{n}_{1}\text{n}_{2}\text{n}_{3} & \varepsilon & \text{I} &
\text{q}_{\Sigma}\text{(m)}\\
\text{E}_{P}\text{=3/4 \ } & \text{{\small (101,011,110)}} &
\text{{\small 1200}} & \text{1} & \text{q}_{\Sigma}\text{(1200)}\\
\text{E}_{\Gamma}\text{=2 \ \ \ \ } & \text{{\small (1-10,-110,01-1,}} &
\text{{\small 1650}} & \text{1} & \text{q}_{\Sigma}\text{(1650)}\\
& \text{{\small \ \ 0-11,10-1,-101})} & \text{{\small 1650}} & \text{1} &
\text{q}_{\Sigma}\text{(1650)}\\
\text{E}_{\Gamma}\text{=2 \ \ \ \ } & \text{{\small (-10-1,0-1-1,-1-10)}} &
\text{{\small 1650}} & \text{1} & \text{q}_{\Sigma}\text{(1650)}\\
\text{E}_{P}\text{=11/4} & \text{{\small (020,002,200) \ \ \ \ \ \ }} &
\text{{\small 1920}} & \text{1} & \text{q}_{\Sigma}\text{(1920)}\\
\text{E}_{P}\text{=11/4} & \text{{\small (121,211,112) }} &
\text{{\small 1920}} & \text{1} & \text{q}_{\Sigma}\text{(1920)}\\
\text{E}_{\Gamma}\text{=4 \ \ \ \ } & \text{{\small (0-20,-200,00-2)}} &
\text{{\small 2370}} & \text{1} & \text{q}_{\Sigma}\text{(2370)}\\
\text{E}_{P}\text{=19/4} & \text{{\small (1-12,-112,21-1,}} &
\text{{\small 2640}} & \text{1} & \text{q}_{\Sigma}\text{(2640)}\\
& \text{{\small \ \ 2-11,12-1,-121))}} & \text{{\small 2640}} & \text{1} &
\text{q}_{\Sigma}\text{(2640)}\\
\text{E}_{P}\text{=19/4} & \text{{\small (202,022,220)}} & \text{{\small 2640}%
} & \text{1} & \text{q}_{\Sigma}\text{(2640)}\\
\text{E}_{\Gamma}\text{=6 \ \ \ \ } & \text{{\small (-211,2-1-1,-1-12,}} &
\text{{\small 3090}} & \text{1} & \text{q}_{\Sigma}\text{(3090)}\\
\text{E}_{\Gamma}\text{=6 \ \ \ \ } & \text{{\small \ \ \ 11-2,-12-1,1-21)}} &
\text{{\small 3091}} & \text{1} & \text{q}_{\Sigma}\text{(3090)}\\
\text{E}_{\Gamma}\text{=6 \ \ \ } & \text{{\small (-1-21,1-2-1,-11-2,}} &
\text{{\small 3091}} & \text{1} & \text{q}_{\Sigma}\text{(3090)}\\
\text{E}_{\Gamma}\text{=6 \ \ \ \ } & \text{{\small \ \ 1-1-2,-21-1,-2-11))}}
& \text{{\small 3091}} & \text{1} & \text{q}_{\Sigma}\text{(3090)}\\
\text{E}_{\Gamma}\text{=6 \ \ \ \ } & \text{{\small (-1-2-1,-1-1-2,-2-1-1)}} &
\text{{\small 3091}} & \text{1} & \text{q}_{\Sigma}\text{(3090)}\\
\text{{\small \ldots\ .}} &  &  &  &
\end{array}
\label{S-Quark}%
\end{equation}

\ \ \ \ \ \ \ \ \ \ \ \ 

\subsubsection{The Quarks on the Axis $\Sigma$($\Gamma$-N)}

The $\Sigma$-axis is a two-fold rotation axis, R = 2. From (\ref{S-Number}), S
= -2 (see Fig. 3(a)).

1. The two-fold degenerate energy bands on the $\Sigma$-axis ($\Gamma$-N)

For the two-fold degenerate energy bands, each of them represents a quark
family q$_{\Xi}$ (q$_{\Xi}^{\frac{1}{2}}$, q$_{\Xi}^{-\frac{1}{2}}$) with B =
1/3, S = -2, I = 1/2 from (\ref{IsoSpin}), I$_{z,B}$ = $\frac{1}{2}$,
-$\frac{1}{2}$.\ Similar to (\ref{D-Quark}), we have [$q_{\Xi}$ (q$_{\Xi
}^{\frac{1}{2}}$, q$_{\Xi}^{-\frac{1}{2}}$)]:
\begin{equation}%
\begin{array}
[c]{llcccc}%
\text{ \ \ \ \ \ E} & \text{ \ \ \ n}_{1}\text{n}_{2}\text{n}_{3} & S_{\Xi} &
\text{I} & \text{m} & \text{q}_{\Xi}\text{(m)}\\
E_{\Gamma}=2 & (\text{1-10,-110)} & -2 & \text{1/2} & \text{1650} &
\text{q}_{\Xi}\text{(1650)}\\
E_{N}=5/2 & \text{(200,020)} & -2 & \text{1/2} & \text{1830} & \text{q}_{\Xi
}\text{(1830)}\\
E_{\Gamma}=4 & \text{(002,00-2)} & -2 & \text{1/2} & \text{2370} &
\text{q}_{\Xi}\text{(2370)}\\
& \text{(-200,0-20)} & -2 & \text{1/2} & \text{2370} & \text{q}_{\Xi
}\text{(2370)}\\
E_{N}=9/2 & \text{(112,11-2)} & -2 & \text{1/2} & \text{2550} & \text{q}_{\Xi
}\text{(2550)}\\
\ldots\text{ .} &  &  &  &  &
\end{array}
\label{2-SIGMA}%
\end{equation}

2. The four-fold degenerate energy bands on the axis $\Sigma$($\Gamma$-N)

According to (\ref{Subdeg}), each four-degenerate energy band on the symmetry
axis $\Sigma$ will be divided into two two-fold degenerate bands. From
(\ref{2-SIGMA}), each of them represents a quark family $q_{\Xi}$ (q$_{\Xi
}^{\frac{1}{2}}$, q$_{\Xi}^{-\frac{1}{2}}$) with B = 1/3, S = -2, I = 1/2,
I$_{z,B}$ = $\frac{1}{2}$, -$\frac{1}{2}$.\ Thus, we have
\begin{equation}%
\begin{array}
[c]{llcc}%
E & \text{(n}_{1}\text{n}_{2}\text{n}_{3}\text{, ... )} & I & \text{q}_{\Xi
}\text{(m)}\\
E_{N}=3/2 & \vec{n}=(\text{101,10-1,011,01-1}) & \text{1/2} & \text{2}%
\times\text{q}_{\Xi}\text{(1470)}\\
E_{\Gamma}=2 & \vec{n}=(\text{-101,-10-1,0-11,0-1-1}) & \text{1/2} &
\text{2}\times\text{q}_{\Xi}\text{(1650)}\\
E_{N}=7/2 & \vec{n}=(\text{121,12-1,211,21-1}) & \text{1/2} & \text{2}%
\times\text{q}_{\Xi}\text{(2190)}\\
\ldots\text{ .} &  &  &
\end{array}
\label{4-fold}%
\end{equation}

3. The single energy bands on the axis $\Sigma$($\Gamma$-N), similarly to
(\ref{D-ONE}) \cite{Q-Mass}, we have:%

\begin{equation}%
\begin{array}
[c]{lllllll}%
\text{ \ \ \ \ E} & \text{\ n}_{1}\text{n}_{2}\text{n}_{3} & \text{ S} &
\text{ \ \ J} & \text{I} & \text{ }\Delta\epsilon & \text{q}_{Name}%
\text{(m)}\\
\text{E}_{N}\text{=1/2} & (\text{1,1,0}) & \text{-1} & \text{J}_{N}\text{=1} &
\text{0} & \text{ \ \ 0} & \text{d}_{S}\text{(1110)}\\
\text{E}_{\Gamma}\text{=2} & (\text{-1,-1,0}) & \text{-3} & \text{J}_{\Gamma
}\text{=1} & \text{0} & \text{ \ \ 0} & \text{d}_{\Omega}\text{(1650)}\\
\text{E}_{N}\text{=9/2} & (\text{2,2,0}) & \text{-1} & \text{J}_{N}\text{=2} &
\text{0} & \text{ \ \ 0} & \text{d}_{S}\text{(2550)}\\
\text{E}_{\Gamma}\text{=8} & (\text{-2,-2,0}) & \text{-3} & \text{J}_{\Gamma
}\text{=2} & \text{0} & \text{ \ \ 0} & \text{d}_{\Omega}\text{(3810)}\\
\text{E}_{N}\text{=25/2} & (\text{3,3,0}) & \text{-1} & \text{J}_{N}\text{=3}
& \text{0} & \text{+100} & \text{d}_{b}\text{(5530)}\\
\text{E}_{\Gamma}\text{=18} & (\text{-3,-3,0}) & \text{-3} & \text{J}_{\Gamma
}\text{=3} & \text{0} & \text{-100} & \text{d}_{\Omega}\text{(7310)}\\
\text{E}_{N}\text{=49/2} & (\text{4,4,0}) & \text{-1} & \text{J}_{N}\text{=4}
& \text{0} & \text{+200} & \text{d}_{b}\text{(9950)}\\
\text{E}_{\Gamma}\text{=32} & \text{(-4,-4,0)} & \text{-3} & \text{J}_{\Gamma
}\text{=4} & \text{0} & \text{-200} & \text{d}_{\Omega}\text{(12250)}\\
\text{E}_{N}\text{=81/2} & (\text{5,5,0}) & \text{-1} & \text{J}_{N}\text{=5}
& \text{0} & \text{+300} & \text{d}_{b}\text{(15810)}\\
...\text{ .} &  &  &  &  &  &
\end{array}
\label{SIGMA-ONE}%
\end{equation}
\ \ \ \ \ \ \ \ \ 

Continuing the above procedure (see Appendix C), we can use Fig. 2-5 to find
the excited states of the quark (the quark spectrum) of lower energies as
shown in the following lists. The I$_{z},_{Baryon}$ of Q$_{\text{Name}%
}^{\text{I}_{Z}\text{, Bareon}}$ is the z-component of the isospin of the
baryon (Q$_{\text{Name}}^{\text{I}_{Z}}$q'$_{1}$q'$_{2}$) from
(\ref{Iz of Baryon}). the I$_{z}$ of Q$_{\text{Ele. uark}}^{\text{I}_{Z}}$ is
the z-component of the isospin of Q$_{\text{uark}}^{\text{I}_{Z}}$ from
I$_{z}$ = Q -$\frac{1}{2}$(B+S+C+b)); the Q$_{\text{Ele}.\text{ Quark}}$ of
Q$_{\text{Ele. uark}}^{\text{I}_{Z}}$ is the excited quark of the elementary
quark (u(0) or d(0)).%

\begin{equation}%
\begin{tabular}
[c]{|l|l|l|l|l|l|l|l|l|l|l|l|}\hline
q$_{\text{Name}}^{\text{I}_{Z}\text{,Bary}}$ & q$_{N}^{\frac{1}{2}}$ &
q$_{N}^{\frac{-1}{2}}$ & q$_{\Delta}^{\frac{3}{2}}$ & q$_{\Delta}^{\frac{1}%
{2}}$ & q$_{\Delta}^{\frac{-1}{2}}$ & q$_{\Delta}^{\frac{-3}{2}}$ & q$_{S}%
^{0}$ & q$_{\Sigma}^{1}$ & q$_{\Sigma}^{0}$ & q$_{\Sigma}^{-1}$ & q$_{\Omega
}^{0}$\\\hline
S & \ 0 & \ 0 & \ 0 & \ 0 & \ \ 0 & \ \ 0 & \ -1 & \ -1 & -1 & \ -1 &
-3\\\hline
C & \ 0 & \ 0 & \ 0 & \ 0 & \ \ 0 & \ \ 0 & \ \ 0 & \ \ 0 & \ \ 0 & \ 0 &
\ 0\\\hline
b & \ 0 & \ 0 & \ 0 & \ 0 & \ \ 0 & \ \ 0 & \ \ 0 & \ 0 & \ \ 0 & \ 0 &
\ 0\\\hline
I & 1/2 & 1/2 & 3/2 & 3/2 & 3/2 & 3/2 & \ \ 0 & \ \ 1 & \ \ 1 & \ 1 &
\ 0\\\hline
I$_{Z}^{\text{baryon}}$ & 1/2 & -1/2 & 3/2 & 1/2 & -1/2 & -3/2 & \ \ 0 &
\ \ 1 & \ \ 0 & -1 & \ 0\\\hline
Q$_{\text{Ele. uark}}^{\text{I}_{Z}}$ & u$_{N}^{\frac{1}{2}}$ & d$_{N}%
^{\frac{-1}{2}}$ & u$_{\Delta}^{\frac{1}{2}}$ & u$_{\Delta}^{\frac{1}{2}}$ &
d$_{\Delta}^{\frac{-1}{2}}$ & d$_{\Delta}^{\frac{-1}{2}}$ & \ \ d$_{S}^{0}$ &
u$_{\Sigma}^{1}$ & \ d$_{\Sigma}^{0}$ & d$_{\Sigma}^{0}$ & d$_{\Omega}^{1}%
$\\\hline
$\text{Q}_{q}$ & $\frac{2}{3}$ & $\frac{-1}{3}$ & $\frac{2}{3}$ & $\frac{2}%
{3}$ & $\frac{-1}{3}$ & $\frac{-1}{3}$ & $\frac{-1}{3}$ & $\frac{2}{3}$ &
$\frac{-1}{3}$ & $\frac{-1}{3}$ & $\frac{-1}{3}$\\\hline
&  &  &  &  &  &  &  &  &  &  & \\\hline
q$_{\text{Name}}^{\text{I}_{Z}\text{,Baryon}}$ & q$_{\Xi}^{\frac{1}{2}}$ &
q$_{\Xi}^{\frac{-1}{2}}$ & \ q$_{C}^{0}$ & \ q$_{b}^{0}$ & q$_{\Omega_{C}}%
^{0}$ & \ q$_{\Xi_{C}}^{\frac{1}{2}}$ & q$_{\Xi_{C}}^{\frac{-1}{2}}$ &
q$_{\Sigma_{C}}^{1}$ & q$_{\Sigma_{C}}^{0}$ & q$_{\Sigma_{C}}^{-1}$ & \\\hline
S & -2 & -2 & \ 0 & \ 0 & \ -2 & \ -1 & \ -1 & \ 0 & \ 0 & \ 0 & \\\hline
C & \ 0 & \ 0 & \ 1 & \ 0 & \ 1 & \ 1 & \ 1 & \ 1 & \ 1 & \ 1 & \\\hline
b & \ 0 & \ 0 & \ \ 0 & -1 & \ 0 & \ 0 & \ 0 & \ 0 & \ 0 & \ 0 & \\\hline
I & 1/2 & 1/2 & \ 0 & \ 0 & \ 0 & 1/2 & 1/2 & \ 1 & \ 0 & \ -1 & \\\hline
I$_{Z}^{\text{baryon}}$ & 1/2 & -1/2 & \ 0 & \ 0 & \ 0 & 1/2 & -1/2 & \ 1 &
\ 0 & \ -1 & \\\hline
Q$_{\text{Ele. uark}}^{\text{I}_{Z}}$ & u$_{\Xi}^{\frac{3}{2}}$ & d$_{\Xi
}^{\frac{1}{2}}$ & u$_{C}^{\frac{1}{2}}$ & d$_{b}^{\frac{-1}{2}}$ &
d$_{\Omega_{C}}^{0}$ & u$_{\Xi_{C}}^{\frac{1}{2}}$ & d$_{\Xi_{C}}^{\frac
{-1}{2}}$ & u$_{\Sigma_{C}}^{0}$ & u$_{\Sigma_{C}}^{0}$ & d$_{\Sigma_{C}}%
^{-1}$ & \\\hline
$\text{Q}_{q}$ & $\frac{2}{3}$ & $\frac{-1}{3}$ & $\frac{2}{3}$ & $\frac
{-1}{3}$ & $\frac{-1}{3}$ & $\frac{2}{3}$ & $\frac{-1}{3}$ & $\frac{2}{3}$ &
$\frac{2}{3}$ & $\frac{-1}{3}$ & .\\\hline
\end{tabular}
\ \ \ \ \label{Q-Number}%
\end{equation}
where \ Name of q$_{\text{Name}}^{\text{I}_{Z},bary}$ is the name of the
quark, I$_{Z},bary$ is the z-component of the isospin of the baryon
(q$_{\text{Name}}^{\text{I}_{Z},Bary}$q'$_{1}$q'$_{2}$);%

\begin{equation}%
\begin{tabular}
[c]{|l|l|l|l|l|l|}\hline
& \ \ \ \ \ \ \ \ \ \ The & \ Quark & Spectrum &  & \\\hline
Elememtary & \ quarks, \ u(0) & \ and d(0), & in vacuum, & $\ $m$_{u}$=Q$_{u}%
$=0, & $\ $m$_{d}$=Q$_{d}$=0.\\\hline
& \ \ \ \ \ \ The two & accompany & \ excited & quarks & \\\hline
u'-quark & S=C=b=0, & I=s=$\frac{1}{2},$ & I$_{z}$=$\frac{1}{2}$, &
Q=$\frac{2}{3}$\  & m$_{u\text{'}}$= 3\\\hline
d'-quark & S=C=b=0, & I=s=$\frac{1}{2},$ & I$_{z}$=$\frac{-1}{2}$, &
Q=$\frac{-1}{3}$ & m$_{d\text{'}}$= 7\\\hline
& The energy & \ \ band & excited & \ quarks & \\\hline
\textbf{q}$_{N}$\textbf{(930)} &  & \textbf{d}$_{S}$\textbf{(1110)} &  &
\textbf{u}$_{C}$\textbf{(2270)} & $\ $\textbf{d}$_{b}$\textbf{(5530)}\\\hline
1q$_{N}$(1200) & 1d$_{S}^{0}$(1390) & 1q$_{\Sigma}$(1200) & 2q$_{\Xi}$(1290) &
u$_{C}$(2550) & d$_{b}$(9950)\\\hline
1q$_{N}$(1470) & 1d$_{S}^{0}$(1490) & 3q$_{\Sigma}$(1650) & 3q$_{\Xi}$(1470) &
2u$_{C}$(2750) & d$_{b}$(15810)\\\hline
2q$_{N}$(1830) & 2d$_{S}^{0}$(1830) & 2q$_{\Sigma}$(1920) & 3q$_{\Xi}$(1650) &
u$_{C}$(4140) & d$_{b}$(23010)\\\hline
3q$_{N}$(1920) & 3d$_{S}^{0}$(1920) & 1q$_{\Sigma}$(2370) & 1q$_{\Xi}$(1830) &
u$_{C}$(6490) & d$_{b}$(31850)\\\hline
1q$_{N}$(2010) & 2d$_{S}^{0}$(2010) & 3q$_{\Sigma}$(2640) & 2q$_{\Xi}$(1920) &
u$_{C}$(13590) & \\\hline
2q$_{N}$(2190) & 2d$_{S}^{0}$(2120) & 5q$_{\Sigma}$(3090) & 3q$_{\Xi}$(2010) &
u$_{C}$(23570) & \\\hline
2q$_{N}$(2640) & 2d$_{S}$(2550) &  & 4q$_{\Xi}$(2190) & u$_{C}$(36430) &
\\\hline
1q$_{N}$(2910) & 1d$_{S}^{0}$(2640) &  & 2q$_{\Xi}$(2370) &  & \\\hline
1q$_{N}$(3360) & 4d$_{S}^{0}$(2730) &  & 3q$_{\Xi}$(2550) &  & \\\hline
\ \  &  &  &  &  & \\\hline
1q$_{\Delta}$(1290) & 1d$_{S}$(4370) & d$_{\Omega}$(1650) & d$_{\Omega_{c}}%
$(2730) & 1q$_{\Xi_{c}}$(2440) & \\\hline
2q$_{\Delta}$(1650) & d$_{S}$(10230)\  & d$_{\Omega}$(2350) & d$_{\Omega_{c}}%
$(2750) & 1q$_{\Xi_{c}}$(2530) & \\\hline
1q$_{\Delta}$(2010) & d$_{S}$(18970) & d$_{\Omega}$(2730) & d$_{\Omega_{c}}%
$(3670) & 1q$_{\Xi_{c}}$(2640) & \\\hline
1q$_{\Delta}$(2370) & d$_{S}$(30590) & d$_{\Omega}$(2870) &  & 1q$_{\Xi_{c}}%
$(2730) & \\\hline
4q$_{\Delta}$(2730) & d$_{S}$(45090) & d$_{\Omega}$(3810) &  & 1q$_{\Xi_{c}}%
$(2960) & \\\hline
3q$_{\Delta}$(3090) &  & d$_{\Omega}$(7310) &  &  & \\\hline
\end{tabular}
\ \ \ \ \ \ \ \ \label{Quarks}%
\end{equation}

\ \ \ \ \ \ \ 

\section{The Spectrum of the Baryons}

We have already found the quantum numbers (\ref{Q-Number}) and the rest masses
(\ref{Quarks}) of the excited quarks and the accompanying excited quarks u'
and d' \cite{Confine}. Now we will recognize the three-quark systems
(qq'$_{1}$q'$_{2}$). Using the sum laws, we can find the quantum numbers and
masses of the three quark (qq'$_{1}$q'$_{2}$) systems first. Then, from the
quantum numbers and the masses of the systems, we recognize the baryons.

\subsection{The Three Quark (qq'$_{1}$q'$_{2}$) Systems Are the
Baryons\ \ \ \ \ \ \ \ \ \ \ \ \ \ \ \ \ \ \ \ \ \ \ \ \ \ \ \ \ \ \ \ \ \ \ \ \ \ \ \ \ \ \ \ \ \ \ \ \ \ \ \ \ \ \ \ \ \ \ \ \ \ \ \ \ \ \ \ \ \ \ \ \ \ \ \ \ \ \ \ \ \ \ \ \ \ \ \ \ \ \ \ \ \ \ \ \ \ \ \ \ \ \ \ \ \ \ \ \ \ \ \ \ \ \ \ \ \ \ \ \ \ \ \ \ \ \ \ \ \ \ \ \ \ \ \ \ \ \ \ \ \ \ \ \ \ \ \ \ \ \ \ \ \ \ \ \ \ \ \ \ \ \ \ \ \ \ \ \ \ \ \ \ \ \ \ \ \ \ \ \ \ \ \ \ \ \ \ \ \ \ \ \ \ \ \ \ \ \ \ \ \ \ \ \ \ \ \ \ \ \ \ \ \ \ \ \ \ \ \ \ \ \ \ \ \ \ \ \ \ \ \ \ \ \ \ \ \ \ \ \ \ \ \ \ \ \ \ \ \ \ \ \ \ \ \ \ \ \ \ \ \ \ \ \ \ \ \ \ \ \ \ \ \ \ \ \ \ \ \ \ \ \ \ \ \ \ \ \ \ \ \ \ \ \ \ \ \ \ \ \ \ \ \ \ \ \ \qquad
\ \ \ \ \ \ \ \ \ \ \ \ \ \ \ \ \ \ \ \ \ \ \ \ \ \ \ \ \ \ \ \ \ \ \ \ \ \ \ \ \ \ \ \ \ \ \ \ \ \ \ \ \ \ \ \ \ \ \ \ \ \ \ \ \ \ }%

The baryon number of the three-quark system (qq'$_{1}$q'$_{2}$), from
(\ref{1/3}),
\begin{equation}
\text{B = B}_{\text{q}}\text{+\ B}_{\text{q'}_{1}}\text{+ B}_{\text{q'}_{2}%
}\text{= 1.} \label{B = 1}%
\end{equation}
Thus the three-quark system is a baryon.

\subsection{The Rest Masses and the Quantum Numbers of the Three Quark
(qq'$_{1}$q'$_{2}$)
Systems\ \ \ \ \ \ \ \ \ \ \ \ \ \ \ \ \ \ \ \ \ \ \ \ \ \ \ \ \ \ \ \ \ \ \ \ \ \ \ \ \ \ \ \ \ \ \ \ \ \ \ \ \ \ \ \ \ \ \ \ \ \ \ \ \ \ \ \ \ \ \ \ \ \ \ \ \ \ \ \ \ \ \ \ \ \ \ \ \ \ \ \ \ \ \ \ \ \ \ \ \ \ \ \ \ \ \ \ \ \ \ \ \ \ \ \ \ \ \ \ \ \ \ \ \ \ \ \ \ \ \ \ \ \ \ \ \ \ \ \ \ \ \ \ \ \ \ \ \ \ \ \ \ \ \ \ \ \ \ \ \ \ \ \ \ \ \ \ \ \ \ \ \ \ \ \ \ \ \ \ \ \ \ \ \ \ \ \ \ \ \ \ \ \ \ \ \ \ \ \ \ \ \ \ \ \ \ \ \ \ \ \ \ \ \ \ \ \ \ \ \ \ \ \ \ \ \ \ \ \ \ \ \ \ \ \ \ \ \ \ \ \ \ \ \ \ \ \ \ \ \ \ \ \ \ \ \ \ \ \ \ \ \ \ \ \ \ \ \ \ \ \ \ \ \ \ \ \ \ \ \ \ \ \ \ \ \ \ \ \ \ \ \ \ \ \ \ \ \ \ \ \ \ \ \ \ \ \ \ \ \ \ \ \ \ \ \ \ \ \ \ \ \ \ \ \ \ \ \ \ \ \ \ \ \ \ \ \ \ \ \ \ \ \ \ \ \ \ \ \ \qquad
\ \ \ \ \ \ \ \ \ \ \ \ \ \ \ \ \ \ \ \ \ \ \ \ \ \ \ \ \ \ \ \ \ \ \ \ \ \ \ \ \ \ \ \ \ \ \ \ \ \ \ \ \ \ \ \ \ \ \ \ \ \ \ \ \ \ }%

Now, we give some phenomenological formulas with which we can find the quantum
numbers and rest masses of the three quark (qq'$_{1}$q'$_{2}$) systems, as
shown in the following:

1.The quantum numbers of the system (qq'$_{1}$q'$_{2}$) are the sums of the
constituent quarks (q, q'$_{1}$ and q'$_{2}$). Since S$_{q\text{'}_{1}}$=
C$_{q\text{'}_{1}}$= b$_{q\text{'}_{1}}$= S$_{q\text{'}_{2}}$= C$_{q\text{'}%
_{2}}$= b$_{q\text{'}_{2}}$= 0 from Hypothesis I,
\begin{equation}
\text{S}_{qu\text{'}d\text{'}}\text{ = S(q), C}_{qu\text{'}d\text{'}}\text{ =
C(q), b}_{qu\text{'}d\text{'}}\text{ = b(q), and Q}_{qu\text{'}d\text{'}%
}\text{ =Q}_{q}\text{+ Q}_{q\text{'}_{1}}\text{+ Q}_{q\text{'}_{2}}\text{.}
\label{B-Q-Num}%
\end{equation}

2. The Isospin I$_{B}$ of the system (qq'$_{1}$q'$_{2}$) is found by
\begin{equation}
\overrightarrow{I_{B}}\text{ = }\overrightarrow{I_{q}}\text{ + }%
\overrightarrow{I_{q_{1}^{^{\prime}}}}\text{ + }\overrightarrow{I_{q_{2}%
^{^{\prime}}}}\text{.} \label{I of qqq}%
\end{equation}
Then, the z-componet of the isospin of the baryon (qq'$_{1}$q'$_{2}$):%

\begin{equation}%
\begin{tabular}
[c]{|l|}\hline
$\text{for I}_{B}\text{ = 3/2, \ \ \ I}_{z,B}\text{ = 3/2, 1/2, -1/2, -3/2,}%
$\\\hline
$\text{for I}_{B}\text{ = 1, \ \ \ \ \ \ I}_{z,B}\text{ = 1, 0, -1,}$\\\hline
$\text{for I}_{B}\text{ = 1/2, \ \ \ I}_{z,B}\text{ = 1/2, -1/2,}$\\\hline
$\text{for I}_{B}\text{ = 0, \ \ \ \ \ \ I}_{z,B}\text{ = 0.}$\\\hline
\end{tabular}
\ \label{Iz of Baryon}%
\end{equation}

3. The top limit (I$_{\max}$) of the isospin of the baryons on the symmetry
axis is determined by
\begin{equation}
\text{2I}_{\max}\text{(axis) + 1 = 4 + S.} \label{I-T-Axis}%
\end{equation}
From(\ref{S-Number}), we have:
\begin{equation}
\text{for the axis }\Delta\text{ and the axis D, S = 0, I}_{\max}\text{ =
3/2;} \label{I-T-S=0}%
\end{equation}%
\begin{equation}
\text{for the axis }\Lambda\text{ and the axis F, S = -1, I}_{\max}\text{=1;}
\label{I-T-S=-1}%
\end{equation}%
\begin{equation}
\text{for the axis }\Sigma\text{ and the axis G, S = -2, I}_{\max}\text{=1/2.}
\label{I-T-S=-2}%
\end{equation}

4. The top limit (I$_{\max}$) of the isospin of the baryons at a symmetry
point is determined by the highest dimension (D$_{high}$) of the irreducible
representations of the double point group \cite{D-Group} at the point
\begin{equation}
\text{2I}_{\max}\text{ + 1 = D}_{high}\text{.} \label{I-T-Point}%
\end{equation}
Since the highest dimension D$_{high}$ = 4 for the $\Gamma$-group$,$ the
H-group, the M-group, and the P-group , we get
\begin{equation}
\text{I}_{\max}\text{ = 3/2, at }\Gamma\text{, H, M, and P;} \label{I-T-HMP}%
\end{equation}
and from that the highest dimension $D_{high}=2$ for the double point group N
\cite{D-Group}, we know
\begin{equation}
\text{I}_{\max}\text{ = 1/2, at the point N.} \label{I-T-N}%
\end{equation}

5. The isospin I$_{B}$ and I$_{B,Z}$ of the baryon (qq'$_{1}$q'$_{2}$) has
already been given by the name of the excited quark q$_{\text{Name}}%
^{\text{I}_{Z}}$ in (\ref{Q-Number}). The electric charge Q$_{q}$ and the
I$_{Z}$ of the excited quark Q$_{uark}^{\text{I}_{Z}}$ have already been given
by (\ref{Q-Number}) also. Thus, the two accompanying excited quarks $q$'$_{1}$
and $q$'$_{2}$ are selected by the excited quark q from u'u', u'd', and d'd',
according to the quantum numbers I$_{Z}$ and Q of the baryon$,$ using
(\ref{Izb}) and (\ref{Q*Q'Q'})%

\begin{equation}
\text{I}_{Z,B}\text{ = I}_{Z,q^{\ast}}\text{+ I}_{Z,}\text{{\small q}'}%
_{1}\text{ + I}_{Z,}\text{{\small q}'}_{2}\text{,} \label{Izb}%
\end{equation}%
\begin{equation}
\text{Q}_{B}\text{ = Q}_{q^{\ast}}\text{ + Q}_{q_{1}^{^{\prime}}}\text{ +
Q}_{q_{2}^{^{\prime}}}\text{.} \label{Q*Q'Q'}%
\end{equation}
We show the results in the following list%

\begin{equation}%
\begin{tabular}
[c]{|l|l|l|l|l|l|l|l|l|l|l|l|}\hline
q$_{Name}^{\text{I}_{z}}$ & q$_{N}^{\frac{1}{2}}$ & q$_{N}^{\frac{-1}{2}}$ &
q$_{\Delta}^{\frac{3}{2}}$ & \ \ \ q$_{\Delta}^{\frac{1}{2}}$ & \ \ q$_{\Delta
}^{\frac{-1}{2}}$ & q$_{\Delta}^{\frac{-3}{2}}$ & q$_{S}^{0}$ & q$_{\Sigma
}^{1}$ & \ q$_{\Sigma}^{0}$ & q$_{\Sigma}^{-1}$ & q$_{\Omega}^{0}$\\\hline
S & \ 0 & \ 0 & \ 0 & \ \ \ \ 0 & \ \ \ 0 & \ \ 0 & \ -1 & \ -1 & \ -1 &
\ -1 & -3\\\hline
C & \ 0 & \ 0 & \ 0 & \ \ \ \ 0 & \ \ \ 0 & \ \ 0 & \ \ 0 & \ \ 0 & \ \ 0 &
\ 0 & \ 0\\\hline
b & \ 0 & \ 0 & \ 0 & \ \ \ \ 0 & \ \ \ 0 & \ \ 0 & \ \ 0 & \ 0 & \ \ 0 &
\ 0 & \ 0\\\hline
I & 1/2 & 1/2 & 3/2 & \ 3/2 & \ \ 3/2 & 3/2 & \ \ 0 & \ \ 1 & \ \ 1 & \ 1 &
\ 0\\\hline
u$^{_{\text{I}_{Z}}}$(d$^{_{\text{I}_{z}}}$) & u$_{N}^{\frac{1}{2}}$ &
d$_{N}^{\frac{-1}{2}}$ & u$_{\Delta}^{\frac{1}{2}}$ & \ \ u$_{\Delta}%
^{\frac{1}{2}}$ & \ \ d$_{\Delta}^{\frac{-1}{2}}$ & \ d$_{\Delta}^{\frac
{-1}{2}}$ & d$_{S}^{0}$ & u$_{\Sigma}^{1}$ & \ d$_{\Sigma}^{0}$ & d$_{\Sigma
}^{0}$ & d$_{\Omega}^{1}$\\\hline
q$_{1}$ & u$^{^{\prime}}$ & u$^{^{\prime}}$ & u$^{^{\prime}}$ &
\ \ u$^{^{\prime}}$ & \ \ u$^{^{\prime}}$ & \ d$^{^{\prime}}$ & u$^{^{\prime}%
}$ & u$^{^{\prime}}$ & \ u$^{^{\prime}}$ & d$^{^{\prime}}$ & d$^{^{\prime}}%
$\\\hline
q$_{2}$ & d$^{^{\prime}}$ & d$^{^{\prime}}$ & u$^{^{\prime}}$ &
\ \ d$^{^{\prime}}$ & \ \ d$^{^{\prime}}$ & \ d$^{^{\prime}}$ & d$^{^{\prime}%
}$ & d$^{^{\prime}}$ & \ d$^{^{\prime}}$ & d$^{^{\prime}}$ & d$^{^{\prime}}%
$\\\hline
I$_{Z,B}$ & $\frac{1}{2}$ & $\frac{-1}{2}$ & $\frac{3}{2}$ & $\frac{1}{2}$ &
$\frac{-1}{2}$ & $\frac{-3}{2}$ & 0 & 1 & 0 & -1 & 0\\\hline
Q$_{B}$ & 1 & 0 & 2 & 1 & 0 & -1 & 0 & 1 & 0 & -1 & -1\\\hline
B$_{aryon}$ & N$^{+}$ & N$^{0}$ & $\Delta^{++}$ & $\ \ \ \Delta^{+}$ &
$\ \ \ \Delta^{0}$ & $\ \Delta^{-}$ & $\Lambda$ & $\Sigma^{+}$ & $\ \Sigma
^{0}$ & $\Sigma^{-}$ & $\Omega^{-}$\\\hline
&  &  &  &  &  &  &  &  &  &  & \\\hline
q$^{\text{I}_{z}}$ & q$_{\Xi}^{\frac{1}{2}}$ & q$_{\Xi}^{\frac{-1}{2}}$ &
\ q$_{c}^{0}$ & \ q$_{b}^{0}$ & q$_{\Omega_{C}}^{0}$ & \ q$_{\Xi_{C}}%
^{\frac{1}{2}}$ & q$_{\Xi_{C}}^{\frac{-1}{2}}$ & q$_{\Sigma_{C}}^{1}$ &
q$_{\Sigma_{C}}^{0}$ & q$_{\Sigma_{C}}^{-1}$ & \\\hline
S & -2 & -2 & \ 0 & \ 0 & \ -2 & \ -1 & \ -1 & \ 0 & \ 0 & \ 0 & \\\hline
C & \ 0 & \ 0 & \ 1 & \ 0 & \ 1 & \ \ 1 & \ \ 1 & \ 1 & \ 1 & \ 1 & \\\hline
b & \ 0 & \ 0 & \ \ 0 & -1 & \ 0 & \ 0 & \ \ 0 & \ 0 & \ 0 & \ 0 & \\\hline
I & 1/2 & 1/2 & \ 0 & \ 0 & \ 0 & 1/2 & 1/2 & \ 1 & \ 0 & \ -1 & \\\hline
I$_{Z}$ & 1/2 & -1/2 & \ 0 & \ 0 & \ 0 & 1/2 & -1/2 & \ 1 & \ 0 & \ -1 &
\\\hline
u$^{_{\text{I}_{Z}}}$(d$^{_{\text{I}_{z}}}$) & u$_{\Xi}^{\frac{3}{2}}$ &
d$_{\Xi}^{\frac{1}{2}}$ & u$_{c}^{0}$ & d$_{b}^{0}$ & d$_{\Omega_{C}}^{0}$ &
u$_{\Xi_{C}}^{\frac{1}{2}}$ & d$_{\Xi_{C}}^{\frac{-1}{2}}$ & u$_{\Sigma_{C}%
}^{0}$ & u$_{\Sigma_{C}}^{0}$ & d$_{\Sigma_{C}}^{-1}$ & \\\hline
q$_{1}$ & d$^{^{\prime}}$ & d$^{^{\prime}}$ & u$^{^{\prime}}$ & u$^{^{\prime}%
}$ & u$^{^{\prime}}$ & u$^{^{\prime}}$ & u$^{^{\prime}}$ & u$^{^{\prime}}$ &
u$^{^{\prime}}$ & u$^{^{\prime}}$ & \\\hline
q$_{2}$ & d$^{^{\prime}}$ & d$^{^{\prime}}$ & d$^{^{\prime}}$ & d$^{^{\prime}%
}$ & d$^{^{\prime}}$ & d$^{^{\prime}}$ & d$^{^{\prime}}$ & u$^{^{\prime}}$ &
d$^{^{\prime}}$ & d$^{^{\prime}}$ & \\\hline
I$_{\text{Z,B}}$ & $\frac{1}{2}$ & $\frac{-1}{2}$ & 0 & 0 & 0 & $\frac{1}{2}$
& $\frac{-1}{2}$ & 1 & 0 & -1 & \\\hline
Q$_{\text{B}}$ & 0 & -1 & 1 & 0 & 0 & 1 & 0 & 2 & 1 & 0 & \\\hline
B$_{ar}$ & $\Xi^{0}$ & $\Xi^{-}$ & $\Lambda_{c}$ & $\Lambda_{b}$ & $\Omega
_{c}^{0}$ & $\Xi_{c}^{+}$ & $\Xi_{c}^{0}$ & $\Sigma_{c}^{++}$ & $\Sigma
_{c}^{+}$ & $\Sigma_{c}^{0}$ & \\\hline
\end{tabular}
\ \label{B-Comp}%
\end{equation}

6. The energy of the system (qq'$_{1}$q'$_{2}$) equals the sum of the energies
of the excited quarks q, q'$_{1}$ and q'$_{2}$%
\begin{equation}
M_{_{(}\text{qq'}_{1}\text{q'}_{2}\text{)}}\text{= m}_{q}\text{ +
m}_{q\text{'}_{1}}\text{ + m}_{q\text{'}_{1}}\text{. } \label{B-Mass}%
\end{equation}

For a quark q with isospin I$_{q}$, adding two accompanying excited quarks
q'$_{1}$ and q'$_{2}$ (one of u'u', u'd', and d'd'), the three-quark system
(qq'$_{1}$q'$_{2}$) may have I$_{baryon}$= I$_{q}$ $\pm$ 1 from
(\ref{I of qqq}). This baryon (qq$_{1}$'q'$_{2}$) may have added energy. For
simplicity, we assume%
\begin{equation}
\delta\varepsilon\text{ = 100S}_{G}\text{(q)[S}_{G}\text{(q)+}\Delta\text{I],}
\label{Sg+dI}%
\end{equation}
where S$_{G}$(q) = S + C + b of q, $\Delta$I = $\left\vert \text{I}%
_{baryon}\text{-I}_{q}\right\vert $= + 1.

Using the formulae (\ref{Q-Number}) - (\ref{Sg+dI}), we can find the baryons
on the symmetry axes.

\subsubsection{The Baryons On the $\Delta$-Axis ($\Gamma$-H)}

1. The four-fold degenerate bands on the $\Delta$-axis ($\Gamma$-H)
\begin{equation}%
\begin{array}
[c]{lllll}%
E_{H}=1 & \vec{n}=(\text{101,-101,011,0-11}) & \text{q}_{\Delta}\text{(1290)}
& \Delta\text{(1300)} & \text{N(1300)}\\
E_{\Gamma}=2\text{ } & \vec{n}=(\text{110,1-10,-110,-1-10}) & \text{q}%
_{\Delta}\text{(1650)} & \Delta\text{(1660)} & \text{N(1660)}\\
E_{\Gamma}=2 & \vec{n}=(\text{10-1,-10-1,01-1,0-1-1}) & \text{q}_{\Delta
}\text{(1650)} & \Delta\text{(1660)} & \text{N(1660)}\\
E_{H}=3 & \vec{n}=(\text{112,1-12,-112,-1-12}) & \text{q}_{\Delta
}\text{(2010)} & \Delta\text{(2020)} & \text{N(2020)}\\
E_{\Gamma}=4 & \vec{n}=(\text{200,-200,020,0-20}) & \text{q}_{\Delta
}\text{(2370)} & \Delta\text{(2380)} & \text{N(2380)}\\
E_{H}=5 & \vec{n}=(\text{121,1-21,-121,--1-21}, & \text{q}_{\Delta
}\text{(2730)} & \Delta\text{(2740)} & \text{N(2740)}\\
& \text{ \ \ \ \ \ \ \ 211,2-11,-211,-2-11}) & \text{q}_{\Delta}\text{(2730)}
& \Delta\text{(2740)} & \text{N(2740)}\\
E_{H}=5 & \vec{n}=(\text{202,-202,022,0-22}) & \text{q}_{\Delta}\text{(2730)}
& \Delta\text{(2740)} & \text{N(2740)}\\
E_{H}=5 & \vec{n}=(\text{013,0-13,103,-103}) & \text{q}_{\Delta}\text{(2730)}
& \Delta\text{(2740)} & \text{N(2740)}\\
\ldots & ... & ... & ... & ...
\end{array}
\label{DELTA_4}%
\end{equation}

2. The single bands on the axis $\Delta$($\Gamma$-H) \cite{Q-Mass}, ( see
Appendix B)%

\begin{equation}%
\begin{array}
[c]{llllll}%
E_{H}=1 & \Delta S=-1 & J_{\text{H}}=1 & \text{q}_{S}\text{(1390)} &
\Lambda\text{(1400)} & \Sigma(1400)\\
E_{\Gamma}=4 & \Delta S=+1 & J_{\Gamma}=1 & \text{q}_{C}\text{(2270)} &
\Lambda_{C}^{+}\text{(2280)} & \Sigma_{C}\text{(2480)}\\
E_{H}=9 & \Delta S=-1 & J_{\text{H}}=2 & \text{q}_{S}\text{(4370)} &
\Lambda\text{(4380)} & \Sigma(4380)\\
E_{\Gamma}=16 & \Delta S=+1 & J_{\Gamma}=2 & \text{q}_{C}\text{(6490)} &
\Lambda_{C}^{+}\text{(6500)} & \Sigma_{C}\text{(6700)}\\
E_{H}=25 & \Delta S=-1 & J_{\text{H}}=3 & \text{q}_{S}\text{(10230)} &
\Lambda\text{(10240)} & \Sigma(10240)\\
E_{\Gamma}=36 & \Delta S=+1 & J_{\Gamma}=3 & \text{q}_{C}\text{(13590)} &
\Lambda_{C}^{+}\text{(13600)} & \Sigma_{C}\text{(13800)}\\
...\text{ .} &  &  &  &  &
\end{array}
\label{B-D-One}%
\end{equation}

\subsubsection{ The Baryons On the Axis $\Lambda$\ ($\Gamma$%
-P)\ \ \ \ \ \ \ \ \ \ \ \ \ \ \ \ \ \ \ \ \ \ \ }%

\begin{equation}%
\begin{array}
[c]{llccc}%
E_{P}=3/4 & \vec{n}=(\text{101,011,110)} & \text{q}_{\Sigma}\text{(1200)} &
\Sigma\text{(1210)} & \Lambda\text{(1210)}\\
E_{\Gamma}=2 & \vec{n}=(\text{1-10,-110,01-1,} & \text{q}_{\Sigma
}\text{(1650)} & \Sigma\text{(1660)} & \Lambda\text{(1660)}\\
& \text{ \ \ \ \ \ \ \ 0-11,10-1,-101)} & \text{q}_{\Sigma}\text{(1650)} &
\Sigma\text{(1660)} & \Lambda\text{(1660)}\\
E_{\Gamma}=2 & \vec{n}=(\text{-10-1,0-1-1,-1-10)} & \text{q}_{\Sigma
}\text{(1650)} & \Sigma\text{(1660)} & \Lambda\text{(1660)}\\
E_{P}=11/4 & \vec{n}=(\text{020,002,200)} & \text{q}_{\Sigma}\text{(1920)} &
\Sigma\text{(1930)} & \Lambda\text{(1930)}\\
E_{P}=11/4 & \vec{n}=(\text{121,211,112)} & \text{q}_{\Sigma}\text{(1920)} &
\Sigma\text{(1930)} & \Lambda\text{(1930)}\\
E_{\Gamma}=4 & \vec{n}=(\text{0-20,-200,00-2)} & \text{q}_{\Sigma
}\text{(2370)} & \Sigma\text{(2380)} & \Lambda\text{(2380)}\\
E_{P}=19/4 & \vec{n}=(\text{1-12,-112,21-1,} & \text{q}_{\Sigma}\text{(2640)}
& \Sigma\text{(2650)} & \Lambda\text{(2650)}\\
& \text{ \ \ \ \ \ \ \ 2-11,12-1,-121)} & \text{q}_{\Sigma}\text{(2640)} &
\Sigma\text{(2650)} & \Lambda\text{(2650)}\\
E_{P}=19/4 & \vec{n}=(\text{202,022,220)} & \text{q}_{\Sigma}\text{(2640)} &
\Sigma\text{(2650)} & \Lambda\text{(2650)}\\
E_{\Gamma}=6 & \vec{n}=(\text{-211,2-1-1,2-1-1}, & \text{q}_{\Sigma
}\text{(3090)} & \Sigma\text{(3100)} & \Lambda\text{(3100)}\\
& \text{\ \ \ \ \ \ \ 11-2,-12-11-21)} & \text{q}_{\Sigma}\text{(3090)} &
\Sigma\text{(3100)} & \Lambda\text{(3100)}\\
E_{\Gamma}=6 & \vec{n}\text{=(-1-21,1-2-1,-11-2,} & \text{q}_{\Sigma
}\text{(3090)} & \Sigma\text{(3100)} & \Lambda\text{(3100)}\\
& \text{\ \ \ \ \ \ 1-1-2,-21-1,-2-11)} & \text{q}_{\Sigma}\text{(3090)} &
\Sigma\text{(3100)} & \Lambda\text{(3100)}\\
E_{\Gamma}=6 & \vec{n}\text{=(-1-2-1,-1-1-2,-2-1-1)} & \text{q}_{\Sigma
}\text{(3090)} & \Sigma\text{(3100)} & \Lambda\text{(3100)}\\
\ldots. &  &  &  &
\end{array}
\label{B-SEGMA}%
\end{equation}

\subsubsection{The baryons On the Axis $\Sigma$($\Gamma$-N)}

1. The two-fold energy bands on the axis $\Sigma(\Gamma-N)$%

\begin{equation}%
\begin{array}
[c]{llll}%
E_{\Gamma}=2 & \vec{n}=(\text{1-10,-110}) & q_{\Xi}(1650) & \Xi(1660)\\
E_{N}=5/2 & \vec{n}=(\text{200,020}) & q_{\Xi}(1830) & \Xi(1840)\\
E_{\Gamma}=4 & \vec{n}=(\text{002,00-2}) & q_{\Xi}(2370) & \Xi(2380)\\
& \vec{n}=(\text{-200,0-20}) & q_{\Xi}(2370) & \Xi(2380)
\end{array}
\label{B-Sigema2}%
\end{equation}

2. The four-fold degenerate energy bands on the axis $\Sigma(\Gamma-N)$
\begin{equation}%
\begin{array}
[c]{llcc}%
E_{N}=3/2 & \vec{n}=(\text{101,10-1,011,01-1}) & 2\times q_{\Xi}(1470) &
2\text{ }\Xi(1480)\\
E_{\Gamma}=2 & \vec{n}=(\text{-101,-10-1,0-11,0-1-1}) & 2\times q_{\Xi
}(1650) & 2\text{ }\Xi(1660)\\
E_{N}=7/2 & \vec{n}=(\text{121,12-1,211,21-1}) & 2\times q_{\Xi}(2190) &
2\text{ }\Xi(2200)
\end{array}
\label{B-Sigema4}%
\end{equation}

3. The single energy bands on the axis $\Sigma$($\Gamma$-N) \cite{Q-Mass},
(see Appendix B)
\begin{equation}%
\begin{array}
[c]{llll}%
E_{N}=1/2 & \vec{n}=(\text{110}) & \text{q}_{S}\text{(1110)} & \Lambda
\text{(1120)}\\
E_{\Gamma}=2 & \vec{n}=(\text{-1-10}) & \text{q}_{\Omega}\text{(1650)} &
\Omega^{-}\text{(1660)}\\
E_{N}=9/2 & \vec{n}=(\text{220}) & \text{q}_{S}\text{(2550)} & \Lambda
\text{(2560)}\\
E_{\Gamma}=8 & \vec{n}=(\text{-2-20}) & \text{q}_{\Omega}\text{(3810)} &
\Omega^{-}\text{(3820)}\\
E_{N}=25/2 & \vec{n}=(\text{330}) & \text{q}_{b}\text{(5530)} & \Lambda
_{b}^{0}\text{(5540)}\\
E_{\Gamma}=18 & \vec{n}=(\text{-3-30}) & \text{q}_{\Omega}\text{(7310)} &
\Omega^{-}\text{(7320)}\\
E_{N}=49/2 & \vec{n}=(\text{440}) & \text{q}_{b}\text{(9950)} & \Lambda
_{b}^{0}\text{(9960)}%
\end{array}
\label{B-Sigma1}%
\end{equation}

The baryons of the D-axis, the F-axis and the G-axis are shown in Appendix C.

\subsection{Comparing the Results\ \ \ \ \ \ \ \ \ \ \ \ \ \ \ \ \ \ \ \ \ \ }

Using Tables 1-6, we compare the theoretical baryon spectrum of the BCC Quark
Lattice Model with the experimental results \cite{Baryon02}. In the
comparison, we do not take into account the angular momenta of the
experimental results. We assume that the small differences of the masses in
the same group of baryons with the same quantum numbers and similar masses are
from their different angular momenta. If we ignore this effect, their masses
would be essentially the same. In the comparison, we use the baryon name to
represent the intrinsic quantum numbers as shown in the second column of Table
1. \ \ \ \ 

\ \ \ \ 

\qquad\qquad Table 1. \ The Ground States of the Baryons.%

\begin{tabular}
[c]{|l|l|l|l|l|}\hline
{\small Theory} & Quantum. No & {\small Experiment} & R & Life Time\\\hline
Name({\small M}) & \ S, \ C,\ b, \ \ \ I, \ \ Q & Name({\small M}) &  &
\\\hline
N$^{+}$(940) & \ 0,\ \ 0, \ 0, \ {\small 1/2, \ \ }1 & p(938) & 0.2 &
$>$%
10$^{25}years$\\\hline
N$^{0}$(940) & 0, \ 0, \ 0, \ {\small 1/2, \ \ }0 & n(940) & 0.0 & 885.7
s\\\hline
$\Lambda^{0}(1120)$ & -1, \ 0, \ 0, \ \ 0, \ \ 0 & $\Lambda^{0}(1116)$ & 0.4 &
2.6$\times$ \ 10$^{-10}$s\\\hline
$\Sigma^{+}(1210)$ & -1, \ 0, \ 0, \ \ 1, \ \ 1 & $\Sigma^{+}(1189)$ & 1.8 &
.80$\times$ \ 10$^{-10}$s\\\hline
$\Sigma^{0}(1210)$ & -1, \ 0, \ 0, \ \ 1, \ \ 0 & $\Sigma^{0}(1193)$ & 1.4 &
7.4$\times$ \ 10$^{-20}$s\\\hline
$\Sigma^{-}(1210)$ & -1, \ 0, \ 0, \ \ 1, \ -1 & $\Sigma^{-}(1197)$ & 1.1 &
1.5$\times$10$^{-10}$s\\\hline
$\Xi^{0}(1300)$ & -2, \ 0, \ 0, \ {\small 1/2}, \ 0 & $\Xi^{0}(1315)$ & 1.2 &
2.9$\times$10$^{-10}$s\\\hline
$\Xi^{-}(1300)$ & -2, \ 0, \ 0, \ {\small 1/2}, -1 & $\Xi^{-}(1321)$ & 1.6 &
1.6$\times$10$^{-10}$s\\\hline
$\Omega^{-}(1660)$ & -3, \ 0, \ 0. \ \ 0,\ \ -1 & $\Omega^{-}(1672)$ & 0.7 &
.82$\times$10$^{-10}$s\\\hline
$\Lambda_{c}^{+}(2280)$ & 0, \ 1, \ 0, \ \ 0, \ \ \ 1 & $\Lambda_{c}%
^{+}(2285)$ & 0.2 & 200$\times$10$^{-15}$s\\\hline
$\Xi_{c}^{+}(2450)$ & -1, \ 1, \ 0, \ {\small 1/2}, \ 1 & $\Xi_{c}^{+}(2466)$
& 0.6 & 442$\times$10$^{-15}$\\\hline
$\Xi_{c}^{0}(2450)$ & -1, \ 1, \ 0, \ {\small 1/2}, \ 0 & $\Xi_{c}^{0}(2470)$
& 0.8 & 98$\times$10$^{-15}$s\\\hline
$\Omega_{c}^{0}(2750)$ & 0, \ 0, -1, \ \ 0, \ \ 0 & $\Omega_{c}(2698)$ & 1.9 &
64$\times$10$^{-15}s$\\\hline
$\Lambda_{b}^{0}(5540)$ & 0, \ 0, -1, \ \ 0, \ \ 0 & $\Lambda_{b}^{0}(5641)$ &
1.8 & 1.2310$^{-12}s$\\\hline
$\Sigma_{C}^{++}(2480)$ & -1, \ 1, 0, \ \ 1, \ \ 2 & $\Sigma_{C}^{++}(2453)$ &
1.1 & $\Gamma$=2.0 Mev\\\hline
$\Sigma_{C}^{+}(2480)$ & -1, \ 1, 0, \ \ 1, \ \ 1 & $\Sigma_{C}^{++}(2451)$ &
1.2 & $\Gamma$%
$<$%
4.6 Mev\\\hline
$\Sigma_{C}^{0}(2480)$ & -1, \ 1, \ 0, \ \ 1, \ \ 0 & $\Sigma_{C}^{++}(2452)$
& 1.2 & $\Gamma$=1.6 Mev\\\hline
$\Delta^{++}(1240)$ & 0,\ \ \ 0, \ 0, \ {\small 3/2, \ \ }$2$ & $\Delta
^{++}(1232)$ & 0.7 & $\Gamma$=120 Mev\\\hline
$\Delta^{+}(1240)$ & 0,\ \ \ 0, \ 0, \ {\small 3/2, \ \ }1 & $\Delta
^{+}(1232)$ & 0.7 & $\Gamma$=120 Mev\\\hline
$\Delta^{0}(1240)$ & 0,\ \ \ 0, \ 0, \ {\small 3/2, \ \ }0 & $\Delta
^{0}(1232)$ & 0.7 & $\Gamma$=120 Mev\\\hline
$\Delta^{-}(1240)$ & 0,\ \ \ 0, \ 0, \ {\small 3/2, \ -}1 & $\Delta^{-}(1232)$
& 0.7 & $\Gamma$=120 Mev\\\hline
\end{tabular}

In the fourth column, R =($\frac{\Delta\text{M}}{\text{M}}$){\small \%.}

\ \ The most importamt baryons are shown in Table 1. \ These baryons have
relatively long lifetimes. They are the most important experimental results of
the baryons. From Table 1, we can see that all theoretical intrinsic quantum
numbers ($I$, $S$, $C$, $b$ and $Q$) are the same as those in the experimental
results. Also the theoretical mass values agree well with the experimental values.

\ \ \ \ \ \ \ \ \ \ \ \ Table 2. Two Kinds of Strange Baryons $\Lambda$ and
$\Sigma$ ($S=-1$)%

\begin{tabular}
[c]{|l|l|l||l|l|l|}\hline
Theory & Experiment & $\frac{\Delta\text{M}}{\text{M}}\%$ & Theory &
Experiment & $\frac{\Delta\text{M}}{\text{M}}\%$\\\hline
$\mathbf{\Lambda(1120)}$ & $\mathbf{\Lambda(1116)}$ & \textbf{0.36} &
$\mathbf{\Sigma(1210)}$ & $\mathbf{\Sigma(1193)}$ & \textbf{1.4}\\\hline%
\begin{tabular}
[c]{l}%
$\Lambda(1400)$\\
$\Lambda(1500)$%
\end{tabular}
& $%
\begin{tabular}
[c]{l}%
$\Lambda(1405$\\
$\Lambda(1520)$%
\end{tabular}
\ \ \ $ &  & $\Sigma(1400)\ \ \ \ \ \ $ & $\Sigma(1385)$ & \\\hline
$\overline{\mathbf{\Lambda(1450)}}$ & $\overline{\mathbf{\Lambda(1463)}}$ &
\textbf{0.9} & $\mathbf{\Sigma(1400)}$ & $\mathbf{\bar{\Sigma}(1385)}$ &
\textbf{1.1}\\\hline%
\begin{tabular}
[c]{l}%
$\Lambda(1660)$\\
$\Lambda(1660)$\\
$\Lambda(1660)$%
\end{tabular}
&
\begin{tabular}
[c]{l}%
$\Lambda(1600)$\\
$\Lambda(1670)$\\
$\Lambda(1690)$%
\end{tabular}
&  &
\begin{tabular}
[c]{l}%
$\Sigma(1660)$\\
$\Sigma(1660)$\\
$\Sigma(1660)$%
\end{tabular}
&
\begin{tabular}
[c]{l}%
$\Sigma(1660)$\\
$\Sigma(1670)$\\
$\Sigma(1750)$\\
$\Sigma(1775)$%
\end{tabular}
& \\\hline
$\mathbf{\bar{\Lambda}(1660)}$ & $\mathbf{\bar{\Lambda}(1653)}$ & \textbf{0.4}
& $\mathbf{\bar{\Sigma}(1660)}$ & $\mathbf{\bar{\Sigma}(1714)}$ &
\textbf{3.2}\\\hline%
\begin{tabular}
[c]{l}%
$\Lambda(1840)$\\
$\Lambda(1840)$\\
$\Lambda(1930)$\\
$\Lambda(1930)$\\
$\Lambda(1930)$%
\end{tabular}
&
\begin{tabular}
[c]{l}%
$\Lambda(1800)$\\
$\Lambda(1810)$\\
$\Lambda(1820)$\\
$\Lambda(1830)$\\
$\Lambda(1890)$%
\end{tabular}
&  &
\begin{tabular}
[c]{l}%
$\Sigma(1930)$\\
$\Sigma(1930)$\\
$\Sigma(1930)$%
\end{tabular}
&
\begin{tabular}
[c]{l}%
$\Sigma(1915)$\\
$\Sigma(1940)$%
\end{tabular}
& \\\hline
$\mathbf{\bar{\Lambda}(1894}$ & $\mathbf{\bar{\Lambda}(1830)}$ & \textbf{3.5.}
& $\mathbf{\bar{\Sigma}(1930)}$ & $\mathbf{\bar{\Sigma}(1928)}$ &
\textbf{.10}\\\hline%
\begin{tabular}
[c]{l}%
$\Lambda(2020)$\\
$\Lambda(2020)$\\
$\Lambda(2130)$\\
$\Lambda(2130)$%
\end{tabular}
&
\begin{tabular}
[c]{l}%
$\Lambda(2100)$\\
$\Lambda(2110)$%
\end{tabular}
&  & $\mathbf{\Sigma(2020)}$ & $\mathbf{\Sigma(2030)}$ & \textbf{.50}\\\hline
$\mathbf{\bar{\Lambda}(2071)}$ & $\mathbf{\Lambda(2105)}$ & \textbf{1.4} &  &
& \\\hline
$\Lambda(2380)\ $ & $\Lambda(2350)$ &  &
\begin{tabular}
[c]{l}%
$\Sigma(2130)$\\
$\Sigma(2130)$%
\end{tabular}
&
\begin{tabular}
[c]{l}%
$\Sigma(2080)^{\ast}$\\
$\Sigma(2250)$%
\end{tabular}
& \\\hline
$\mathbf{\bar{\Lambda}(2380)}$ & $\mathbf{\bar{\Lambda}(2350)}$ & \textbf{1.3}
& $\mathbf{\bar{\Sigma}(2130)}$ & $\mathbf{\bar{\Sigma}(2165)}$ &
\textbf{1.6}\\\hline
2$\mathbf{\Lambda(2560)}$ & $\mathbf{\Lambda(2585)}^{\ast}$ & \textbf{1.0} &
$\Sigma(2380)$ & $\Sigma(2455)^{\ast}$ & \textbf{3.0}\\\hline
\textbf{6}$\mathbf{\Lambda(2650)}$ &  &  & \textbf{6}$\mathbf{\Sigma(2650)}$ &
$\mathbf{\Sigma(2620)}^{\ast}$ & \textbf{1.1}\\\hline
\end{tabular}
\ \ \ \ \ \ \ \ \ \ \ \ \ \ \ \ \ \ \ \ \ \ \ \ \ \ \ \ \ \ \ 

{\small \ *Evidences of existence for these baryons are only fair, they are
not listed in the Baryon Summary Table \cite{Baryon02}.}

\ \ Two kinds of the strange baryons $\Lambda$ and $\Sigma$ are compared in
Table 2. Their theoretical and experimental intrinsic quantum numbers are the
same.\ The theoretical masses of the baryons $\Lambda$ and $\Sigma$ agree well
with the experimental results.\newpage

\qquad\ \ \ \ \ \ \ Table 3. The Unflavored Baryons $N$ and $\Delta$ ($S$=
$C$=$b$ = 0) \ %

\begin{tabular}
[c]{|l|l|l||l|l|l|}\hline
Theory & Experiment & $\frac{\Delta\text{M}}{\text{M}}\%$ & Theory &
Experiment & $\frac{\Delta\text{M}}{\text{M}}\%$\\\hline%
\begin{tabular}
[c]{l}%
{\small 2}$N(1210)$\\
$N(1300)$%
\end{tabular}
&  &  & $%
\begin{array}
[c]{c}%
2\Delta(1210)\\
1\Delta(1300)
\end{array}
$ & $\Delta(1232)$ & \\\hline
$\mathbf{\bar{N}(1240)}$ &  &  & $\mathbf{\bar{\Delta}(1240)}$ &
$\mathbf{\bar{\Delta}(1232)}$ & \textbf{0.7}\\\hline
$N(1480)$ &
\begin{tabular}
[c]{l}%
$N(1440)$\\
$N(1520)$\\
$N(1535)$%
\end{tabular}
&  &  &  & \\\hline
$\mathbf{\bar{N}(1480)}$ & $\mathbf{\bar{N}(1498)}$ & \textbf{1.2} &  &  &
\\\hline%
\begin{tabular}
[c]{l}%
$N(1660)$\\
$N(1660)$%
\end{tabular}
&
\begin{tabular}
[c]{l}%
$N(1650)$\\
$N(1675)$\\
$N(1680)$\\
$N(1700)$\\
$N(1710)$\\
$N(1720)$%
\end{tabular}
&  &
\begin{tabular}
[c]{l}%
$\Delta(1660)$\\
$\Delta(1660)$%
\end{tabular}
&
\begin{tabular}
[c]{l}%
$\Delta(1600)$\\
$\Delta(1620)$\\
$\Delta(1700)$%
\end{tabular}
& \\\hline
$\mathbf{\bar{N}(1660)}$ & $\mathbf{\bar{N}(1689)\ \ }$ & \textbf{1.7} &
$\mathbf{\bar{\Delta}(1660)}$ & $\mathbf{\bar{\Delta}(1640)}$ & \textbf{1.2}%
\\\hline%
\begin{tabular}
[c]{l}%
2$N(1840)$\\
3$N(1930)$\\
2$N(2020)$%
\end{tabular}
&
\begin{tabular}
[c]{l}%
$N(1900)\ast$\\
$N(1990)\ast$\\
$N(2000)\ast$\\
$N(2080)\ast$%
\end{tabular}
&  &
\begin{tabular}
[c]{l}%
1$\Delta(1930)$\\
1$\Delta(1930)$\\
1$\Delta(1930)$\\
1$\Delta(2020)$\\
1$\Delta(2020)$%
\end{tabular}
&
\begin{tabular}
[c]{l}%
$\Delta(1905)$\\
$\Delta(1910)$\\
$\Delta(1920)$\\
$\Delta(1930)$\\
$\Delta(1950)$%
\end{tabular}
& \\\hline
$\mathbf{\bar{N}(1930)}$ & $\mathbf{\bar{N}(1923)}$ & \textbf{0.4} &
$\mathbf{\bar{\Delta}(1966)}$ & $\mathbf{\bar{\Delta}(1923)}$ & \textbf{2.3}%
\\\hline%
\begin{tabular}
[c]{l}%
$N(2200)$\\
$N(2200)$%
\end{tabular}
&
\begin{tabular}
[c]{l}%
$N(2190)$\\
$N(2220)$\\
$N(2250)$%
\end{tabular}
&  &  &  & \\\hline
$\mathbf{\bar{N}(2200)}$ & $\mathbf{\bar{N}(2220)}$ & \textbf{0.9} &  &  &
\\\hline
$\mathbf{N(2380)}$ &  &  & $\mathbf{\Delta(2380)}$ & $\mathbf{\Delta(2420)}$ &
\textbf{1.7}\\\hline
$3N(2650)$ & $\mathbf{N(2600)}$ & \textbf{1.9} & \textbf{3}$\Delta(2650)$ &  &
\\\hline
5$N(2740)$ & $\mathbf{N(2700)}^{\ast}$ & \textbf{1.5} & \textbf{4}%
$\Delta(2740)$ & $\mathbf{\Delta(2750)}^{\ast}$ & \textbf{0.4}\\\hline
1N(2920) &  &  & 3$\mathbf{\Delta(3100)}$ & $\mathbf{\Delta(2950)}^{\ast}$ &
5.1\\\hline
3N(3100) &  &  &  &  & \\\hline
&  &  &  &  & \\\hline
\end{tabular}

*{\small Evidences are fair, they are not listed in the Baryon Summary Table
}\cite{Baryon02}.

\ A comparison of the theoretical results with the experimental results of the
unflavored baryons $N$ and $\Delta$ is made in Table 3. From Table 3, we can
see that the intrinsic quantum numbers of the theoretical results are exactly
the same as those of the experimental results.\ Also the theoretical masses of
the baryons $N$ and $\Delta$\ agree well with the experimental results. The
theoretical results N(1210) and N(1300) are not found in the experiment. We
believe that they are covered up by the experimental baryon $\Delta(1232)$
because of the following reasons: (1) they are unflavored baryons with the
same S, C and b; (2) the width (120 Mev) of $\Delta(1232)$ is very large, and
the baryons $N(1210)$ and $N(1300)$ both fall within the width region of
$\Delta(1232)$; and (3) the average mass (1255 Mev) of $N(1210)$ and $N(1300)$
is essentially the same as the mass (1232 Mev) of $\Delta(1232)$ $(\Gamma$ =
120 Mev).

\ \ \ \ \ \ \ \ \ \ \ \ \ \ \ \ \ \ \ \ Table 4. The Baryons $\Xi$ and the
Baryons $\Omega$%

\begin{tabular}
[c]{|l|l|l||l|l|l|}\hline
Theory & Experiment & $\frac{\Delta\text{M}}{\text{M}}\%$ & Theory &
Experiment & $\frac{\Delta\text{M}}{\text{M}}\%$\\\hline
\textbf{2}$\mathbf{\Xi(1300)}$ & $\mathbf{\Xi(1318)}$ & \textbf{1.4} &
$\mathbf{\Omega(1660)}$ & $\mathbf{\Omega(1672)}$ & \textbf{0.7}\\\hline
\textbf{3}$\mathbf{\Xi(1480)}$ & $\mathbf{\Xi(1530)}$ & \textbf{3.3} &
$\Omega(2360)$ & $\Omega(2250)$ & \\\hline
\textbf{3}$\mathbf{\Xi(1660)}$ & $\mathbf{\Xi(1690)}$ & \textbf{1.5} &
$\Omega(2360)$ & $\Omega(2380)^{\ast}$ & \\\hline
\textbf{1}$\mathbf{\Xi(1840)}$ & $\mathbf{\Xi(1820)}$ & \textbf{1.1} &
$\Omega(2360)$ & $\Omega(2470)^{\ast}$ & \\\hline
\textbf{2}$\mathbf{\Xi(1930)}$ & $\mathbf{\Xi(1950)}$ & \textbf{1.0} &
$\overline{\mathbf{\Omega(2360)}}$ & $\overline{\mathbf{\Omega(2367)}}$ &
\textbf{0.3}\\\hline
\textbf{3}$\mathbf{\Xi(2020)}$ & $\mathbf{\Xi(2030)}$ & \textbf{0.5} &
$\mathbf{\Omega(2740)}$ &  & \\\hline
\textbf{4}$\mathbf{\Xi(2200)}$ & $\mathbf{\Xi(2250)}^{\ast}$ & \textbf{2.2} &
$\mathbf{\Omega(2880)}$ &  & \\\hline
\textbf{2}$\mathbf{\Xi(2380)}$ & $\mathbf{\Xi(2370)}^{\ast}$ & \textbf{0.4} &
$\mathbf{\Omega(3820)}$ &  & \\\hline
3$\Xi(2560)$ & $\mathbf{\Xi(2500)}^{\ast}$ & \textbf{0.3} &  &  & \\\hline
5$\Xi(2740)$ &  &  &  &  & \\\hline
\end{tabular}

{\small *Evidences of existence for these baryons are only fair, they are not
listed }

\ {\small in the Baryon Summary Table \cite{Baryon02}.}

\ \ The theoretical intrinsic quantum numbers of the baryons $\Xi$ and
$\Omega$ are the same as the experimental results (see Table 4). The
theoretical masses of the baryons $\Xi$ and $\Omega$ are compatible with the
experimental results.\newpage

\qquad\qquad Table 5. Charmed \ $\Lambda_{c}^{+}$\ and Bottom $\Lambda
_{b\text{ }}^{0}$ Baryons \ %

\begin{tabular}
[c]{|l|l|l||l|l|l|}\hline
Theory & Experiment & $\frac{\Delta\text{M}}{\text{M}}\%$ & Theory &
Experiment & $\frac{\Delta\text{M}}{\text{M}}\%$\\\hline
1$\mathbf{\Lambda}_{c}^{+}\mathbf{(2280)}$ & $\mathbf{\Lambda}_{c}%
^{+}\mathbf{(2285)}$ & \textbf{0.22} & 1$\mathbf{\Lambda}_{b}^{0}%
$\textbf{(5540)} & $\mathbf{\Lambda}_{b}^{0}$\textbf{(5641)} & 1.8\\\hline
1$\mathbf{\Lambda}_{c}^{+}\mathbf{(2560)}$ & $\mathbf{\Lambda}_{c}%
^{+}\mathbf{(2593)}$ & \textbf{1.2} & 1$\Lambda_{b}^{0}${\small (9960)} &  &
\\\hline
1$\Lambda_{c}^{+}(2760)$ & $\Lambda_{c}^{+}(2625)$ &  & 1$\Lambda_{b}^{0}%
${\small (15820)} &  & \\\hline
1$\Lambda_{c}^{+}(2760)$ & $\Lambda_{c}^{+}(2880)^{\ast}$ &  &  &  & \\\hline
$\overline{\mathbf{\Lambda}_{c}^{+}\mathbf{(2760)}}$ & $\overline
{\mathbf{\Lambda}_{c}^{+}\mathbf{(2723)}}$ & \textbf{1.4} &  &  & \\\hline
1$\mathbf{\Lambda}_{c}^{+}\mathbf{(6600)}$ &  &  &  &  & \\\hline
\end{tabular}

\ The charmed and bottom baryons $\Lambda_{c}^{+}$ and $\Lambda_{b}^{0}$ can
be found in Table 5. \ The experimental masses of the charmed baryons
($\Lambda_{c}^{+}$) and bottom baryons ($\Lambda_{b}^{0}$) coincide with the
theoretical results.

{\small \bigskip}

\qquad\qquad Table 6. Charmed Strange Baryon $\Xi_{c}$, $\Sigma_{c}$ and
$\Omega_{C}$ \ %

\begin{tabular}
[c]{|l|l|l||l|l|l|}\hline
Theory & Experiment & $\frac{\Delta\text{M}}{\text{M}}\%$ & Theory &
Experiment & $\frac{\Delta\text{M}}{\text{M}}\%$\\\hline
1$\mathbf{\Xi}_{c}$\textbf{(2450)} & $\mathbf{\Xi}_{c}$\textbf{(2469)} &
\textbf{0.8} & 1$\mathbf{\ \Sigma}_{c}$\textbf{(2480)} &
\begin{tabular}
[c]{|l|}\hline
$\Sigma_{c}$(2455)$\mathbf{\ }$\\\hline
$\Sigma_{c}$(2520)\\\hline
\end{tabular}
& \\\hline
1$\Xi_{c}$(2540) & $\Xi_{c}^{\prime}$(2577) & \textbf{1.4} &
1$\mathbf{\ \Sigma}_{c}$\textbf{(2480)} & $\mathbf{\ \Sigma}_{c}%
$\textbf{(2488)} & \textbf{0.3}\\\hline
1$\Xi_{c}$(2650) & $\Xi_{c}$(2645) & \textbf{0.2} &  &  & \\\hline%
\begin{tabular}
[c]{|l|}\hline
$\Xi_{C}$(2740)\\\hline
$\Xi_{C}$(2970)\\\hline
\end{tabular}
&
\begin{tabular}
[c]{l}%
$\Xi_{c}$(2790)\\
$\Xi_{c}$(2815)
\end{tabular}
&  &
\begin{tabular}
[c]{|l|}\hline
1$\Omega_{C}$(2740)\\\hline
1$\Omega_{C}$(2760)\\\hline
\end{tabular}
&  & \\\hline
$\overline{\Xi_{C}(2855)}$ & $\overline{\Xi_{c}(2803)}$ & \textbf{1.9} &
1$\mathbf{\Omega}_{C}$\textbf{(2750)} & $\mathbf{\Omega}_{C}$\textbf{(2698)} &
\textbf{1.9}\\\hline
1$\Xi_{C}$(3670) &  &  & 1$\Omega_{C}$\textbf{(2880)} &  & \\\hline
\end{tabular}
\ \ \ 

{\small *Evidences of existence for these baryons are only fair, they are not}

{\small listed in the Baryon Summary Table \cite{Baryon02}.}

\ \ \ Finally, we compare the theoretical results with the experimental
results for the charmed strange baryons $\Omega_{C}$, $\Xi_{c}$ and
$\Sigma_{c}$ in Table 6.\ Their intrinsic quantum numbers are all matched
completely, and their masses also agree well.

\ \ In summary, the BCC Quark Lattice Model explains all baryon experimental
intrinsic quantum numbers and masses. Virtually no experimentally confirmed
baryon is not included in the model. The angular momenta and the parities of
the baryons, however, are not included in this paper. They depend on the wave
functions of the energy bands. We will discuss them in some later
papers.$\ \ $%
\ \ \ \ \ \ \ \ \ \ \ \ \ \ \ \ \ \ \ \ \ \ \ \ \ \ \ \ \ \ \ \ \ \ \ \ \ \ \ \ \ \ \ \ \ \ \ \ \ \ \ \ \ \ \ \ \ \ \ \ \ \ \ \ \ \ \ \ \ \ \ \ \ \ \ \ \ \ \ \ \ \ \ \ \ \ \ \ \ \ \ \ \ \ \ \ \qquad
\ \ \ \ \ \ \ \ \ \ \ \ \ \ \ \ \ \ \ \ \ \ \ \ \ \ \ \ \ \ \ \ \ \ \ \ \ \ \ \ \ \ \ \ \ \ \ \ \ \ \ \ \ \ \ \ \ \ \ \ \ \ \ \ \ \ \ \ \ \ \ \ \ \ \ \ \ \ \ \ \ \ \ \ \ \ \ \ \ \ \ \ \ \ \ \ \ \ \ \ \ \ \ \ \ \ \ \ \ \ \ \ \ \ \ \ \ \ \ \ \ \ \ \ \ \ \ \ \ \ \ \ \ \ \ \ \ \ \ \ \ \ \ \ \ \ \ \ \ \ \ \ \ \ \ \ \ \ \ \ \ \ \ \ \ \ \ \ \ \ \ \ \ \ 

\section{Predictions}

\subsection{New Quarks}

1. There is an energy band excited quark spectrum (\ref{Quarks}). Some of the
quarks have already been discovered: q$_{N}$(930), d$_{S}$(1110), u$_{C}%
$(2270), d$_{b}$(5530), q$_{\Sigma}$(1200), q$_{\Xi}$(1290), $\Omega
$(1650),\ d$_{S}$(1390), q$_{\Sigma}$(1650) and so forth.

2. There is always an excited q in a baryon. It has more than 98\% of the mass
of the baryon. We hope experimental physasts to find them.

3. The new quarks d$_{S}$(4370), u$_{C}$(6490) and d$_{b}$(9950) still need to
be found. \ \ \ \ 

\subsection{New Baryons}%

\begin{equation}%
\begin{array}
[c]{cccc}%
\text{I=0} & \text{C=1} & \text{Q=1} & \Lambda_{C}^{+}\text{(6500)}\\
\text{I=0} & \text{S=-1} & \text{Q=0} & \Lambda^{0}\text{(4380)}\\
\text{I=0} & \text{b=-1} & \text{Q=0} & \Lambda_{b}^{0}\text{(9960)}%
\end{array}
\label{N-Baryon}%
\end{equation}

\section{Discussion}

1. The constant (\ref{360})%

\[
\alpha\text{ = h}^{2}\text{/2m}_{q}\text{\textit{a}}^{2}\text{ = 360 Mev,}%
\]
with a lattice constant \textit{a} $\leq$ 10$^{-18}$m. Thus the bare mass
(m$_{q}$) of the elementary quarks
\begin{equation}%
\begin{tabular}
[c]{l}%
$\text{m}_{q}\text{{}}=\text{{}}\frac{\text{h}^{2}}{\text{\ \ 720Mev}%
\times\text{\textit{a}}^{2}}\text{ }\geqslant\text{ }\frac{\text{43.90}%
\times\text{ 10}^{-68}\text{(J s)}^{2}}{\text{720}\times\text{ 10}^{6}%
\times\text{\ 1.602}\times\text{ 10}^{-19}\text{J\ }\times\text{\ 10}%
^{-36}\text{m}^{2}}$\\
$\text{\ m}_{q}\text{ }\geqslant\text{ 3.8 }\times\text{\ \ 10}^{-21}\text{kg
= 2.27 }\times\text{ \ 10}^{6}\text{m}_{p},$%
\end{tabular}
\ \ \ \ \label{Bare-M}%
\end{equation}
is much larger than the excited quark masses. This ensures that the
Schr\"{o}dinger equation (\ref{Schrod}) is a good approximation of the special
quark Dirac equation (\ref{S-Q-Dir}).

2. The paper has shown that the u-quark and the c-quark are the excited states
of the elementary u(0)-quark and that the d-quark, the s-quark and the b-quark
are the excited states of the elementary d(0)-quark. The u(0)-quark and the
d(0)-quark have a SU(2) symmetry (u(0) and d(0)). Therefore SU(3) (u, d and
s), SU(4) (u, d, s and c) and SU(5) (u, d, s, c and b) are correct although
there are large differences in masses between the quarks. In fact, the SU(3),
SU(4) and SU(5) are natural expansions of the SU(2). Since the bare masses of
the elementary quarks u(0) and d(0) (\ref{Bare-M}) are huge
\begin{equation}
\text{\ m}_{\text{u(0)}}\text{(or m}_{\text{d(0)}})\text{ }\geqq\text{
2.27}\times\text{10}^{6}\text{m}_{p}=\text{2.13}\times\text{10}^{9}\text{Mev.}
\label{BareM}%
\end{equation}
Thus the bare masses (taking the absolutely empty space as the zero energy
point) of the quarks (u, d, s, c, b, u' and d') are huge too:%

\begin{equation}%
\begin{tabular}
[c]{|l|l|}\hline
$\text{m}_{u}^{bare}\text{=\ m}_{u(0)}$+930, & $\text{m}_{d}^{bare}%
\text{=\ m}_{d(0)}$+930,\\\hline
$\text{m}_{c}^{bare}\text{=\ m}_{u(0)}$+2270, & $\text{m}_{s}^{bare}%
\text{=\ m}_{d(0)}$+1110,\\\hline
$\text{m}_{b}^{bare}\text{=\ m}_{d(0)}$+5530, & $\text{m}_{u\text{'}}%
^{bare}\text{=\ m}_{u(0)}$+ 3,\\\hline
$\text{m}_{d}^{bare}\text{=\ m}_{d(0)}$+ 7. & \ \ \\\hline
\end{tabular}
\ \ \label{QbM}%
\end{equation}
From (\ref{BareM}) and (\ref{QbM}), we can see that the bare masses of the
quarks (u, d, s, c, b, u' and d') are essentially the same. This is a rigorous
physical basis of the SU(3), SU(4), SU(5) and so forth symmetries. We had long
thought that SU(4) and SU(5) do not have a rigorous physical basis since the
differences of the quark masses are too large. Considering the bare masses of
these quarks, we now believe that SU(4) and SU(5) symmetries really exist.

3. At distance scales $\leq$ 10$^{-18}$ m, we can see the BCC quark lattice
and can deduce the rest masses (\ref{Quarks}) and the intrinsic quantum
numbers (I, S, C, b and q) (\ref{Q-Number}) of the quarks. The BCC Quark
Lattice Model not only provides a rigorous physical basis for the Quark Model,
but also opens a door to study more fundamental structure than the Standard Model.

4. At the distance scales
$>$%
10$^{-18}m$, although we cannot see the quark lattice, we can see the
u(0)-quark Dirac sea and the d(0)-quark Dirac sea (the lattice looks like
Dirac sea). Sometimes we can also see the s-quark, the c-quark and the b-quark
(inside baryons and mesons); and from the Dirac sea concept, we guess that
there will be an s-quark Dirac sea, a c-quark Dirac sea and a b-quark Dirac
sea also. Since we cannot see the quark lattice (we can only see the Dirac
seas), we cannot deduce the masses and the intrinsic quantum numbers (we can
measurement them by experiment). Naturally we think that the quarks (u, d, s,
c and b) are all independent elementary particles. The Standard Model is a
reasonably excellent approximation to nature at distance scales as small as
10$^{-18}$m \cite{S-Model}. Thus there is no contradiction between the
Standard Model and the BCC Quark lattice Model; however the BCC Quark Lattice
Model does provide a physical basis for the Standard Model.

5. After the discovery of superconductors, we now understand the vacuum
material. In a sense, the vacuum material (skeleton-- the BCC quark lattice)
works like a superconductor. Since the transition temperature is much higher
than the temperature at the center of the sun,\ all phenomena that we can see
are under the transition temperature. Thus there are no electric or mechanical
resistances to any particle or to any physical body moving inside the vacuum
material. Moving inside it, they look as if they are moving in completely
empty space. The vacuum material is a super superconductor.

6. A baryon is composed of three quarks in the Quark Model. The masses of the
quarks (m$_{u}$ = 1 to 4.5 Mev, m$_{d}$ = 5 to 8.5 Mev, m$_{s}$= 80 to 155
Mev, m$_{c}$= 1.0 to 1.4 Gev and m$_{b}$= 4.0 to 4.5 Gev) \cite{Q-Mass02} are
too small to build a stable baryon. Using the sum laws and the quark masses of
the Quark Model, we find the theoretical masses of the most important baryons.
We find that the theoretical baryon masses of the Quark Model are too small to
match the experimental masses. Thus there may be one quark with a large mass
inside the baryon. According to the BCC Quark Lattice Model, this quark with a
large mass really exists in a baryon-- it is excited quark (q) (\ref{Baryon-M}%
); the quarks with small masses are the accompanying excited quarks (u' and
d') \cite{Confine}$.$ The theoretical masses of the baryons of the BCC Quark
Lattice Model agree well with the experimental results. We list the
theoretical results and the experimental results \cite{Baryon02} for both the
Quark Model and the BCC Quark Latice Model as follows list (\ref{Baryon-M}%
):\ \ \ \ \
\begin{equation}%
\begin{tabular}
[c]{|l|l|l|l|l|l|l|l|l|}\hline
{\small Baryon} & {\small p({\tiny 938})} & {\small n({\tiny 940})} &
$\Lambda${\small ({\tiny 1116})} & $\Sigma^{0}${\small ({\tiny 1193})} &
$\Xi^{0}${\small ({\tiny 1315})} & $\Omega${\small ({\tiny 1672})} &
$\Lambda_{C}${\small ({\tiny 2285})} & $\Lambda_{b}${\small ({\tiny 5624}%
)}\\\hline
{\small Quark.} & {\small uud} & {\small udd} & {\small uds} & {\small uds} &
{\small uss} & {\small sss} & {\small udc} & {\small udb}\\\hline
{\small Mass } & {\small 13} & {\small 16} & {\small 128} & {\small 128} &
{\small 238} & {\small 353} & {\small 1210} & {\small 4260}\\\hline
{\small BCC } & {\small uu'd'} & {\small du'd'} & {\small d}$_{s}%
${\small u'd'} & {\small q}$_{\Sigma^{0}}${\small u'd'} & {\small q}$_{\Xi
^{0}}${\small u'd'} & {\small q}$_{\Lambda}${\small u'd'} & {\small q}$_{C}%
${\small u'd'} & {\small q}$_{b}${\small u'd'}\\\hline
{\small Mass } & {\small 940} & {\small 940} & {\small 1120} & {\small 1210} &
{\small 1300} & {\small 1660} & {\small 2280} & {\small 5540}\\\hline
{\small Exper.} & {\small 938} & {\small 940} & {\small 1116} & {\small 1193}
& {\small 1315} & {\small 1672} & {\small 2285} & {\small 5624 .}\\\hline
\end{tabular}
\ \ \ \ \ \ \label{Baryon-M}%
\end{equation}
Thus, the experimental results support the baryon model ({\small qu}%
'{\small d}') of the BCC Quark Lattice Model.

\section{Conclusions\ \ \ }

1. There are only two kinds of the elementary quarks (u(0) and d(0)) in the
vacuum state; other quarks (u, d, s(d$_{s}$), c(u$_{c}$), b(d$_{b}$), u', d'
and so forth) are all the excited states (from the vacuum) of the elementary
quarks (u(0) or d(0)).

2. The rest masses of the quarks are the energy minima of the energy bands that
represent the quarks. The baryon spectrum agrees well with experimental
results. This shows that the theoretical rest masses of the quarks are correct.

3. The strange number, the charmed number and the bottom number are the
products of the body center cubic periodic symmetries and the fluctuation of
the BCC quark lattice.

4. The BCC Quark Lattice Model not only provide a physical basis for the Quark
Model \cite{QuarkModel}, but also opens a door to study the more nature at
distance scales $\leq$ 10$^{-18}$ m and to look for the more fundamental
theory than the Standard Model.

5. Due to the existence of the vacuum material, all observable particles are
constantly affected by the vacuum material (the vacuum state quark lattice).
Thus some laws of statistics (such as fluctuation) cannot be ignored.

6. The classic approximation (\ref{Schrod}) is really a good approximation of
the special quark Dirac equation (\ref{S-Q-Dir}) for deducing the rest masses
and the intrinsic quantum numbers of the quarks, but it cannot deduce the spin
angular momentums of the quarks. Thus we need to resolve the special quark
Dirac equation (\ref{S-Q-Dir}) to find the angular momentum and to improve the
mass spectrum of the quarks that we have deduced using the classic approximation.

\textbf{Acknowledgment}

I would like to express my heartfelt gratitude to Dr. Xin Yu for checking the
calculations of the energy bands and for helping to write this paper. I
sincerely thank Professor Robert L. Anderson for his valuable advice. I also
acknowledge\textbf{\ }my indebtedness to Professor D. P. Landau for his help.
I thank Professor W. K. Ge very much for all of his help. I thank my classmate
J. S. Xie very much for checking the calculations of the energy bands. I thank
Professor Y. S. Wu, H. Y. Guo, and S. Chen very much for very useful
discussions. I also thank Dr. Fugao Wang very much for his help.

\section{Appendix A: The Body Center Cubic Quark Lattice in the Vacuum}

According to Dirac's sea concept \cite{D-Sea}, there is an electron-Dirac sea,
a $\mu$-lepton Dirac sea, a $\tau$-lepton Dirac sea, a $u$-quark Dirac sea, a
$d$-quark Dirac sea, an $s$-quark Dirac sea, a $c$-quark Dirac sea, a
$b$-quark Dirac sea and so forth in the vacuum. All of these Dirac seas are in
the same space, at any location, that is, at any physical space point. These
particles will interact with one another and form the perfect physical vacuum
material. Some kinds of particles, however, do not play an important role in
forming the physical vacuum material. First, the main force which makes and
holds the structure of the physical vacuum material must be the strong
interactions, not the electromagnetic interactions. Hence, in considering the
structure of the vacuum material, we leave out the Dirac seas of those
particles which do not have strong interactions ($e$, $\mu$ and $\tau$).
Secondly, the physical vacuum material is super stable, hence we also omit the
Dirac seas which can only make unstable baryons (the $s$-quark, the $c$-quark
and the $b$-quark). Finally, there are only two kinds of possible particles
left: the vacuum state $u(0)$-quarks and the vacuum state $d(0)$-quarks. There
are super strong attractive forces between the $u(0)$-quarks and the
$d(0)$-quarks (colors) that will make and hold the densest structure of the
vacuum material.

According to solid state physics \cite{Solid}, if two kinds of particles (with
radius $R_{1}<R_{2}$) satisfy the condition $1>R_{1}/R_{2}>0.73$, the densest
structure is the body center cubic crystal \cite{BCC}. We know the following:
first, the $u(0)$-quarks and the $d(0)$-quarks are not exactly the same, thus
$R_{u}\neq R_{d}$; second, they are very close to each other (the same isospin
with different $I_{z}$), thus $R_{u}\approx R_{d}$. Hence, if $R_{u}<R_{d}$
(or $R_{d}<R_{u}$), we have 1
$>$
R$_{u}$/R$_{d}$
$>$
0.73 (or 1%
$>$
R$_{d}$/R$_{u}$
$>$
0.73). Therefore, we conjecture that the vacuum state u(0)-quarks and
d(0)-quarks construct a body center cubic quark lattice in the vacuum.

\newpage

\section{Appendix B: The Energy Bands}

In Table 7-12, E$_{Start}$ (E$_{end}$) means the starting (end) energy
($\varepsilon$) of the energy band, it is the lowest (the highest) energy of
the energy band (bands). d (in the third column) is the number of the energy
bands with the same energy. We mark the energy band using n$_{1}$n$_{2}$%
n$_{3}$ (n$_{1}$, n$_{2}$, n$_{3}$ of\ (\ref{L-N})) and $\overline
{\text{n}_{i}}$ = - n$_{i}$.

\ \ \ \ \ \ \ \ \ \ \ \ \ \ \ \ \ \ \ \ \ \ \ \ \ \ \ \ \ \ \ \ \ \ \ \ \ \ \ \ \ \ \ \ \ \ \ \ \ \ \ \ \ \ \ \ \ \ \ \ \ \ \ \ \ \ \ \ \ \ \ \ \ \ \ \ \ \ \ \ \ \ \ \ \ \ \ \ \ \ \ \ \ \ \ \ \ \ \ \ \ \ \ \ \ \ \ \ \ \ \ \ \ \ \ \ \ \ \ \ \ \ \ \ \ \ \ \ \ \ \ \ \ \ \ \ \ \ \ \ \ \ \ \ \ \ \ \ \ \ \ \ \ \ \ \ \ \ \ \ \ \ \ \ \ \ \ \ \ \ \ 

\ \ \ \ \ \ \ \ \ \ \ \ \ \ Table 7. The Energy Bands of the $\Delta$-Axis

$%
\begin{tabular}
[c]{|l|l|l|l|l|}\hline
$\text{E}_{Start}$ & $\text{\ }{\small \varepsilon}$ & d & $\text{Energy\ Band
(n}_{1}\text{n}_{2}\text{n}_{3}$, $\text{...)}$ & $\text{E}_{end}$\\\hline
$\text{E}_{\Gamma}\text{=0}$ & 930 & 1 & (000) & $\text{E}_{H}\text{=1}%
$\\\hline
$\text{E}_{H}\text{=1}$ & 1290 & 4 & $\text{(101,}\overline{\text{1}%
}\text{01,011,0}\overline{\text{1}}\text{1)}$ & $\text{E}_{\Gamma}\text{=2}%
$\\\hline
$\text{E}_{H}\text{=1}$ & 1290 & 1 & (002) & $\text{E}_{\Gamma}\text{=4}%
$\\\hline
$\text{E}_{\Gamma}\text{=2}$ & 1650 & 4 & $\text{(110,1}\overline{\text{1}%
}\text{0,}\overline{\text{1}}\text{10,}\overline{\text{1}}\overline{\text{1}%
}\text{0)}$ & $\text{E}_{H}\text{=3}$\\\hline
$\text{E}_{\Gamma}\text{=2}$ & 1650 & 4 & $\text{(10}\overline{\text{1}%
}\text{,}\overline{\text{1}}\text{0}\overline{\text{1}}\text{,01}%
\overline{\text{1}}\text{,0}\overline{\text{1}}\overline{\text{1}}\text{)}$ &
$\text{E}_{H}\text{=5}$\\\hline
$\text{E}_{H}\text{=3}$ & 2010 & 4 & $\text{(112,1}\overline{\text{1}%
}\text{2,}\overline{\text{1}}\text{12,}\overline{\text{1}}\overline{\text{1}%
}\text{2)}$ & $\text{E}_{\Gamma}\text{=6}$\\\hline
$\text{E}_{\Gamma}\text{=4}$ & 2370 & 4 & $\text{(200,}\overline{\text{2}%
}\text{00,020,0}\overline{\text{2}}\text{0)}$ & $\text{E}_{H}\text{=5}%
$\\\hline
$\text{E}_{\Gamma}\text{=4}$ & 2370 & 1 & (00$\overline{2}$) & $\text{E}%
_{H}\text{=9}$\\\hline
$\text{E}_{H}\text{=5}$ & 2730 & 8 & $%
\begin{array}
[c]{l}%
\text{(121,1}\overline{\text{2}}\text{1,}\overline{\text{1}}\text{21,}%
\overline{\text{1}}\overline{\text{2}}\text{1, {\small \ }\ }\\
\text{211,2}\overline{\text{1}}\text{1,}\overline{\text{2}}\text{11,}%
\overline{\text{2}}\overline{\text{1}}\text{1)}%
\end{array}
$ & $\text{E}_{\Gamma}\text{=6}$\\\hline
$\text{E}_{H}\text{=5}$ & 2730 & 4 & $\text{(202,}\overline{\text{2}%
}\text{02,022,0}\overline{\text{2}}\text{2)}$ & $\text{E}_{\Gamma}\text{=8}%
$\\\hline
$\text{E}_{H}\text{=5}$ & 2730 & 4 & $\text{(013,0}\overline{\text{1}%
}\text{3,103,}\overline{\text{1}}\text{03)}$ & $\text{E}_{\Gamma}\text{=10}%
$\\\hline
$\text{E}_{\Gamma}\text{=6}$ & 3090 & 8 & $%
\begin{array}
[c]{l}%
\text{(12}\overline{\text{1}}\text{,1}\overline{\text{2}}\overline{\text{1}%
}\text{,}\overline{\text{1}}\text{21,}\overline{\text{1}}\overline{\text{2}%
}\overline{\text{1}}{\small ,}\text{ }\\
\text{21}\overline{\text{1}}\text{,2}\overline{\text{1}}\overline{\text{1}%
}\text{,}\overline{\text{2}}\text{1}\overline{\text{1}}\text{,}\overline
{\text{2}}\overline{\text{1}}\overline{\text{1}}\text{)}%
\end{array}
$ & $\text{E}_{H}\text{=9}$\\\hline
$\text{E}_{\Gamma}\text{=6}$ & 3090 & 4 & $\text{(11}\overline{\text{2}%
}\text{,1}\overline{\text{1}}\overline{\text{2}}\text{,}\overline{\text{1}%
}\text{1}\overline{\text{2}}\text{,}\overline{\text{1}}\overline{\text{1}%
}\overline{\text{2}}\text{)}$ & $\text{E}_{H}\text{=11}$\\\hline
$\text{E}_{H}\text{=9}$ & 4170 & 1 & ({\small 004)} & $\text{E}_{\Gamma
}\text{=16}$\\\hline
$\text{E}_{\Gamma}\text{=16}$ & 6690 & 1 & ({\small 00}$\overline
{\text{{\small 4}}}$) & $\text{E}_{H}\text{=25}$\\\hline
$\text{E}_{H}\text{=25}$ & 9930 & 1 & ($\text{{\small 006)}}$ & $\text{E}%
_{\Gamma}\text{=36}$\\\hline
$\text{E}_{\Gamma}\text{=36}$ & 13890 & 1 & ({\small 00}$\overline
{\text{{\small 6}}}$) & $\text{E}_{H}\text{=49}$\\\hline
$\text{E}_{H}\text{=49}$ & 17640 & 1 & (008) & $\text{E}_{\Gamma}\text{=64}%
$\\\hline
$\text{E}_{\Gamma}\text{=64}$ & 23970 & 1 & (00$\overline{8}$) & $\text{E}%
_{H}\text{=81}$\\\hline
$\text{E}_{H}\text{=81}$ & 31020 & 1 & (009) & $\text{E}_{\Gamma}\text{=100}%
$\\\hline
\end{tabular}
\ \ \ \ \ $

\ \ \ \ \ \ \ \ \ \ \ \ \ \ \ \ \ \ \ \ \ \ \ \ \ \ \ \ \ \ \ \ \ \ \ \ \ \ \ \ \ \ \ \ \ \ \ \ \ \ \ \ \ \ \ \ \ \ \ \ \ \ \ \ \ \ \ \ \ \ \ \ \newpage

\qquad\ \ \ \ \ \ \ \ Table 8. The Energy Bands of the $\Sigma$-Axis%

\begin{tabular}
[c]{|l|l|l|l|l|}\hline
$\text{E}_{\text{Start}}$ & $\text{\ \ }\varepsilon$ & d & $\text{Energy\ Band
(n}_{1}\text{n}_{2}\text{n}_{3})$ & $\text{E}_{end}$\\\hline
$\text{E}_{\Gamma}\text{= 0}$ & $\text{930}$ & 1 & $\text{(000)}$ &
$\text{E}_{N}\text{=1/2}$\\\hline
$\text{E}_{N}\text{=1/2}$ & 1110 & 2 & $(\text{1,1,0})$ & $\text{E}_{\Gamma
}\text{= 2}$\\\hline
$\text{E}_{N}\text{=3/2}$ & 1470 & 4 & $(\text{101,10}\overline{\text{1}%
}\text{,011,01}\overline{\text{1}}$ & $\text{E}_{\Gamma}\text{= 2}$\\\hline
$\text{E}_{\Gamma}\text{= 2}$ & 1650 & 2 & $(\text{1}\overline{\text{1}%
}\text{0,}\overline{\text{1}}\text{10)}$ & $\text{E}_{N}\text{=5/2}$\\\hline
$\text{E}_{\Gamma}\text{= 2}$ & 1650 & 4 & $(\overline{\text{1}}%
\text{01,}\overline{\text{1}}\text{0}\overline{\text{1}}\text{,0}%
\overline{\text{1}}\text{1,0}\overline{\text{1}}\overline{\text{1}}$ &
$\text{E}_{N}\text{=7/2}$\\\hline
$\text{E}_{\Gamma}\text{= 2}$ & 1650 & 2 & $(\overline{\text{1}}%
\text{,}\overline{\text{1}}\text{,0})$ & $\text{E}_{N}\text{=9/2}$\\\hline
$\text{E}_{N}\text{=5/2}$ & 1830 & 2 & $\text{(200,020)}$ & $\text{E}_{\Gamma
}\text{= 4}$\\\hline
$\text{E}_{N}\text{=7/2}$ & 2190 & 4 & (121,12$\overline{\text{1}}%
$,211,21$\overline{\text{1}}$) & $\text{E}_{\Gamma}\text{= 6}$\\\hline
$\text{E}_{\Gamma}\text{= 4}$ & 2370 & 2 & $\text{(002,00}\overline{\text{2}%
}\text{)}$ & $\text{E}_{N}\text{=9/2}$\\\hline
$\text{E}_{\Gamma}\text{= 4}$ & 2370 & 2 & $\text{(}\overline{\text{2}%
}\text{00,0}\overline{\text{2}}\text{0)}$ & $\text{E}_{N}\text{=13/2}$\\\hline
$\text{E}_{N}\text{=9/2}$ & 2550 & 2 & $\text{(112,11}\overline{\text{2}%
}\text{)}$ & $\text{E}_{\Gamma}\text{= 6}$\\\hline
$\text{E}_{N}\text{=9/2}$ & 2550 & 1 & (2,2,0) & $\text{E}_{\Gamma}\text{= 8}%
$\\\hline
$\text{E}_{\Gamma}\text{= 8}$ & 3810 & 1 & ($\overline{\text{2}}$%
,$\overline{\text{2}}$,0) & $\text{E}_{N}\text{=25/2}$\\\hline
$\text{E}_{N}\text{=25/2}$ & 5430 & 1 & (3,3,0) & $\text{E}_{\Gamma}\text{=
18}$\\\hline
$\text{E}_{\Gamma}\text{= 18}$ & 7410 & 1 & ($\overline{\text{3}}$%
,$\overline{\text{3}}$,0) & $\text{E}_{N}\text{=49/2}$\\\hline
$\text{E}_{N}\text{=49/2}$ & 9750 & 1 & (4,4,0) & $\text{E}_{\Gamma}\text{=
32}$\\\hline
$\text{E}_{\Gamma}\text{= 32}$ & 12450 & 1 & $\text{(}\overline{\text{4}%
}\text{,}\overline{\text{4}}\text{,0)}$ & $\text{E}_{N}\text{=81/2}$\\\hline
$\text{E}_{N}\text{=81/2}$ & 15510 & 1 & (5,5,0) & $\text{E}_{\Gamma}\text{=
50}$\\\hline
$\text{E}_{\Gamma}\text{= 50}$ & 18930 & 1 & ($\overline{\text{5}}%
\text{,}\overline{\text{5}}\text{,0})$ & $\text{E}_{N}\text{=121/2}$\\\hline
$\text{E}_{N}\text{=121/2}$ & 22710 & 1 & (6,6,0) & $\text{E}_{\Gamma}\text{=
72}$\\\hline
$\text{E}_{\Gamma}\text{= 72}$ & 26850 & 1 & ($\overline{6}$,$\overline{6}%
$,0) & $\text{E}_{N}\text{=169/2}$\\\hline
... & ... & ... & ... & ...\\\hline
\end{tabular}

\newpage

$\qquad\ \ $\qquad\ \ \ \ \ Table 9. The Energy Bands of the $\Lambda$-Axis%

\begin{tabular}
[c]{|l|l|l|l|l|}\hline
$\text{E}_{\text{Start}}$ & $\varepsilon$ & d & $\text{Energy\ Band (n}%
_{1}\text{n}_{2}\text{n}_{3}\text{, ...)}$ & $\text{E}_{\text{End}}$\\\hline
$\text{E}_{\Gamma}\text{=0}$ & $\text{930}$ & 1 & $\text{(000)}$ &
$\text{E}_{\text{P}}\text{=3/4 \ }$\\\hline
$\text{E}_{\text{P}}\text{=3/4 \ }$ & $\text{1200}$ & 3 &
$\text{(101,011,110)}$ & $\text{E}_{\Gamma}\text{= 2 \ \ }$\\\hline
$\text{E}_{\Gamma}\text{= 2 \ \ }$ & $\text{1650}$ & 6 & $\text{(1}%
\overline{\text{1}}\text{0,}\overline{\text{1}}\text{10,01}\overline{\text{1}%
}\text{, 0}\overline{\text{1}}\text{1,10}\overline{\text{1}}\text{,}%
\overline{\text{1}}\text{01)}$ & $\text{E}_{\text{P}}\text{=11/4}$\\\hline
$\text{E}_{\Gamma}\text{= 2 \ \ }$ & $\text{1650}$ & 3 & $\text{(}%
\overline{\text{1}}\text{0}\overline{\text{1}}\text{,0}\overline{\text{1}%
}\overline{\text{1}}\text{,}\overline{\text{1}}\overline{\text{1}}\text{0)}$ &
$\text{E}_{\text{P}}\text{=19/4}$\\\hline
$\text{E}_{\text{P}}\text{=11/4}$ & $\text{1920}$ & 3 & $\text{(020,002,200)}$
& $\text{E}_{\Gamma}\text{= 4\ }$\\\hline
$\text{E}_{\text{P}}\text{=11/4}$ & $\text{1920}$ & 3 & $\text{(121,211,112)}$
& $\text{E}_{\Gamma}\text{= 6 \ }$\\\hline
$\text{E}_{\Gamma}\text{= 4 \ \ }$ & $\text{2370}$ & 3 & (0$\overline{2}%
$0,$\overline{2}$00,00$\overline{2}$) & $\text{E}_{\text{P}}\text{=27/4}%
$\\\hline
$\text{E}_{\text{P}}\text{=19/4}$ & $\text{2640}$ & 6 & $\text{(1}%
\overline{\text{1}}\text{2,}\overline{\text{1}}\text{12,21}\overline{\text{1}%
}\text{, 2}\overline{\text{1}}\text{1,12}\overline{\text{1}}\text{,}%
\overline{\text{1}}\text{21)}$ & $\text{E}_{\Gamma}\text{= 6 \ }$\\\hline
$\text{E}_{\text{P}}\text{=19/4}$ & $\text{2640}$ & 3 & $\text{(202,022,220)}$
& $\text{E}_{\Gamma}\text{= 8 \ }$\\\hline
$\text{E}_{\Gamma}\text{= 6 \ \ }$ & $\text{3090}$ & 6 & $\text{(}\overline
{2}\text{11,2}\overline{\text{1}}\overline{\text{1}}\text{,}\overline
{\text{1}}\overline{\text{1}}\text{2,11}\overline{2}\text{,}\overline
{\text{1}}\text{2}\overline{\text{1}}\text{,1}\overline{2}\text{1)}$ &
$\text{E}_{\text{P}}\text{=27/4}$\\\hline
$\text{E}_{\Gamma}\text{= 6 \ \ }$ & $\text{3090}$ & 6 & $\text{(}%
\overline{\text{1}}\overline{2}\text{1,1}\overline{2}\overline{\text{1}%
}\text{,}\overline{\text{1}}\text{1}\overline{2}\text{, 1}\overline{\text{1}%
}\overline{2}\text{,}\overline{2}\text{1}\overline{\text{1}}\text{,}%
\overline{2}\overline{\text{1}}\text{1)}$ & $\text{E}_{\text{P}}\text{=35/4}%
$\\\hline
$\text{E}_{\Gamma}\text{= 6 \ \ }$ & $\text{3090}$ & 3 & $\text{(}%
\overline{\text{1}}\overline{2}\overline{\text{1}}\text{,}\overline{\text{1}%
}\overline{\text{1}}\overline{2}\text{,}\overline{2}\overline{\text{1}%
}\overline{\text{1}}\text{)}$ & $\text{E}_{\text{P}}\text{=43/4}$\\\hline
$\text{E}_{\text{P}}\text{=27/4}$ & $\text{3360}$ & 1 & $\text{(222)}$ &
$\text{E}_{\Gamma}\text{= 12 \ }$\\\hline
... & ... & ... & ... & ...\\\hline
\end{tabular}

\ \ \ \ \ \ \ \ \ 

\ \ \ \ \ \ \ \ \ \ \ \ \ \ \ \ \ \ \ \ \ \ \ \ \ \ \ \ \ \ \ \ \ \ \ \ \ \ \ \ \ \ \ \ \ \ \ \ \ \ \ \ \ \ \ \ \ \ \ \ \ \ \ \ \ \ \ \ \ \ \ \ \ \ \ \ \ \ \ \ \ \ \ \ \ \ \ \ \ \ \ \ \ \ \ \ \ \ \ \ \ \ \ \ \ \ \ \ \ \ \ \ \ \ \ \ \ \ \ 

\ \ \ \ \ \ \ \ \ \ \qquad\ Table 10. The Energy Bands of the $D$-Axis

$%
\begin{tabular}
[c]{|l|l|l|l|l|}\hline
$\text{E}_{Start}$ & $\varepsilon$ & d & $\text{Energy\ Band (n}_{1}%
\text{n}_{2}\text{n}_{3}\text{, ...)}$ & $\text{E}_{end}$\\\hline
E$_{\text{N}}$\ = 1/2 & 1110 & 2 & (000,110) & E$_{\text{P}}$\ =3/4\\\hline
E$_{\text{P}}$\ = 3/4 & 1200 & 2 & (101,011) & E$_{\text{N}}$\ =3/2\\\hline
E$_{\text{N}}$\ = 3/2 & 1470 & 2 & (10$\overline{1}$, 01$\overline{1}$) &
E$_{\text{P}}$\ =11/4\\\hline
E$_{\text{N}}$\ = 5/2 & 1830 & 4 & (1$\overline{1}$0,$\overline{1}%
$10,020,200) & E$_{\text{P}}$\ =11/4\\\hline
E$_{\text{P}}$\ = 11/4 & 1920 & 4 & ($\overline{1}$01,0$\overline{1}%
$1,211,121) & E$_{\text{N}}$\ =7/2\\\hline
E$_{\text{P}}$\ = 11/4 & 1920 & 2 & (002,112) & E$_{\text{N}}$\ =9/2\\\hline
E$_{\text{N}}$\ = 7/2 & 2190 & 4 & (12$\overline{1}$,21$\overline{1}%
$,$\overline{1}$0$\overline{1}$,0$\overline{1}\overline{1}$) & E$_{\text{P}}%
$\ =19/4\\\hline
E$_{\text{N}}$\ = 9/2 & 2550 & 2 & (220,$\overline{1}\overline{1}$0) &
E$_{\text{P}}$\ =19/4\\\hline
E$_{\text{N}}$\ = 9/2 & 2550 & 2 & (11$\overline{2}$,00$\overline{2}$) &
E$_{\text{P}}$\ =19/4\\\hline
E$_{\text{p}}$\ = 19/4 & 2640 & 2 & ($\overline{1}$21,2$\overline{1}$1) &
E$_{\text{N}}$\ =11/2\\\hline
E$_{\text{P}}$\ = 19/4 & 2640 & 4 & ($\overline{1}$12,1$\overline{1}%
$2,202,022) & E$_{\text{N}}$\ =13/2\\\hline
E$_{\text{N}}$\ = 11/2 & 2910 & 2 & (2$\overline{1}\overline{1}$,$\overline
{1}$2$\overline{1}$) & E$_{\text{P}}$\ =27/4\\\hline
... & ... & ... & ... & ...\\\hline
\end{tabular}
\ \ $

\ \ \ \ \ \ \ \ 

\newpage

\ \ \ \ \ \ \ \ \ \ \ Table 11. The Energy Bands of the $F$-Axis (the P-H axis)

$%
\begin{tabular}
[c]{|l|l|l|l|l|}\hline
\ \ \ $\text{E}_{Start}$ & $\text{\ }{\small \varepsilon}$ & d &
$\text{Energy\ Band (n}_{1}\text{n}_{2}\text{n}_{3}\text{...)}$ &
$\text{E}_{end}$\\\hline
E$_{\text{P}}$=3/4 & 1200 & 3 & (000,011,101) & E$_{\text{H}}$= 1\\\hline
E$_{\text{P}}$=3/4 & 1200 & 1 & (110) & E$_{\text{H}}$= 3\\\hline
E$_{\text{H}}$= 1 & 1290 & 3 & (002,$\overline{1}$01,0$\overline{1}$1) &
E$_{\text{P}}$=11/4\\\hline
E$_{\text{P}}$=11/4 & 1920 & 3 & (112,1$\overline{1}$0,$\overline{1}$10) &
E$_{\text{H}}$=\ \ 3\\\hline
E$_{\text{P}}$=11/4 & 1920 & 6 & (01$\overline{1}$,10$\overline{1}%
$,121,211,020,200)$\ \ $ & E$_{\text{H}}$= 5\\\hline
E$_{\text{H}}$= 3 & 2010 & 3 & ($\overline{1}\overline{1}$0,$\overline{1}%
$12,1$\overline{1}$2) & E$_{\text{P}}$=19/4\\\hline
E$_{\text{H}}$= 3 & 2010 & 1 & ($\overline{1}\overline{1}$2) & E$_{\text{P}}%
$=27/4\\\hline
E$_{\text{P}}$=19/4 & 2640 & 6 & (202,022,$\overline{1}$21,2$\overline{1}%
$1,0$\overline{1}\overline{1}$,$\overline{1}$0$\overline{1}$)$\ $ &
E$_{\text{H}}$= \ 5\\\hline
E$_{\text{P}}$=19/4 & 2640 & 3 & (220,21$\overline{1}$,12$\overline{1}$) &
E$_{\text{H}}$= 9\\\hline
E$_{\text{H}}$= 5 & 2730 & 6 & (0$\overline{2}$0,$\overline{2}$00,$\overline
{2}$11,1$\overline{2}$1,013,103)$\ \ $ & E$_{\text{P}}$=27/4\\\hline
E$_{\text{H}}$= 5 & 2730 & 6 & (0$\overline{2}$2,$\overline{2}$02,$\overline
{2}\overline{1}$1,$\overline{1}\overline{2}$1,0$\overline{1}$3,$\overline{1}%
$03) & E$_{\text{P}}$=35/4\\\hline
... & ... & ... & ... & ...\\\hline
\end{tabular}
\ \ \ \ \ \ $

\ \ \ \ \ \ \ \ \ \ \ \ \ \ \ \ \ \ \ \ \ \ \ \ \ \ \ \ \ \ \ \ \ \ \ \ \ \ \ \ \ \ \ \ \ \ \ \ \ \ \ \ \ \ \ \ \ \ \ 

\ \ \ \ \ \ \ \ \ \ \ \ \ \ \ \ \ \ \ \ \ \ \ \ \ \ \ \ \ \ \ \ \ \ \ \ \ \ \ \ \ \ \ \ \ \ \ \ \ \ \ \ \ \ \ \ \ \ \ \ \ \ \ \ \ \ \ \ \ \ \ \ \ \ \ \ \ \ \ \ \ \ \ \ \ \ \ \ \ 

$\ \ \ \qquad\ \ \ \ \ \ \ \ \ $Table 12. The Energy Bands of the G-Axis

$%
\begin{tabular}
[c]{|l|l|l|l|l|}\hline
$\text{E}_{Start}$ & $\varepsilon$ & d & $\text{Energy\ Band (n}_{1}%
\text{n}_{2}\text{n}_{3}\text{, ...)}$ & E$_{end}$\\\hline
E$_{\text{N}}$=1/2 & 1110 & 2 & (000, 110) & E$_{\text{M}}$= 1\\\hline
E$_{\text{M}}$= 1 & 1290 & 2 & (101, 10$\overline{1}$) & E$_{\text{N}}%
$=3/2\\\hline
E$_{\text{M}}$= 1\  & 1290 & 2 & (200, 1$\overline{1}$0) & E$_{\text{N}}%
$=5/2\\\hline
E$_{\text{N}}$=3/2 & 1470 & 2 & (011, 01$\overline{1}$) & E$_{\text{M}}$=
3\\\hline
E$_{\text{N}}$=5/2 & 1830 & 2 & (020, $\overline{1}$10) & E$_{\text{M}}$=
5\\\hline
E$_{\text{M}}$= 3 & 2010 & 4 & (0$\overline{1}$1, 0$\overline{1}\overline{1}$,
211, 21$\overline{1}$) & E$_{\text{N}}$=7/2\\\hline
E$_{\text{M}}$= 3 & 2010 & 2 & (2$\overline{1}$1, 2$\overline{1}\overline{1}%
$) & E$_{\text{N}}$=11/2\\\hline
E$_{\text{N}}$=7/2 & 2190 & 4 & ($\overline{1}$01,$\overline{1}0\overline{1}$,
121, 12$\overline{1}$) & E$_{\text{M}}$= 5\\\hline
E$_{\text{N}}$=9/2 & 2550 & 6 & (112, 11$\overline{2}$, 002,\ 00$\overline{2}%
$,\ 220,$\overline{1}\overline{1}$0) & E$_{\text{M}}$= 5\\\hline
E$_{\text{M}}$= 5 & 2730 & 6 & (202, 20$\overline{2}$, 1$\overline{1}$2,
1$\overline{1}\overline{2}$, 310, 0$\overline{2}$0) & E$_{\text{N}}%
$=13/2\\\hline
E$_{\text{M}}$= 5 & 2730 & 3 & (301,30$\overline{1}$,1$\overline{2}%
$1,1$\overline{2}\overline{1}$) & E$_{\text{N}}$=15/2\\\hline
E$_{\text{M}}$= 5 & 2730 & 2 & (310.2$\overline{2}$0) & E$_{\text{N}}%
$=17/2\\\hline
E$_{\text{N}}$=11/2 & 2910 & 2 & ($\overline{1}$21,$\overline{1}$%
2$\overline{1}$) & E$_{\text{M}}$= 9\\\hline
... & ... & ... & ... & ...\\\hline
\end{tabular}
\ \ \ \ $

\newpage

\section{Appendix C: The Quarks and the Baryons on the D-axis, the F-axis and
the G-axis}

For the three symmetry axes (the D-axis (P-N), the F-axis (P-H) and the G-axis
(M-N)) that are on the surfaces of the first Brillouin zone (see Fig. 1), the
energy bands with the same energy may have asymmetric $\overrightarrow{n}$
values (see Fig. (3b), (4a) and (4b)). For symmetric $\vec{n}$, we give a
definition:\ a group of $\vec{n}$\ = ($n_{1},n_{2},n_{3}$) values is said to
be symmetric if any two $\vec{n}$\ values in the group can transform into each
other by various permutations (change component order) and by changing the
sign \textquotedblleft$\pm"$\ (multiplied by \textquotedblleft%
-1\textquotedblright\ ) of the components (one, two or three). Otherwise they
are asymmetric. For example, $(-2,-1,3)$ and $(-3,2,1)$ are symmetric;
$(-3,0,2)$ and $(-3,0,1)$ are asymmetric. For these energy bands
(`degeneracy') with the same energy but asymmetric\ $\vec{n}$\ values, if the
`deg '
$>$
the rotary fold ($R)$\ of the symmetry axis,
\begin{equation}
\text{`deg'}>\text{ R,} \label{Deg > R}%
\end{equation}
the `degeneracy' will be divided to $\gamma$-subdegeneracies first
(\textbf{the first kind of division, K = 0}),
\begin{equation}
\gamma\text{ = `deg '/R.} \label{D-S-Group}%
\end{equation}
There is not a change of the strange number and energy for the first kind of
division. Each subgroup of the $\gamma$-subgroups has R energy bands and the
same strange number with the symmetrary axis. If the $\overrightarrow{n}$
values are symmetric, using (\ref{IsoSpin}), we can find the isospin values.
If the $\overrightarrow{n}$ values are asymmetric, the subgroup will be
divided into two sub-subgroups with symmetric (or single) $\overrightarrow{n}$
values (\textbf{the second kind of division, K = 1}) again. Then, using
(\ref{IsoSpin}), we can find the isospin values of the sub-subgroups. For the
each sub-subgroup, using (\ref{S-bar})
\begin{equation}
\text{S = S}_{axis}\text{+ }\Delta\text{S,} \label{S+D-S}%
\end{equation}
we can find the strange numbers. The $\Delta$S can be found using
(\ref{DaltaS}) and (\ref{Sign(n)}).\ If Sign($\overrightarrow{n}$) = 0%

\begin{equation}
\Delta S=(-1)^{S_{Axis}}\text{.} \label{dalta(S)}%
\end{equation}
\ 

The fluctuation of the strange number will be accompanied by an energy change
(Hypothesis II). We assume that the change of the energy (perturbation energy)
is proportional to $\Delta$S\ and a number, J,\ representing the energy level
with an asymmetric $\overrightarrow{n}$ values, as a phenomenological
formula:
\begin{equation}
\Delta\varepsilon\text{=(-1)}^{\text{S}}\text{200(J-2-K)}\Delta
\text{S,\ \ \ J=\ (R-SK-2)+(1,2,3,...) } \label{K+S}%
\end{equation}
where K\ is the division number of the energy bands and R is the symmetric
rotation number of the symmetry axis. For a single energy band, J will take 1,
2, and so forth from the lowest energy band to higher ones for each of the two
end points of the axes respectively.

There are three energy bands ($\overrightarrow{n}$ = (000), $\overrightarrow
{n}$ = (100) and $\overrightarrow{n}$ = (200))that have already been
recognized inside the Brillouin zones. The bands ($\overrightarrow{n}$ =
(000), $\overrightarrow{n}$ = (100) and $\overrightarrow{n}$ = (200)) on the
surfaces of the Brillouin zones are the same quarks:
\begin{equation}%
\begin{tabular}
[c]{|l|l|l|l|}\hline
$\overrightarrow{n}$ & {\small Bands (Inside Brillouin zone)} & {\small Bands
(on Surface )} & {\small Quark}\\\hline
{\small (000)} & {\small E}$_{\Gamma}${\small (0)}$\rightarrow${\small E}%
$_{N}${\small (}$\frac{1}{2}${\small ), E}$_{p}${\small (}$\frac{3}{4}%
${\small ), E}$_{H}${\small (1)} & {\small E}$_{M}${\small (1)}$\leftarrow
${\small E}$_{N}${\small (}$\frac{1}{2}${\small )}$\rightarrow${\small E}%
$_{p}${\small (}$\frac{3}{4}${\small )}$\rightarrow${\small E}$_{H}%
${\small (1)} & {\small q}$_{N}${\small (930)}\\\hline
{\small (100)} & {\small E}$_{N}${\small (}$\frac{1}{2}${\small )}%
$\rightarrow${\small E}$_{\Gamma}${\small (2)} & {\small E}$_{M}$%
{\small (1)}$\leftarrow${\small E}$_{N}${\small (}$\frac{1}{2}${\small )}%
$\rightarrow${\small E}$_{p}${\small (}$\frac{3}{4}${\small )}$\rightarrow
${\small E}$_{H}${\small (3)} & {\small d}$_{S}${\small (1110)}\\\hline
{\small (200)} & {\small E}$_{H}${\small (1)}$\rightarrow${\small E}$_{\Gamma
}${\small (2)} & {\small E}$_{M}${\small (1)}$\rightarrow${\small E}$_{N}%
${\small (}$\frac{5}{2}${\small ), E}$_{H}${\small (1)}$\rightarrow$%
{\small E}$_{p}${\small (}$\frac{3}{4}${\small )} & {\small d}$_{S}%
${\small (1390)}\\\hline
\end{tabular}
\ \ \label{3Band}%
\end{equation}
\ 

\subsection{The Axis D(P-N)}

From (\ref{S(D)}), the D-axis has S = 0. For low energy level, there are
four-fold degenerate energy bands and two-fold degenerate energy bands on the
axis (see Fig. (3b)).

\subsubsection{The Quarks and the Baryons on the Four-fold Energy Bands}

We can see that each four-fold degenerate energy band has four symmetric
$\overrightarrow{\text{n}}$ values. They can be divided into two groups
(\ref{Subdeg}). Each of them has two symmetric $\overrightarrow{\text{n}}$
values. Using (\ref{IsoSpin}), for the two-fold degenerate energy bands, we
get I = 1/2; I$_{\text{Z,Baryon}}$ = 1/2, -1/2; Q = 2/3, -1/3 from\ (\ref{2/3}%
) and (\ref{-1/3}). Thus, for the four-fold degenerate energy bands, we have
{\small q}$_{\text{N}}$(m)=[u$_{N}^{\frac{1}{2}}$(m), d$_{N}^{\frac{-1}{2}}%
$(m)]. Using (\ref{B-Comp}), for a {\small q}$_{\text{N}}$(m), we have a
N(m+m$_{q_{1}^{\prime}}$+m$_{q_{2}^{\prime}}$) and a $\Delta$(m+m$_{q_{1}%
^{\prime}}$+m$_{q_{2}^{\prime}}$) from (\ref{I-T-HMP}) at the point P; we have a 
N(m+m$_{q_{1}^{\prime}}$+m$_{q_{2}^{\prime}}$) only from (\ref{I-T-N}) at the point N:

\begin{equation}%
\begin{tabular}
[c]{lllll}%
{\small E}$_{\text{N}}${\small \ = 5/2} & {\small (1}$\overline{1}$%
{\small 0,}$\overline{1}${\small 10,020,200)} & {\small 1830} & {\small J}%
$_{\text{N}}${\small \ = 2} & {\small K = 0}\\
{\small \ }$\ \Delta${\small S = 0} & {\small (1}$\overline{1}${\small 0,}%
$\overline{1}${\small 10)} & {\small q}$_{\text{N}}${\small (1830)} &
{\small N(1840)} & \\
{\small \ \ }$\Delta${\small S = 0} & {\small (020,200)} & {\small q}%
$_{\text{N}}${\small (1830)} & {\small N(1840)} & \\
{\small E}$_{\text{P}}${\small \ = 11/4} & {\small (}$\overline{1}%
${\small 01,0}$\overline{1}${\small 1,211,121)} & {\small 1920} &
{\small J}$_{\text{P}}$ = 1 & {\small K = 0}\\
{\small \ \ }$\Delta${\small S = 0} & {\small (}$\overline{1}${\small 01,0}%
$\overline{1}${\small 1)} & {\small q}$_{\text{N}}${\small (1920)} &
{\small N(1930)} & $\Delta${\small (1930)}\\
{\small \ \ }$\Delta${\small S = 0} & {\small (211,121)} & {\small q}%
$_{\text{N}}${\small (1920)} & {\small N(1930)} & $\Delta${\small (1930)}\\
{\small E}$_{\text{N}}${\small \ = 7/2} & {\small (12}$\overline{1}%
${\small ,21}$\overline{1}${\small ,}$\overline{1}${\small 0}$\overline{1}%
${\small ,0}$\overline{1}\overline{1}${\small )} & {\small 2190} &
{\small J}$_{\text{N}}${\small \ = 3} & {\small K = 0}\\
{\small \ \ }$\Delta${\small S = 0} & {\small (12}$\overline{1}$%
{\small ,21}$\overline{1}${\small )} & {\small q}$_{\text{N}}${\small (2190)}
& {\small N(2200)} & \\
{\small \ \ }$\Delta${\small S = 0} & {\small (}$\overline{1}${\small 0}%
$\overline{1}${\small ,0}$\overline{1}\overline{1}${\small )} & {\small q}%
$_{\text{N}}${\small (2190)} & {\small N(2200)} & \\
{\small E}$_{\text{P}}${\small \ = 19/4} & {\small (}$\overline{1}%
${\small 12,1}$\overline{1}${\small 2,202,022)} & {\small 2640} &
{\small J}$_{\text{P}}$ = 2 & {\small K = 0}\\
{\small \ \ }$\Delta${\small S = 0} & {\small (}$\overline{1}${\small 12,1}%
$\overline{1}${\small 2)} & {\small q}$_{\text{N}}${\small (2640)} &
{\small N(2650)} & $\Delta${\small (2650)}\\
{\small \ \ }$\Delta${\small S = 0} & {\small (202,022)} & {\small q}%
$_{\text{N}}${\small (2640)} & {\small N(2650)} & $\Delta${\small (2650)}\\
... & ... & ... & ... & ...
\end{tabular}
\label{4-D}%
\end{equation}

\subsubsection{The Quarks and Baryons on the Two-fold Energy Bands}

From (\ref{n=000}), (\ref{D-S-D}), (\ref{dalta(S)}) and (\ref{K+S}), we have
($\Delta\varepsilon$ = 200(J-2)$\Delta S$ \ J = 1, 2, 3, ...)$:$%

\begin{equation}%
\begin{tabular}
[c]{llllll}%
{\small E}$_{\text{N}}${\small \ = 1/2} & {\small (000,110)} & {\small 1110} &
{\small \ J}$_{\text{N}}${\small = 1} & {\small K=0} & \\
{\small \ } & {\small (000)} from (\ref{3Band}) &  & {\small q}$_{\text{N}}%
${\small (930)} & {\small N(940)} & \\
{\small \ \ }$\Delta${\small S= -1} & {\small (110)} from (\ref{3Band}) &
{\small 1110} & {\small q}$_{\text{S}}${\small (1110)} & $\Lambda
${\small (1120)} & \\
{\small E}$_{\text{P}}${\small \ = 3/4} & {\small (101,011)} & {\small 1200} &
{\small q}$_{\text{N}}${\small (1200)} & {\small N(1210)} & $\Delta(1210)$\\
{\small E}$_{\text{N}}${\small \ = 3/2} & {\small (10}$\overline{1}${\small ,
01}$\overline{1}${\small )} & {\small 1470} & {\small q}$_{\text{N}}%
${\small (1470)} & {\small N(1480)} & \\
{\small E}$_{\text{P}}${\small \ = 11/4} & {\small (002,112)} & {\small 1920}
& {\small J}$_{\text{P}}${\small = 1} & {\small K = 0} & \\
{\small \ \ }$\Delta${\small S= -1} & {\small (002)} & {\small 2120} &
{\small q}$_{\text{S}}${\small (2120)} & $\Lambda${\small (2130)} & $\Sigma
${\small (2130)}\\
{\small \ \ }$\Delta${\small S= -1} & {\small (112)} & {\small 2120} &
{\small q}$_{\text{S}}${\small (2120)} & $\Lambda${\small (2130)} & $\Sigma
${\small (2130)}\\
{\small E}$_{\text{N}}${\small \ = 9/2}$^{1}$ & {\small (220,}$\overline
{1}\overline{1}${\small 0)} & {\small 2550} & {\small J}$_{N}${\small = 2} &
{\small K = 0} & \\
$\ \ \Delta${\small S = +1} & {\small (}$\overline{1}\overline{1}$%
{\small 0)}$\ ${\small \ }$\ $ & {\small 2550} & {\small q}$_{\text{C}}%
${\small (2550)} & $\Lambda_{c}${\small (2560)} & \\
$\ \ \Delta${\small S = -1} & {\small (220)} & {\small 2550} & {\small q}%
$_{\text{S}}${\small (2550)} & $\Lambda${\small (2560)} & \\
{\small E}$_{\text{N}}${\small \ = 9/2}$^{2}$ & {\small (11}$\overline{2}%
${\small ,00}$\overline{2}${\small )} & {\small 2550} & {\small \ J}$_{N}%
${\small = 3} & {\small K = 0} & \\
$\ \ \Delta${\small S = +1} & {\small (00}$\overline{2}${\small )\ \ \ } &
{\small 2750} & {\small q}$_{\text{C}}${\small (2750)} & $\Lambda_{c}%
${\small (2760)} & \\
{\small \ }$\ \Delta${\small S = +1} & {\small (11}$\overline{2}${\small )} &
{\small 2750} & {\small q}$_{\text{C}}${\small (2750)} & $\Lambda_{c}%
${\small (2760)} & \\
{\small E}$_{\text{p}}${\small \ = 19/4} & {\small (}$\overline{1}%
${\small 21,2}$\overline{1}${\small 1)} & {\small 2640} & {\small q}%
$_{\text{N}}${\small (2640)} & {\small N(2650)} & $\Delta${\small (2650)}\\
{\small E}$_{\text{N}}${\small \ = 11/2} & {\small (2}$\overline{1}%
\overline{1}${\small ,}$\overline{1}${\small 2}$\overline{1}${\small )} &
{\small 2910} & {\small q}$_{\text{N}}${\small (2910)} & {\small N(2920)} & \\
{\small E}$_{\text{p}}${\small \ = 27/4} & {\small (103,013)} & {\small 3360}
& {\small q}$_{\text{N}}${\small (3360)} & {\small N(3370)} & $\Delta
${\small (3370)}%
\end{tabular}
\label{2-D}%
\end{equation}

\subsection{The Axis F(P-H) \ \ }

The axis is a three-fold symmetric axis, S = -1 from (\ref{S(F)}). There are
six-fold energy bands, three-fold energy bands and single energy bands on the
axis (see Fig. 4(a)). Using Fig. 4(a), we get:\qquad

\subsubsection{The Quarks and Baryons on the Single Energy Bands}

For the single bands, the strange number S = -1, the isospin I = 0 from
(\ref{IsoSpin}), and electric charge Q = -1/3 from (\ref{-1/3}). Each single
energy band represents an excited quark q$_{S}$ (d$_{S}^{0}$) with S = -1, I =
0, and Q = -1/3:
\begin{equation}%
\begin{tabular}
[c]{lllll}%
{\small E}$_{\text{P}}${\small =3/4} & $\overline{n}${\small =(110)
}(\ref{3Band}) & \textit{E}{\small =1110} & {\small d}$_{S}^{0}$%
{\small (1110)} & $\Lambda${\small (1120)}\\
{\small E}$_{\text{H}}${\small = 3} & $\overline{n}${\small =(}$\overline
{1}\overline{1}${\small 2)} & \textit{E}{\small =2010} & {\small d}$_{S}^{0}%
${\small (2010)} & $\Lambda${\small (2020)}%
\end{tabular}
\label{Sing-F}%
\end{equation}

\subsubsection{The Quarks and Baryons on the Three-fold Energy Bands}

From (\ref{n=000}), (\ref{D-S-S}), (\ref{dalta(S)}) and (\ref{K+S}), we have
($\Delta\varepsilon$ = -200(J-2)$\Delta$S\ \ \qquad J =\ 2, 3, 4, ...):%

\begin{equation}%
\begin{tabular}
[c]{llllll}%
E$_{\text{P}}$=3/4 & (000,011, 101) & 1200 & J$_{\text{P}}$ = 1 & {\small K =
0} & \\
\  & (000) from (\ref{n=000}) &  & q$_{\text{N}}$(930) & N(940) & \\
\ \ $\Delta S$=+1 & (011, 101) & 1400 & q$_{\text{N}}$(1200) & N(1210) &
$\Delta$(1210)\\
E$_{\text{H}}$= 1 & (002,$\overline{1}$01,0$\overline{1}$1) & 1290 &
J$_{\text{H}}$ = 1 & {\small K = 0} & \\
\ $\ \Delta$S= 0 & (002) from (\ref{3Band}) & 1390 & q$_{\text{S}}$(1390) &
$\Lambda$(1400) & $\Sigma$(1400)\\
$\ \Delta$S=-1 & ($\overline{1}$01,0$\overline{1}$1) & 1290 & q$_{\Xi}%
$(1290) & $\Xi$(1300) & \\
E$_{\text{P}}$=11/4 & (112,1$\overline{1}$0,$\overline{1}$10) & 1920 &
\ J$_{\text{P}}$ = 2 & {\small K = 0} & \\
\ \ $\Delta$S= 0 & (112) & 1920 & q$_{\text{S}}$(1920) & $\Lambda$(1930) &
$\Sigma$(1930)\\
$\ \ \Delta S$=-1 & (1$\overline{1}$0,$\overline{1}$10) & 1920 & q$_{\Xi}%
$(1920) & $\Xi$(1930) & \\
E$_{\text{H}}$= 3 & ($\overline{1}\overline{1}$0,$\overline{1}$12,1$\overline
{1}$2) & 2010 & \ J$_{\text{H}}$ = 2 & {\small K = 0} & \\
$\ \Delta$S= 0 & ($\overline{1}\overline{1}$0) & 2010 & q$_{\text{S}}$(2010) &
$\Lambda$(2020) & $\Sigma$(2020)\\
$\ \Delta$S=+1 & (21$\overline{1}$,12$\overline{1}$) & 2010 & q$_{\text{N}}%
$(2010) & N(2020) & $\Delta$(2020)\\
E$_{\text{P}}$=19/4 & (220,21$\overline{1}$,12$\overline{1}$) & 2640 &
\ J$_{\text{P}}$ = 3 & {\small K = 0} & \\
$\ \Delta$S= 0 & (220) & 2640 & q$_{\text{S}}$(2640) & $\Lambda$(2650) &
$\Sigma$(2650)\\
$\ \Delta$S=+1 & (21$\overline{1}$,12$\overline{1}$) & 2440 & q$_{\Xi_{C}}%
$(2440) & $\Xi_{C}$(2450) & \\
E$_{\text{P}}$=27/4 & (130,310,11$\overline{2}$) & 3360 & J$_{\text{P}}$ = 4 &
K = 0 & \\
$\Delta$S= 0 & (11$\overline{2}$) & 3360 & q$_{\text{S}}$(3360) & $\Lambda
$(3370) & $\Sigma$(2650)\\
$\Delta$S=+1 & (130,310) & 2960 & q$_{\Xi_{C}}$(2960) & $\Xi_{C}$(2970) & \\
&  &  &  &  & \\
... . &  &  &  &  &
\end{tabular}
\label{3-F}%
\end{equation}

\subsubsection{The Quarks and Baryons on the Six-fold Energy Bands}

From (\ref{D-S-Group}), (\ref{D-S-S}), (\ref{dalta(S)}) and (\ref{K+S}), we
have ($\Delta\varepsilon$ = -200(J-3)$\Delta$S\ \ \ J =\ 3,4,5,...) $\qquad$%

\begin{equation}%
\begin{tabular}
[c]{llllll}%
{\small E}$_{\text{P}}${\small =11/4} & $%
\begin{tabular}
[c]{l}%
(01$\overline{1}$,10$\overline{1}$,121,\\
211, 020, 200)
\end{tabular}
\ \ $ & {\small 1920} & {\small \ \ J}$_{\text{P}}${\small = 2} &
{\small \ \ K=0} & \\
$\Delta${\small S= 0\ } & {\small (01}$\overline{1}${\small ,10}$\overline{1}%
${\small ,121)} & {\small 1920} & {\small \ J}$_{\text{P}}${\small = 2} &
{\small \ K=1} & \\
{\small \ \ }$\Delta${\small S= 0\ } & {\small (121)} & {\small 1920} &
{\small q}$_{S}${\small (1920)} & $\Lambda${\small (1930)} & $\Sigma
${\small (1930)}\\
$\ \Delta${\small S= -1} & {\small (01}$\overline{1}${\small ,10}$\overline
{1}${\small )} & {\small 1920} & {\small q}$_{\Xi}${\small (1920)} & $\Xi
$(1930) & \\
$\Delta${\small S= 0} & {\small (211, 020, 200)} & {\small 1920} &
{\small J}$_{\text{P}}${\small = 2} & {\small \ K=1} & \\
{\small \ \ }$\Delta${\small S= 0} & {\small (211)} & {\small 1920} &
{\small q}$_{S}${\small (1920)} & $\Lambda${\small (1930)} & $\Sigma
${\small (1930)}\\
{\small \ }$\ \Delta${\small S= 1} & {\small (020, 200)} & {\small 1920} &
{\small q}$_{N}${\small (1920)} & {\small N(1930)} & $\Delta${\small (1930)}\\
{\small E}$_{\text{P}}${\small =19/4} & $%
\begin{tabular}
[c]{l}%
(202,022,$\overline{1}$21,\\
2$\overline{1}$1,0$\overline{1}\overline{1}$,$\overline{1}$0$\overline{1}$)
\end{tabular}
\ \ $ & {\small 2640} & {\small \ \ J}$_{\text{P}}${\small = 3} &
{\small \ \ K=0} & \\
$\Delta${\small S= 0} & {\small (202,022,}$\overline{1}${\small 21)} &
{\small 2640} & {\small \ \ J}$_{\text{P}}${\small = 3} & {\small \ \ K=1} &
\\
{\small \ \ }$\Delta${\small S=+1} & {\small (202,022)} & {\small 2640} &
{\small q}$_{\Xi_{C}}${\small (2640)} & {\small q}$_{\Xi_{C}}${\small (2650)}
& \\
{\small \ }$\ \Delta${\small S= 0} & {\small (}$\overline{1}${\small 21)} &
{\small 2640} & {\small q}$_{S}${\small (2640)} & $\Lambda${\small (2650)} &
$\Sigma${\small (2650)}\\
$\Delta${\small S= 0} & {\small (2}$\overline{1}${\small 1,0}$\overline
{1}\overline{1}${\small ,}$\overline{1}${\small 0}$\overline{1}${\small )} &
{\small 2640} & {\small \ J}$_{\text{P}}${\small = 3} & {\small \ \ K=1} & \\
{\small \ }$\ \Delta${\small S=-1} & {\small (0}$\overline{1}\overline{1}%
${\small ,}$\overline{1}${\small 0}$\overline{1}${\small )} & {\small 2640} &
{\small q}$_{\Xi}${\small (2640)} & $\Xi${\small (2650)} & \\
{\small \ }$\ \Delta${\small S= 0} & {\small (2}$\overline{1}${\small 1)} &
{\small 2640} & {\small q}$_{\text{S}}${\small (2640)} & $\Lambda
${\small (2650)} & $\Sigma${\small (2650)}\\
{\small E}$_{\text{H}}${\small = 5} & $%
\begin{tabular}
[c]{l}%
(0$\overline{2}$0,$\overline{2}$00,$\overline{2}$11,\\
1$\overline{2}$1,013,103)
\end{tabular}
\ \ $ & {\small 2730} & {\small \ \ J}$_{\text{H}}${\small = 3} &
{\small \ \ K=0} & \\
$\Delta${\small S= 0} & {\small (0}$\overline{2}${\small 0,}$\overline{2}%
${\small 00,}$\overline{2}${\small 11)} & {\small 2730} & {\small \ \ J}%
$_{\text{H}}${\small = 3} & {\small \ K=1} & \\
{\small \ }$\Delta${\small S= 0} & {\small (}$\overline{2}${\small 11)} &
{\small 2730} & {\small q}$_{S}${\small (2730)} & $\Lambda${\small (2740)} &
$\Sigma${\small (2740)}\\
{\small \ }$\Delta${\small S= -1} & {\small (0}$\overline{2}${\small 0,}%
$\overline{2}${\small 00)} & {\small 2730} & {\small q}$_{\Xi}${\small (2730)}
& $\Xi${\small (2740)} & \\
$\Delta${\small S= 0} & {\small (1}$\overline{2}${\small 1,013,103)} &
{\small 2730} & {\small \ \ J}$_{\text{H}}${\small = 3} & {\small \ K=1} & \\
{\small \ }$\Delta${\small S= 0} & {\small (1}$\overline{2}${\small 1)} &
{\small 2730} & {\small q}$_{S}${\small (2730)} & $\Lambda${\small (2740)} &
$\Sigma${\small (2740)}\\
{\small \ }$\Delta${\small S=+1} & {\small (013,103)} & {\small 2730} &
{\small q}$_{\Xi_{C}}${\small (2730)} & $\Xi_{C}${\small (2730)} & \\
{\small E}$_{\text{H}}${\small = 5}$^{2}$ & $%
\begin{tabular}
[c]{l}%
(0$\overline{2}$2,$\overline{2}$02,$\overline{2}\overline{1}$1,\\
$\overline{1}\overline{2}$1,0$\overline{1}$3,$\overline{1}$03)
\end{tabular}
\ \ $ & {\small 2730} & {\small \ \ J}$_{\text{H}}${\small = 4} &
{\small \ K=0} & \\
$\Delta${\small S= 0} & {\small (0}$\overline{2}${\small 2,}$\overline{2}%
${\small 02,}$\overline{2}\overline{1}${\small 1)} & {\small 2730} &
{\small \ \ J}$_{\text{H}}${\small = 4} & {\small \ \ K=1} & \\
$\ \Delta${\small S= -1} & {\small (0}$\overline{2}${\small 2,}$\overline{2}%
${\small 02)} & {\small 2930} & {\small q}$_{\Xi}${\small (2930)} & $\Xi
${\small (2940)} & \\
$\Delta${\small S= \ 0} & {\small (}$\overline{2}\overline{1}${\small 1)} &
{\small 2730} & {\small q}$_{S}${\small (2730)} & $\Lambda${\small (2740)} &
$\Sigma${\small (2740)}\\
$\Delta${\small S= \ 0} & {\small (}$\overline{1}\overline{2}${\small 1,0}%
$\overline{1}${\small 3,}$\overline{1}${\small 03)} & {\small 2730} &
{\small \ \ J}$_{\text{H}}${\small = 3} & {\small \ \ K=1} & \\
$\ \Delta${\small S= +1} & {\small (0}$\overline{1}${\small 3,}$\overline{1}%
${\small 03)} & {\small 2530} & {\small q}$_{\Xi_{C}}${\small (2530)} &
$\Xi_{C}${\small (2540)} & \\
{\small \ }$\Delta${\small S= \ 0} & {\small (}$\overline{1}\overline{2}%
${\small 1)} & {\small 2730} & {\small q}$_{S}${\small (2730)} & $\Lambda
${\small (2740)} & $\Sigma${\small (2740)}%
\end{tabular}
\ \label{6-F}%
\end{equation}

\ \ \ \ \ \ \ \ \ \ \ \ \ \ \ \ \ \ \ \ \ \ \ \ \ \ \ \ \ \ \ \ \ \ \ \ \ \ \ \ \ \ \ \ \ 

\subsection{The Axis G(M-N) \ \ }

The axis G(M-N) is a two-fold symmetric axis, S = -2 from (\ref{S(G)}). There
are two-fold, four-fold and six-fold energy bands on the axis (see Fig. 4(b)).
Using Fig. 4(b), we get:

\subsubsection{The Quarks and Baryons on the Two-fold Energy Bands}

From (\ref{D-S-Group}), (\ref{D-S-S}), (\ref{dalta(S)}) and (\ref{K+S}), we
have ($\Delta\varepsilon$ = 200(J-2)$\Delta$S, \ J = \ 1, 2, 3, 4,
...):$\qquad$%

\begin{equation}%
\begin{tabular}
[c]{llllll}%
E$_{\text{N}}$=1/2 & (000,110) &  & J$_{\text{N}}$ =1 & {\small K = 0} & \\
& (000) from (\ref{3Band}) & 930 &  & q$_{\text{N}}$(930) & N(940)\\
\ $\ \Delta$S=+1 & (110) from (\ref{3Band}) & 1110 &  & q$_{S}$(1110) & $\Lambda
$(1120)\\
E$_{\text{M}}$= 1 & (101,10$\overline{1}$) & 1290 &  & q$_{\Xi}$(1290) & $\Xi
$(1300)\\
E$_{\text{M}}$= 1\  & (200,1$\overline{1}$0) & 1290 & J$_{\text{M}}$ =1 &
K=0 & \\
$\ \Delta$S=+1 & (200) from (\ref{3Band}) & 1390 & J$_{\text{M}}$ =1 & q$_{S}%
$(1390) & $\Lambda$(1400)\\
$\ \Delta$S=+1 & (1$\overline{1}$0) & 1490 & J$_{\text{M}}$ =1 & q$_{S}%
$(1490) & $\Lambda$(1500)\\
E$_{\text{N}}$=3/2 & (011,01$\overline{1}$) & 1470 &  & q$_{\Xi}$(1470) &
$\Xi$(1480)\\
E$_{\text{N}}$=5/2 & (020,$\overline{1}$10) & 1830 & J$_{\text{N}}$ =2 & K=0 &
\\
$\ \Delta$S=+1 & (020) & 1830 & J$_{\text{N}}$ =2 & q$_{S}$(1830) & $\Lambda
$(1840)\\
$\ \Delta$S=+1 & ($\overline{1}$, 1, 0) & 1830 & J$_{\text{N}}$ =2 & q$_{S}%
$(1830) & $\Lambda$(1840)\\
E$_{\text{M}}$= 3 & (2$\overline{1}$1.2$\overline{1}\overline{1}$) & 2010 &
J$_{\text{M}}$ =2 & q$_{\Xi}$(2010) & $\Xi$(2020)\\
E$_{\text{M}}$= 5 & (310.2$\overline{1}$0) & 2730 & J$_{\text{M}}$ =3 & K=0 &
\\
$\ \Delta$S=+1 & (310) & 2930 & J$_{\text{M}}$ =3 & q$_{S}$(2930) & $\Lambda
$(2940)\\
$\ \Delta$S=+1 & (2$\overline{1}$0) & 2930 & J$_{\text{M}}$ =3 & q$_{S}%
$(2930) & $\Lambda$(2940)\\
E$_{\text{N}}$=11/2 & ($\overline{1}$21,$\overline{1}$2$\overline{1}$) &
2910 &  & q$_{\Xi}$(2910) & $\Xi$(2920)\\
... & ... & ... & ... & ... & ....
\end{tabular}
\label{2-G}%
\end{equation}

\subsubsection{The Quarks and Baryons on the Four-fold Energy Bands}

\ \ From (\ref{D-S-Group}), (\ref{D-S-S}), (\ref{dalta(S)}) and (\ref{K+S}),
we have:\ \ \ \ \ \ \ \ \ \ \ \ \ \ \ \ %

\begin{equation}%
\begin{tabular}
[c]{lllll}%
E$_{\text{M}}$= 3 & {\small (0}$\overline{1}${\small 1,0}$\overline
{1}\overline{1}${\small ,211,21}$\overline{1}${\small )} & {\small 2010} &
J$_{\text{M}}$ = 2 & {\small K = 0}\\
{\small \ }$\Delta${\small S=0} & {\small \ \ \ \ (0}$\overline{1}%
${\small 1,0}$\overline{1}\overline{1}${\small )} & {\small 2010} &
{\small q}$_{\Xi}${\small (2010)} & $\Xi(2020)$\\
$\ \Delta${\small S=0} & {\small \ \ \ \ (211,21}$\overline{1}${\small )} &
{\small 2010} & {\small q}$_{\Xi}${\small (2010)} & $\Xi(2020)$\\
E$_{\text{N}}$=7/2 & {\small (}$\overline{1}${\small 01,}$\overline
{1}0\overline{1}${\small , 121, 12}$\overline{1}${\small )} & {\small 2190} &
J$_{\text{N}}$ = 3 & {\small K = 0}\\
$\ \Delta${\small S=0} & {\small \ \ \ \ (}$\overline{1}${\small 01,}%
$\overline{1}0\overline{1}$ & {\small 2190} & {\small q}$_{\Xi}$%
{\small (2190)} & $\Xi(2200)$\\
$\ \Delta${\small S=0} & {\small \ \ \ \ (121, 12}$\overline{1}${\small )} &
{\small 2190} & {\small q}$_{\Xi}${\small (2190)} & $\Xi(2200)$\\
E$_{\text{M}}$= 5 & {\small (301,30}$\overline{1}${\small ,1}$\overline{2}%
${\small 1,1}$\overline{2}\overline{1}${\small )} & {\small 2730} &
J$_{\text{M}}$ = 3 & {\small K = 0}\\
{\small \ }$\Delta${\small S=0} & {\small \ \ \ \ (301,30}$\overline{1}%
${\small )} & {\small 2730} & {\small q}$_{\Xi}${\small (2730)} & $\Xi
(2740)$\\
{\small \ }$\Delta${\small S=0} & {\small \ \ \ \ (1}$\overline{2}%
${\small 1,1}$\overline{2}\overline{1}${\small )} & {\small 2730} &
{\small q}$_{\Xi}${\small (2730)} & $\Xi(2740)$\\
... & ... & ... & ... & ....
\end{tabular}
\label{4-G}%
\end{equation}
\ \ \ \ \ \ \ \ \ \ \ \ \ \ \ \ \ \ \ \ \ \ \ \ \ \ \ \ \ \ 

\subsubsection{The Quarks and Baryons on the Six-fold Energy Bands}

From (\ref{D-S-Group}), , (\ref{D-S-S}), (\ref{dalta(S)}) and
(\ref{K+S}), we have:%

\begin{equation}%
\begin{tabular}
[c]{lllll}%
{\small E}$_{\text{N}}${\small =9/2} & $\left[
\begin{tabular}
[c]{l}%
{\small 112,11}$\overline{2}${\small ,002,00}$\overline{2}${\small ,}\\
{\small \ \ \ \ \ \ \ \ 220, }$\overline{1}\overline{1}${\small 0}%
\end{tabular}
\ \ \right]  $ & {\small 2550} & {\small \ \ J}$_{\text{N}}${\small \ =4} &
{\small \ \ K = 0}\\
{\small \ }$\Delta${\small S= 0} & {\small (112,11}$\overline{2}${\small )} &
{\small 2550} & {\small q}$_{\Xi}${\small (2550)} & $\Xi${\small (2560)}\\
{\small \ }$\Delta${\small S= 0} & {\small (002,00}$\overline{2}${\small )} &
{\small 2550} & {\small q}$_{\Xi}${\small (2550)} & $\Xi${\small (2560)}\\
$\Delta${\small S= 0} & {\small (220,}$\overline{1}\overline{1}${\small 0)} &
{\small 2550} & {\small \ \ J}$_{\text{N}}${\small \ =4} & {\small \ \ K =
1}\\
{\small \ \ }$\Delta${\small S=+1} & {\small (220)} & {\small 2750} &
{\small q}$_{\Omega_{C}}${\small (2750)} & $\Omega_{C}${\small (2760)}\\
{\small \ \ }$\Delta${\small S=-1} & {\small (}$\overline{1}\overline{1}%
${\small 0)\ } & {\small 2350} & {\small q}$_{\Omega}${\small (2350)} &
$\Omega${\small (2360)}\\
{\small E}$_{\text{M}}${\small = 5} & $\left[
\begin{tabular}
[c]{l}%
{\small 202,20}$\overline{2}${\small ,1}$\overline{1}${\small 2,1}%
$\overline{1}\overline{2}${\small ,}\\
{\small \ \ \ \ \ \ \ \ 310, 0}$\overline{2}${\small 0}%
\end{tabular}
\ \ \right]  $ & {\small 2730} & {\small \ \ J}$_{\text{M}}${\small \ =3} &
{\small \ K = 0}\\
{\small \ }$\Delta${\small S= \ 0} & {\small (1}$\overline{1}${\small 2,1}%
$\overline{1}\overline{2})$ & {\small 2730} & {\small q}$_{\Xi}$%
{\small (2730)} & $\Xi${\small (2740)}\\
{\small \ }$\Delta${\small S= \ 0} & {\small (202,20}$\overline{2}${\small )}
& {\small 2730} & {\small q}$_{\Xi}${\small (2730)} & $\Xi${\small (2740)}\\
$\Delta${\small S= \ 0} & {\small (310.0}$\overline{2}${\small 0)} &
{\small 2730} & {\small \ \ J}$_{\text{M}}${\small \ =3} & {\small K = 1}\\
{\small \ \ \ }$\Delta${\small S= +1} & {\small (310)} & {\small 2730} &
{\small q}$_{\Omega_{C}}${\small (2730)} & $\Omega_{C}${\small (2740)}\\
{\small \ \ \ }$\Delta${\small S= -1} & {\small \ (0}$\overline{2}${\small 0)}
& {\small 2730} & {\small q}$_{\Omega}${\small (2730)} & $\Omega
${\small (2740)}\\
{\small E}$_{\text{N}}${\small =13/2} & $\left[
\begin{tabular}
[c]{l}%
{\small 112,112,022,02}$\overline{2}${\small ,}\\
{\small \ \ \ \ \ \ 130,}$\overline{2}${\small 00}%
\end{tabular}
\ \ \ \right]  $ & {\small 3270} & {\small \ J}$_{\text{N}}${\small \ = 5} &
{\small \ \ K = 0}\\
$\Delta${\small S= \ 0} & {\small (112,112)} & {\small 3270} & {\small q}%
$_{\Xi}${\small (3270)} & $\Xi${\small (3280)}\\
$\Delta${\small S= \ 0} & {\small (022,02}$\overline{2})$ & {\small 3270} &
{\small q}$_{\Xi}${\small (3270)} & {\small q}$_{\Xi}^{\ast}${\small (3280)}\\
$\Delta${\small S= \ 0} & ({\small 130,}$\overline{2}${\small 00)} &
{\small 3270} & {\small \ \ J}$_{\text{N}}${\small \ = 5} & \ \ {\small K =
1}\\
$\ \ \Delta${\small S= +1} & ({\small 130)} & {\small 3670} & {\small q}%
$_{\Omega_{C}}${\small (3670)} & $\Omega_{C}${\small (3680)}\\
$\ \ \Delta${\small S= -1} & ($\overline{2}${\small 00)} & {\small 2870} &
{\small q}$_{\Omega}${\small (2870)} & $\Omega${\small (2880)}\\
{\small ...} & {\small ...} & {\small ...} & {\small ...} & {\small ....}%
\end{tabular}
\label{6-G}%
\end{equation}

\end{document}